\documentclass[fleqn,usenatbib]{mnras}

\usepackage{newtxtext,newtxmath}
\usepackage[T1]{fontenc}
\usepackage{ae,aecompl}
\usepackage{times}
\usepackage{times}
\usepackage{amsmath}
\usepackage{amssymb}
\usepackage{textcomp}
\usepackage{verbatim}
\usepackage{esdiff}

\newcommand{\msun}{{\,\rm M_\odot}}
\newcommand{\Lsun}{{\,\rm L_\odot}}
\newcommand{\kms}{\,{\rm km}\,{\rm s}^{-1}}
\newcommand{\cm}{\,{\rm cm}}
\newcommand{\erg}{\,{\rm erg}}

\newcommand{\Mpc}{\,{\rm Mpc}}
\newcommand{\cMpc}{\,{\rm cMpc}}

\newcommand{\keV}{\,{\rm keV}}
\newcommand{\mmag}{\,{\rm mag}}
\newcommand{\Hz}{\,{\rm Hz}}

\def\aap{A\&A}
\def\apj{ApJ}

\def\apjl{ApJ}
\def\mnras{MNRAS}
\def\araa{ARA\&A}
\def\aj{AJ}

\def\nat{Nature}

\def\apjs{ApJS}

\usepackage{float}
\usepackage[normalem]{ulem}

\usepackage{graphicx}	% Including figure files
\usepackage{amsmath}	% Advanced maths commands
\usepackage{amssymb}	% Extra maths symbols

\makeatletter

\newcommand{\Rmnum}[1]{\expandafter\@slowromancap\romannumeral #1@}

\renewcommand\paragraph{\@startsection{paragraph}{4}{\z@}{3.25ex\@plus1ex\@minus.2ex}{-1em}{\normalfont\it\normalsize}}

\title[The Bolometric Quasar Luminosity Function at $z=0-7$]{The Bolometric Quasar Luminosity Function at $z=0-7$}

\author[Shen et al.]{\parbox{18.5cm}{
Xuejian Shen$^{1}$\thanks{E-mail: xshen@caltech.edu},
Philip F. Hopkins$^{1}$,
Claude-Andr\'{e} Faucher-Gigu\`{e}re$^{2}$,
D. M. Alexander$^{3}$,
Gordon T. Richards$^{4}$,
Nicholas P. Ross$^{5}$,
R. C. Hickox$^{6}$
}\vspace{0.3cm}\\
$^{1}$TAPIR, California Institute of Technology, Pasadena, CA 91125, USA\\
$^{2}$Department of Physics and Astronomy and Center for Interdisciplinary Exploration and Research in Astrophysics (CIERA),
Northwestern University,\\ 2145 Sheridan Road, Evanston, IL 60208, USA\\
$^{3}$Centre for Extragalactic Astronomy, Department of Physics, Durham University, South Road, Durham DH1 3LE, UK\\
$^{4}$Department of Physics, Drexel University, 32 S. 32nd Street, Philadelphia, PA 19104, USA\\
$^{5}$Institute for Astronomy, University of Edinburgh, Royal Observatory, Blackford Hill, Edinburgh EH9 3HJ, UK\\
$^{6}$Department of Physics and Astronomy, Dartmouth College, 6127 Wilder Laboratory, Hanover, NH 03755, USA\\
}

\date{Accepted XXX. Received YYY; in original form ZZZ}

\pubyear{2020}

\begin{document}

\label{firstpage}
\pagerange{\pageref{firstpage}--\pageref{lastpage}}
\maketitle

\begin{abstract}
In this paper, we provide updated constraints on the bolometric quasar luminosity function (QLF) from $z=0$ to $z=7$. The constraints are based on an observational compilation that includes observations in the rest-frame IR, B band, UV, soft and hard X-ray in past decades. Our method follows Hopkins et al. 2007 with an updated quasar SED model and bolometric and extinction corrections. The new best-fit bolometric quasar luminosity function behaves qualitatively different from the Hopkins et al. 2007 model at high redshift. Compared with the old model, the number density normalization decreases towards higher redshift and the bright-end slope is steeper at $z\gtrsim 2$. Due to the paucity of measurements at the faint end, the faint end slope at $z\gtrsim 5$ is quite uncertain. We present two models, one featuring a progressively steeper faint-end slope at higher redshift and the other featuring a shallow faint-end slope at $z\gtrsim 5$. Further multi-band observations of the faint-end QLF are needed to distinguish between these models. The evolutionary pattern of the bolometric QLF can be interpreted as an early phase likely dominated by the hierarchical assembly of structures and a late phase likely dominated by the quenching of galaxies. We explore the implications of this model on the ionizing photon production by quasars, the CXB spectrum, the SMBH mass density and mass functions. The predicted hydrogen photoionization rate contributed by quasars is subdominant during the epoch of reionization and only becomes important at $z\lesssim 3$. The predicted CXB spectrum, cosmic SMBH mass density and SMBH mass function are generally consistent with existing observations.
\end{abstract}


%%%%%%%%%%%%%%%%%%%%%%%%%%%%%%%%%%%%%%%%%%%%%%%%%%

%%%%%%%%%%%%%%%%% BODY OF PAPER %%%%%%%%%%%%%%%%%%

\section{Introduction}
Luminous quasars and active galactic nuclei (AGN) in general~\footnote{We use the phrase "quasar" across the paper. We are not just referring to the optically bright and unobscured systems but the entire AGN population.} are observable manifestations of accreting supermassive black holes (SMBHs) at galaxy centers. Gas accreted onto the SMBH forms an accretion disk from which thermal emission is generated through dissipative processes~\citep[e.g.,][]{Shakura1973,Rees1984}. Due to their high radiative efficiency, such objects can be extremely luminous and are detected at $z>7$~\citep{Mortlock2011,Venemans2015,Banados2018}. The evolution of quasars is crucial to understand the formation and evolution of SMBHs in the Universe. Apart from that, quasars are one of the most important radiation sources in the Universe. They are luminous in almost all accessible bands and their radiation has a significant impact in the Universe. For example, quasar emission is important for the build-up of cosmic infrared (IR) and X-ray radiation backgrounds. Quasar emission in the extreme ultraviolet (UV) is believed to dominate the reionization of helium in the Universe and may have a non-negligible contribution to the reionization of hydrogen, although star-forming galaxies dominate hydrogen reionization in most current models~\citep[e.g.,][]{FG2008Letter, FG2008,FG2009,Kuhlen2012,Robertson2015,Haardt2015,Giallongo2015,Onoue2017,Parsa2018}. Furthermore, observations have demonstrated that galaxies and SMBHs co-evolve~\citep[see reviews of][and references therein]{Alexander2012,Fabian2012,Kormendy2013,Heckman2014}. For example, the masses of the SMBHs are correlated with the masses, luminosities and velocity dispersions of their host galaxy spheroids~\citep[e.g.,][]{Magorrian1998,Ferrarese2000,Gebhardt2000,Gultekin2009}. AGN are also widely believed to impact star formation in their host galaxies via a "feedback" mechanism that helps quench galaxies~\citep[e.g.,][]{Sanders1996,Springel2005,Bower2006,Croton2006,Sijacki2007,Somerville2008,Hopkins2008,Feruglio2010,Fabian2012,Cicone2014} and solve the classical "cooling flow" problem~\citep[e.g.,][]{Cowie1977,Fabian1977,Fabian1984,Tabor1993,Fabian1994,Croton2006}. Therefore, studying the evolution of quasar populations along cosmic time is of great importance in cosmology and galaxy formation. 

The quasar luminosity function (QLF), which is the comoving number density of quasars as a function of luminosity, is perhaps the most important observational signature of quasar populations. The study of the QLF goes back decades in the rest-frame optical/UV~\citep[e.g.,][]{Schmidt1968,Schmidt1983,Koo1988,Boyle1988,Hartwick1990,Hewett1993,Warren1994,Schmidt1995,Kennefick1995,Pei1995,Boyle2000,Fan2001,Fan2004,Richards2006b,Croom2009,Willott2010,Glikman2011,Ross2013,McGreer2013,Kashikawa2015,Jiang2016}, soft X-ray~\citep[e.g.,][]{Maccacaro1991,Boyle1993,Jones1997,Page1997,Miyaji2000,Hasinger2005}, hard X-ray~\citep[e.g.,][]{Ueda2003,LaFranca2005,Barger2005,Silverman2008,Ebrero2009,Yencho2009,Aird2010,Ueda2014,Aird2015a} and IR~\citep[e.g.,][]{Brown2006,Matute2006,Assef2011,Lacy2015}. These studies have conclusively shown that the observed QLF exhibits a strong redshift evolution. This is not simply an evolution in the normalization~(number density) but also in the slope of the QLF. For instance, the number density of low luminosity AGN peaks at lower redshift than that of bright quasars indicating the "cosmic downsizing" of AGN~\citep[e.g.,][]{Cowie1996,Barger2005,Hasinger2005}. AGN feedback that shuts down the supply of gas for accretion may be responsible for this phenomenon. Both optical and X-ray studies have argued that the faint-end slope of the QLF gets steeper from $z=2$ to $z=0$~\citep[e.g.,][]{Aird2015a,Kulkarni2018}. These investigations of the QLF have also found that both the typical spectral shape~\citep[e.g.,][]{Wilkes1994,Green1995,Vignali2003,Strateva2005,Richards2006a,Steffen2006,Just2007,Lusso2010,Kashikawa2015,Lusso2016} and the obscuring column density distribution of quasars~\citep[e.g.,][]{Hill1996,Simpson1999,Willott2000,Steffen2003,Ueda2003,Grimes2004,Sazonov2004,Barger2005,Hao2005,Ueda2014} have a dependence on quasar luminosity. For example, fainter quasars tend to be more obscured and their emission is more dominated by the X-rays. 

In the last decade, the redshift frontier of the observations of quasars have been pushed up to $z>7$~\citep{Mortlock2011,Banados2018,Wang2018b} and about $40$ quasars are now known at $z\gtrsim 6.5$~\citep[e.g.,][]{Willott2010,Venemans2013,Venemans2015,Jiang2016,Reed2017,Mazzucchelli2017,Matsuoka2018,Ross2019}. These quasars reveal the early growth of SMBHs and also pinpoint the locations for the assembly of massive galaxies in the early Universe. The absorption spectra of these high redshift quasars are important to study the reionization history of the Universe~\citep[e.g.,][]{Miralda1998,Madau2000,Fan2002,Fan2006}. However, due to the rapid decline in the quasar number density at high redshift, detecting quasars and constraining the QLF is currently very difficult at $z\gtrsim 6$. The next generation deep, wide-field infrared surveys will help push the detection of quasars to $z\simeq 9-10$ and deeper optical/UV surveys will provide better constraints on the faint end of the QLF. 

Interpreting the observational findings, however, is complicated by the fact that observations in a single band are always subject to selection effects, host galaxy contamination and reddening and obscuration all in a complicated, wavelength-dependent manner. Although quasars are intrinsically very luminous in the optical/UV, dust extinction along some viewing angles~\citep[e.g.,][]{Antonucci1993,Urry1995} can make quasars much more difficult to detect. Heavily obscured AGN can easily be contaminated with the UV stellar light from their host galaxies~\citep[e.g., see review of][]{Hickox2018}. Even in the X-ray, which is much less affected by dust, the Compton-thick (CTK) AGN, which account for $20\%-50\%$~\citep[e.g.,][]{Burlon2011,Ricci2015} of the total AGN population, are still severely blocked and current observations remain largely incomplete. In the mid-IR, due to the strong absorption in the terrestrial atmosphere, observations are more limited and also can be contaminated by the hot dust emission in star forming galaxies. In far-IR to millimeter wavelengths ($30\micron-10\,{\rm mm}$), the majority of AGN are contaminated by emission from dust heated by star formation in host galaxies, which limits the effectiveness of AGN identification. Furthermore, measurements of the QLF based on a single survey are limited in their luminosity coverage and volume probed and are subjected to various biases and uncertainties in completeness corrections. 

Given these limitations, what physical models for AGN demographics, SMBH growth and AGN feedback, really require is the bolometric QLF over all redshifts. The bolometric quasar luminosity is the quantity tightly related to the accretion rate of the SMBH and is the ideal quantity to study the physical evolution of quasars. \citet{Hopkins2007} developed a bolometric QLF model that simultaneously fitted the accessible measurements at the time, in different bands. The model has been widely used but has several important shortcomings: First, the model was poorly constrained at $z\gtrsim 3$ due to limited available data at the time and has been shown to deviate significantly from recent observations. Second, the integrated bolometric luminosity at the bright end predicted by this model actually diverges when extrapolating to high redshift~($z\sim7-8$). Third, the number density normalization of the QLF was assumed to be a constant over redshifts, which does not agree with newer observations at high redshift. 

In this paper, we provide a new model for the bolometric QLF at $z=0-7$ constrained by emerging observations of the QLF in the optical, UV, IR and X-ray in the last decade. The paper is organized as follows: In Section~\ref{sec:obs}, we introduce our observational data compilation. In Section~\ref{sec:model}, we introduce our model linking the observed QLFs with the bolometric QLF. The model includes new bolometric and extinction corrections. In Section~\ref{sec:LF}, we perform a fit to the data and constrain the bolomeric QLF. In Section~\ref{sec:LFE}, the evolution of the bolometric QLF is analyzed. In Section~\ref{sec:pred}, we present several predictions from our best-fit bolometric QLF model and demonstrate its consistency with observations from independent channels. 

We employ the following cosmological parameters: $\Omega_{\rm m} = 0.30$, $\Omega_{\Lambda} = 0.70$, $H_{0} = 100h \kms  \Mpc^{-1} = 70 \kms \Mpc^{-1}$. The code of all the analysis in this paper along with the observational data compiled are publicly available~(see Appendix~\ref{app:code} for details).  

\begin{table*}
    \centering
    \begin{tabular}{
    p{0.12\textwidth}|p{0.25\textwidth}|p{0.22\textwidth}|p{0.22\textwidth}}
    Band name & Definition of luminosity & Bolometric correction parameters  & Dispersion parameters  \\
    & & ($c_1$,$k_1$,$c_2$,$k_2$) & ($\sigma_1$,\,$\sigma_2$,\,$\log{L_0}$,\,$\sigma_3$)\\
    \hline 
    \hline
    B band &  $\nu_{4400\text{\AA}}L_{\nu_{4400\text{\AA}}}$ & ($3.759$,\,$-0.361$,\,$9.830$,\,$-0.0063$) & ($-0.383$,\,$0.405$,\,$42.39$,\,$2.378$) \\
    UV & the AB magnitude measured in a top-hat filter centering at rest-frame $1450\text{\AA}$ with bandwidth $100\text{\AA}$ or almost equivalently in terms of luminosity $\nu_{1450\text{\AA}}L_{\nu_{1450\text{\AA}}}$ & ($1.862$,\,$-0.361$,\,$4.870$,\,$-0.0063$) & ($-0.372$,\,$0.405$,\,$42.31$,\,$2.310$)\\
    Soft X-ray & the integrated luminosity in $0.5-2\keV$ & ($5.712$,\,$-0.026$,\,$17.67$,\,$0.278$) & ($0.080$,\,$0.180$,\,$44.16$,\,$1.496$)\\
    Hard X-ray & the integrated luminosity in $2-10\keV$ & ($4.073$,\,$-0.026$,\,$12.60$,\,$0.278$) &  ($0.193$,\,$0.066$,\,$42.99$,\,$1.883$)\\
    Mid-IR & $\nu_{15\micron}L_{\nu_{15\micron}}$ & ($4.361$,\,$-0.361$,\,$11.40$,\,$-0.0063$) & ($-0.338$,\,$0.407$,\,$42.16$,\,$2.193$)\\
    \hline
    \end{tabular}
    \caption{Definitions of the luminosities in the bands considered in this paper and best-fit parameters of their bolometric corrections and dispersions.}
    \label{tab:bands}
\end{table*}

\section{Observational Data Sets}
\label{sec:obs}

\begin{figure*}
    \centering
    \includegraphics[width=1.00\textwidth]{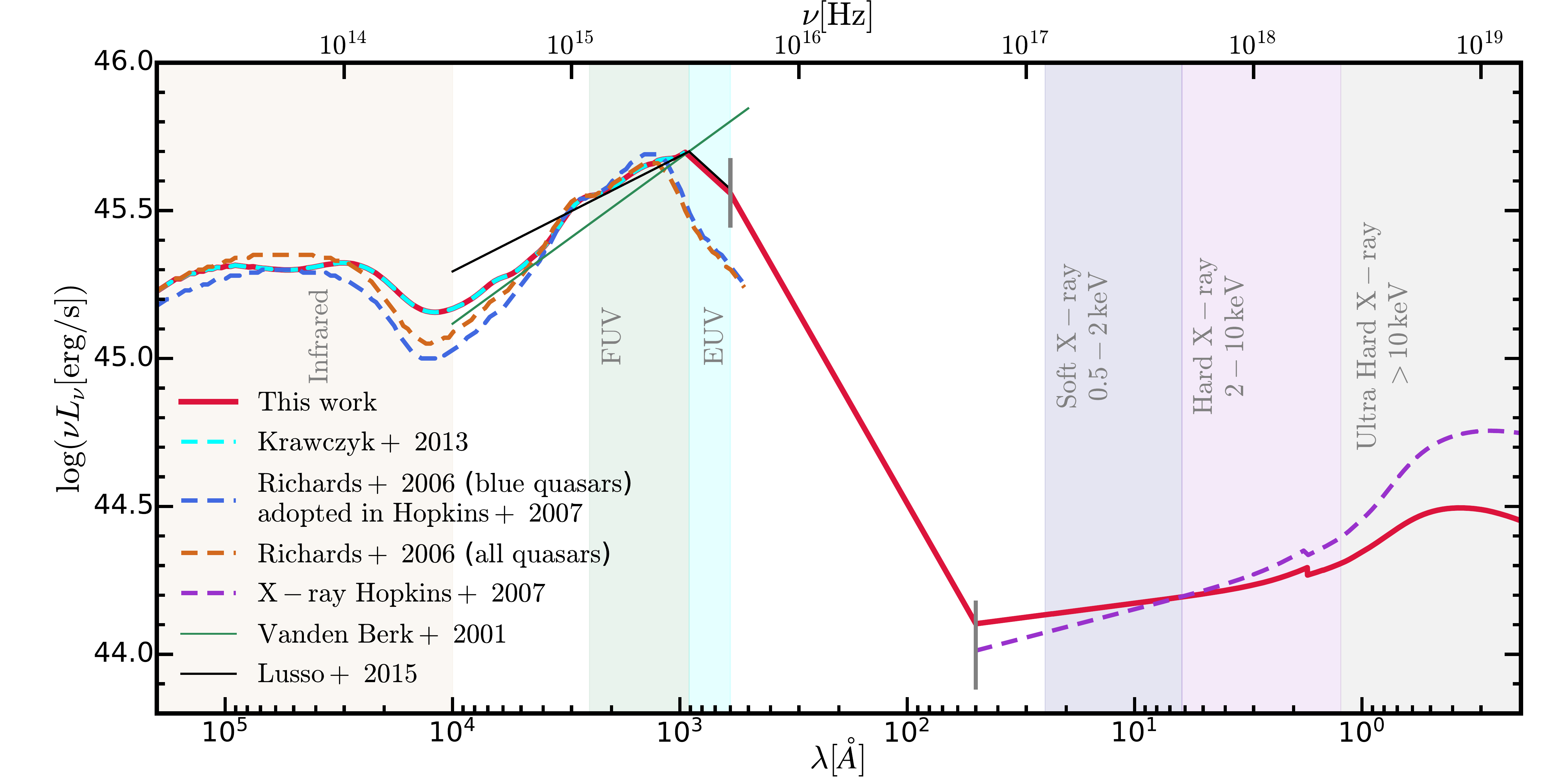}
    \caption{\textbf{Mean SED template of quasars constructed in this work.} The template SED shown here has the normalization $\nu L_{\nu}\simeq45.5 \erg\,{\rm s}^{-1}$ at $2500\text{\AA}$. The solid red line represents our fiducial SED model. SED templates from other works are presented: \citet{Richards2006a}, the blue and orange dashed lines; \citet{Hopkins2007} X-ray SED, the purple dashed line; \citet{Krawczyk2013}, the cyan dashed line. The power-law models for optical/UV SED are shown in the green~\citep{VandenBerk2001} and black~\citep{Lusso2015} thin lines. The common bands for the measurements of quasar luminosities are shown with shaded regions. In this paper, the bolometric luminosity is defined as the integrated luminosity from $30\micron$ to $500\keV$.}
    \label{fig:sed}
\end{figure*}

In this section, we briefly introduce the observations compiled in this work and emphasize the corrections adopted. A full list of the observations compiled is shown in Table~\ref{tab:observations}. We note that some observations used overlapping quasar samples in their binned estimations and are thus not fully independent. We do not include older observations if all the quasar samples used there were covered by later work. For all the observational data, we correct all relevant quantities~(distances, luminosities, volumes) to be consistent with our adopted cosmological parameters.

\subsection{Optical/UV}
We define "optical" wavelengths as $2500\text{\AA}\leq \lambda\leq1\micron$ and "UV" wavelengths as $600\text{\AA} \leq \lambda\leq 2500\text{\AA}$~\footnote{The quasar SED at rest-frame $50\text{\AA}\leq \lambda\leq 600\text{\AA}$ is almost inaccessible in optical/UV observations due to strong extinction at these wavelengths. In the construction of our SED model in Section~\ref{sec:sed}, we directly connect the $600\text{\AA}$ flux with the X-ray SED.}. We unify the luminosities measured in rest-frame optical (UV) wavelengths in observations to the B band (UV) luminosity defined in Table~\ref{tab:bands}. The optical/UV QLF observations compiled in this work are largely based on the observations listed in \citet{Hopkins2007,Giallongo2012,Manti2017} and \citet{Kulkarni2018}~(along with their QLF data shared online~\footnote{\href{https://github.com/gkulkarni/QLF/blob/master/Data/allqlfs.dat}{https://github.com/gkulkarni/QLF/blob/master/Data/allqlfs.dat}}). The observational compilation from \citet{Kulkarni2018} includes: \citet{Bongiorno2007,Siana2008,Jiang2009,Willott2010,Glikman2011,Masters2012,Palanque2013,Ross2013,McGreer2013,Kashikawa2015}. In the \citet{Kulkarni2018} compilation, the poisson errors in several works were recomputed using the \citet{Gehrels1986} formula. The K-corrections have been unified to that in \citet{Lusso2015}, which is based on the stacked spectra of $53$ quasars observed at $z\sim 2.4$. In fact, the uncertainty in K-corrections owing to different spectral assumptions was estimated to be within $0.2\mmag$~\citep{Lusso2015} which is smaller than the uncertainties of the binned estimation itself. The uncertainties in conversion factors between luminosities of different rest-frame bands were also estimated to be smaller than other sources of errors. Other specific corrections have been made in the \citet{Kulkarni2018} compilation are: (1) bins with severe incompleteness from \citet{Ross2013} were discarded; (2) binned estimations in \citet{Palanque2013} at $z>2.6$ were discarded since Lyman-alpha forest enters {\it g} band for those redshifts; (3) data from \citet{McGreer2013} was restricted to $M_{\rm 1450}>-26.73$ to avoid overlapping with \citet{Yang2016}; (4) for \citet{Willott2010} and \citet{Kashikawa2015}, the redshift intervals were recomputed using consistent completeness estimations. 

Outside the \citet{Kulkarni2018} compilation, we include measurements from \citet{Fontanot2007,Croom2009,Shen2012,Jiang2016,Palanque2016,Yang2016,Akiyama2018,Matsuoka2018,McGreer2018,Wang2018,Yang2018}. Observed optical band luminosities are all converted to UV luminosity either with corrections made in these papers or with the formula in \citet{Ross2013} if no corrections had already been made. \citet{Matsuoka2018} and \citet{Yang2018} have binned estimations that correspond to only one object in the bin, which were interpreted as upper limits there. However, their Poisson error estimations were not correct and we recalculate the Poisson errors using the table in \citet{Gehrels1986}. After the correction, these data points have proper upper and lower limits and can be included into our standard fitting procedure. For the observations compiled in \citet{Hopkins2007}~\citep{Kennefick1995,Schmidt1995,Fan2001a,Fan2001b,Fan2003,Fan2004,Wolf2003,Cristiani2004,Croom2004,Hunt2004,Richards2005,Richards2006a,Siana2006}, we include only those whose quasar samples are not completely covered by the more recent work discussed above. The details of all the observations compiled in this paper are listed in Table~\ref{tab:observations}.

\subsection{X-ray}
We define "X-ray" wavelengths as $\lambda \leq 50\text{\AA}$~($E\gtrsim 0.25\keV$) which covers the typical soft X-ray and hard X-ray bands defined in Table~\ref{tab:bands}. In the X-ray, in addition to the observations compiled in \citet{Hopkins2007}~\citep{Miyaji2000,Miyaji2001,Ueda2003,Sazonov2004,Barger2005,LaFranca2005,Hasinger2005,Nandra2005,Silverman2005}, we include new observational data from \citet{Ebrero2009,Aird2008,Silverman2008,Yencho2009,Aird2010,Fiore2012,Ueda2014,Aird2015a,Aird2015b,Miyaji2015,Khorunzhev2018}. Among them, \citet{Aird2008} is an update based on \citet{Nandra2005} and \citet{Silverman2008} is an extension to \citet{Silverman2005}. \citet{Aird2015a} and \citet{Ueda2014} derived binned estimation of the hard X-ray luminosity functions separately based on soft or hard X-ray selected samples. We include both of them in our compilation. \citet{Aird2015b} is an observation of the $10-40\keV$ X-ray luminosity function. The luminosities are converted to the hard X-ray luminosities with our SED model which will be discussed in the following section. Some observational works~\citep{Ebrero2009, Ueda2014, Aird2015b, Miyaji2015} have done their own "absorption" corrections and presented the "de-absorbed" compton thin QLFs. This would potentially generate double-counting of the extinction effects since we also intend to do extinction corrections in our model. We address this by reintroducing the extinction effect (only in the compton thin regime) for these data points using our extinction model which will be discussed in Section~\ref{sec:extinction}. 

\subsection{Infrared~(IR)}
We define "IR" wavelengths as $\lambda \geq 1\micron$. We unify the luminosities measured in rest-frame IR wavelengths to the mid-IR luminosity defined in Table~\ref{tab:bands}. In the IR, in addition to the observations compiled in \citet{Hopkins2007}~\citep{Brown2006,Matute2006}, we include new observations from \citet{Assef2011} and \citet{Lacy2015}. The luminosities are converted to the mid-IR~($15\micron$) luminosity with our SED model. These observations have extended the redshift coverage of the IR QLF up to $z=5.8$. However, there is still an apparent deficiency in IR observations compared with other wavelengths. Deep and large field IR surveys are an urgent need in the study of the QLF at high redshift. Though the total number of IR data points are limited and thus they have low statistical significance in the fit of the bolometric QLF, they do provide an independent check for our bolometric QLF model.

\section{Model}
\label{sec:model}
\subsection{SED model and bolometric corrections}
\label{sec:sed}
In this section, we construct the mean SED model for quasars. With the mean SED, we will calculate the bolometric corrections for the rest-frame B band, UV, soft \& hard X-ray and mid-IR, respectively.

\subsubsection{Optical/UV}
In the optical/UV, we start with the SED template in \citet{Krawczyk2013}, which was based on $108184$ luminous broad-lined quasars observed at $0.064<z<5.46$. Among these sources, $11468$ showing sign of dust reddening ($\Delta(g-i)>0.3$) had been discarded by \citet{Krawczyk2013} in deriving the mean SED template. Therefore, this SED template can be considered not strongly affected by reddening and obscuration. The extinction corrections on the quasar luminosities will be considered separately in the next section. This SED template starts at $\sim 30\micron$ and truncates at $912\text{\AA}$. We extend the SED to the extreme UV~(here defined as $\lambda<912\text{\AA}$) using the power-law model $f_{\nu}=\nu^{\alpha_{\nu}}$ with index $\alpha_{\nu}=-1.70$ reported by \citet{Lusso2015}. We truncate this extension at $600\text{\AA}$ where \citet{Lusso2015}'s measurement ended and directly connect the flux at $600\text{\AA}$ with the X-ray template which will be discussed then.

Historically, the optical/UV SED was often modelled as a power-law $f_{\nu}=\nu^{\alpha_{\nu}}$. In the UV, \citet{VandenBerk2001} found that the $1300\text{\AA}$ to $5000\text{\AA}$ continuum roughly has a power-law index $\alpha_{\nu}=-0.44\pm0.10$. \citet{Telfer2002} found $\alpha_{\nu}=-0.69\pm0.06$ at $1200\text{\AA}\lesssim\lambda\leq2200\text{\AA}$. \citet{Shull2012} found $\alpha_{\nu}=-0.68\pm0.14$ at $1200\text{\AA}\leq\lambda\leq2000\text{\AA}$. \citet{Lusso2015} found $\alpha_{\nu}=-0.61\pm0.01$ at $912\text{\AA}\leq\lambda\leq2500\text{\AA}$. The differences between \citet{VandenBerk2001} and other updated measurements arise from different continuum regions used to measure the slope. In the extreme UV, \citet{Telfer2002} found $\alpha_{\nu}=-1.76\pm0.12$ at $500\text{\AA}\lesssim\lambda\leq1200\text{\AA}$. \citet{Scott2004} found $\alpha_{\nu}=-0.56^{+0.38}_{-0.28}$ at $630\text{\AA}\lesssim\lambda\leq1155\text{\AA}$. \citet{Lusso2015} found $\alpha_{\nu}=-1.70\pm0.61$ at $\sim 600\text{\AA}\leq\lambda\leq912\text{\AA}$.
The update of break point from $\sim 1200\text{\AA}$ to $\sim 912\text{\AA}$ mainly attributes to more careful correction on IGM absorption~\citep{Lusso2015}. We do not consider the potential redshift/luminosity dependence of the break point, since it has almost no influence on the bolometric corrections. In Figure~\ref{fig:sed}, we show that our optical/UV SED template is generally consistent with the most recent power-law models.

\subsubsection{IR}
In the IR, we adopt the SED template in \citet{Krawczyk2013}. We extend the template in the long wavelength end to $100\micron$ using the \citet{Richards2006a} SED which behaves almost the same as the \citet{Krawczyk2013} SED at $\lambda>10\micron$. We note that this IR SED has already included dust emission. No additional dust emission model will be required. 

\subsubsection{X-ray}
The X-ray SED template is generated with a cut-off power-law model $f(E) \sim E^{1-\Gamma}\exp(-E/E_{\rm c})$ with the photon index $\Gamma=1.9$ and the cut-off energy $E_{\rm c}=300\keV$~\citep[e.g.,][]{Dadina2008,Ueda2014,Aird2015a}. An additional reflection component is added using the {\sc pexrav} model~\citep{Magdziarz1995} assuming the reflection relative strength $R=1$, the inclination angle $i=60^{\circ}$ and solar abundances. Then, we have to properly normalize the X-ray SED relative to the optical SED. Previous studies have reported a correlation between $L_{\nu}(2\keV)$ and $L_{\nu}(2500$\AA)~(the unit of $L_{\nu}$ is $\erg\,{\rm s}^{-1}\Hz^{-1}$):
\begin{equation}
    \log{L_{\nu}(2\keV)}=\beta \log{L_{\nu}(2500\text{\AA})} + C,
    \label{eq:Lx-L2500}
\end{equation}
where $\beta$ is found to be $0.7-0.8$ suggesting a non-linear correlation between the X-ray and optical luminosities. Defining $\alpha_{\rm ox}$ as:
\begin{equation}
    \alpha_{\rm ox}=\dfrac{\log{L_{\nu}(2\keV)}-\log{L_{\nu}(2500\text{\AA})}}{\log{\nu(2\keV)}-\log{\nu(2500\text{\AA})}}=0.384\log{\Big( \dfrac{L_{\nu}(2\keV)}{L_{\nu}(2500\text{\AA})}\Big)}.
\end{equation}
Then Equation~\ref{eq:Lx-L2500} can be rewritten as: 
\begin{equation}
    \alpha_{\rm ox}= -A\log{\Big( \dfrac{L_{\nu}(2500\text{\AA})}{\erg\,{\rm s}^{-1}\Hz^{-1}}\Big)} + C^{\prime},
\end{equation}
where $A=0.384\,(1-\beta)$ and $C^{\prime}=0.384C$. These prefactors have been measured through observations. However, since there is scatter in this relation, treating $L_{\nu}(2500\text{\AA})$ or $L_{\nu}(2\keV)$ as the independent variable will lead to different results if quasars are not perfectly selected in observations. The bisector of the two fitted relation treating either $L_{\nu}(2500\text{\AA})$ or $L_{\nu}(2\keV)$ as the independent variable is usually adopted. For example, \citet{Steffen2006} measured $\beta=0.721\pm0.011$ and $C=4.531\pm0.688$; \citet{Just2007} measured $\beta=0.709\pm0.010$ and $C=4.822\pm0.627$; \citet{Lusso2010} measured $\beta=0.760\pm0.022$ and $C=3.508\pm0.641$. \citet{Young2010,Xu2011,Lusso2016} found consistent results with previous works though they treated $L_{\nu}(2500\text{\AA})$ as the independent variable. Dependence of $\alpha_{\rm ox}$ on redshift had been reported in \citet{Bechtold2003}, but was not confirmed in the following studies. Given these observational results, we conclude that the relation constrained by \citet{Steffen2006}, which was adopted in \citet{Hopkins2007}, is still consistent with updated observations. We continue to use the parameters measured by \citet{Steffen2006} though varying the parameter choices does not have a significant influence on the bolometric corrections. The X-ray SED is then scaled with the $\alpha_{\rm ox}$ with respect to the optical SED.

\subsubsection{Bolometric corrections}
The direct product of our quasar SED model is the bolometric correction, defined as the ratio between the bolometric luminosity, $L_{\rm bol}$, and the observed luminosity in a certain band, $L_{\rm band}$. The definitions of the luminosities in the bands are presented in Table~\ref{tab:bands}. The bolometric luminosity is defined as the integrated luminosity from $30\micron$ to $500\keV$, which represents all the energy budget generated by the accretion of the SMBH. \footnote{Some studies included the emission beyond $30\micron$ in the bolometric luminosity. But we find that extending the long-wavelength bound to $100\micron$ will only lead to $<0.02\, {\rm dex}$ difference in the bolometric luminosity.} Some studies~\citep[e.g.,][]{Marconi2004,Krawczyk2013} have discussed that the reprocessed emission in the IR and $>2\keV$ X-ray should be excluded in determining the bolometric luminosity to avoid potential double-counting of quasars' intrinsic emission. We have tested that using $1\micron$ to $2\keV$ as the range for integration will systematically decrease the bolometric luminosity by $\sim 0.2\, {\rm dex}$. 

\begin{figure}
    \centering
    \includegraphics[width=0.48\textwidth]{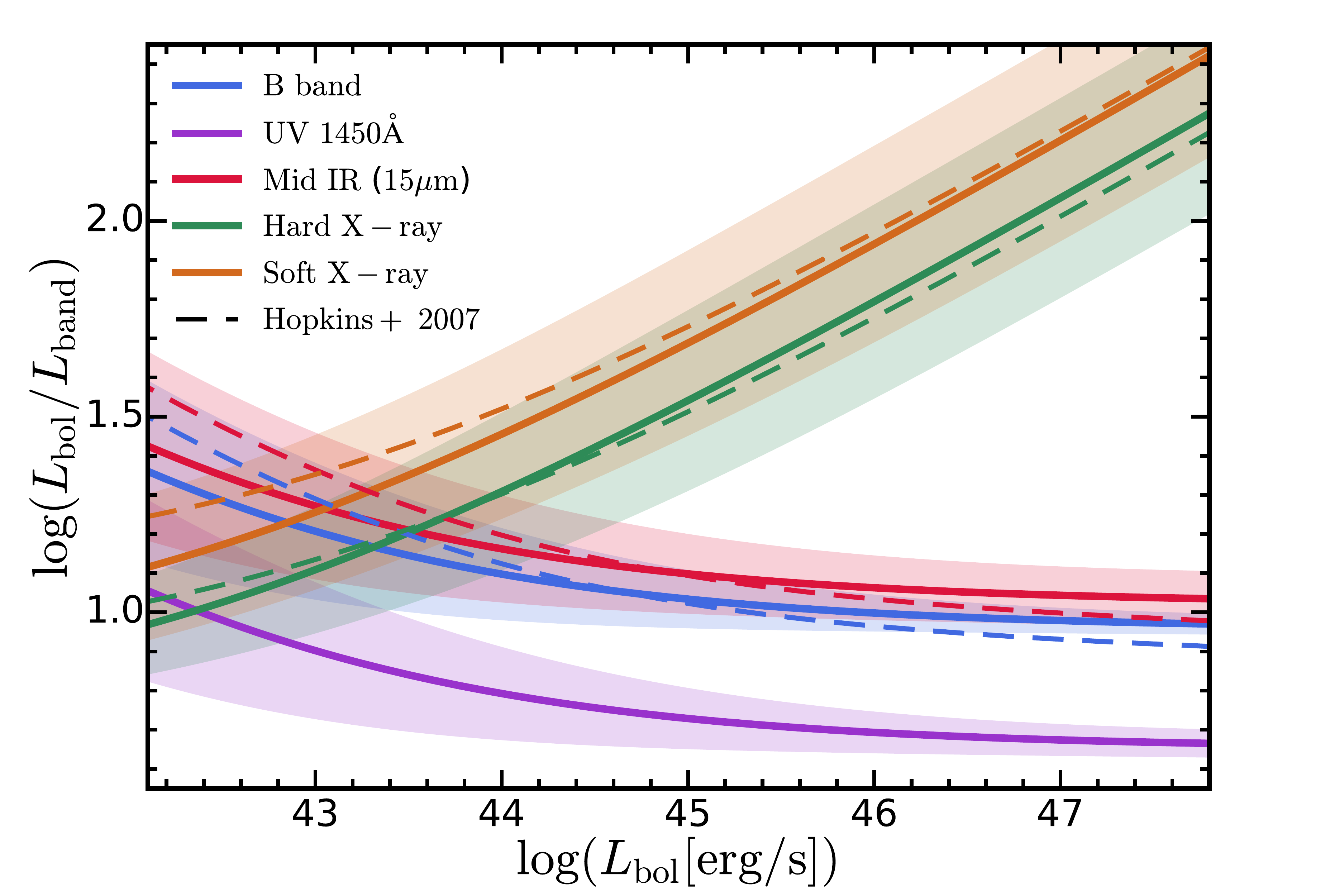}
    \includegraphics[width=0.48\textwidth]{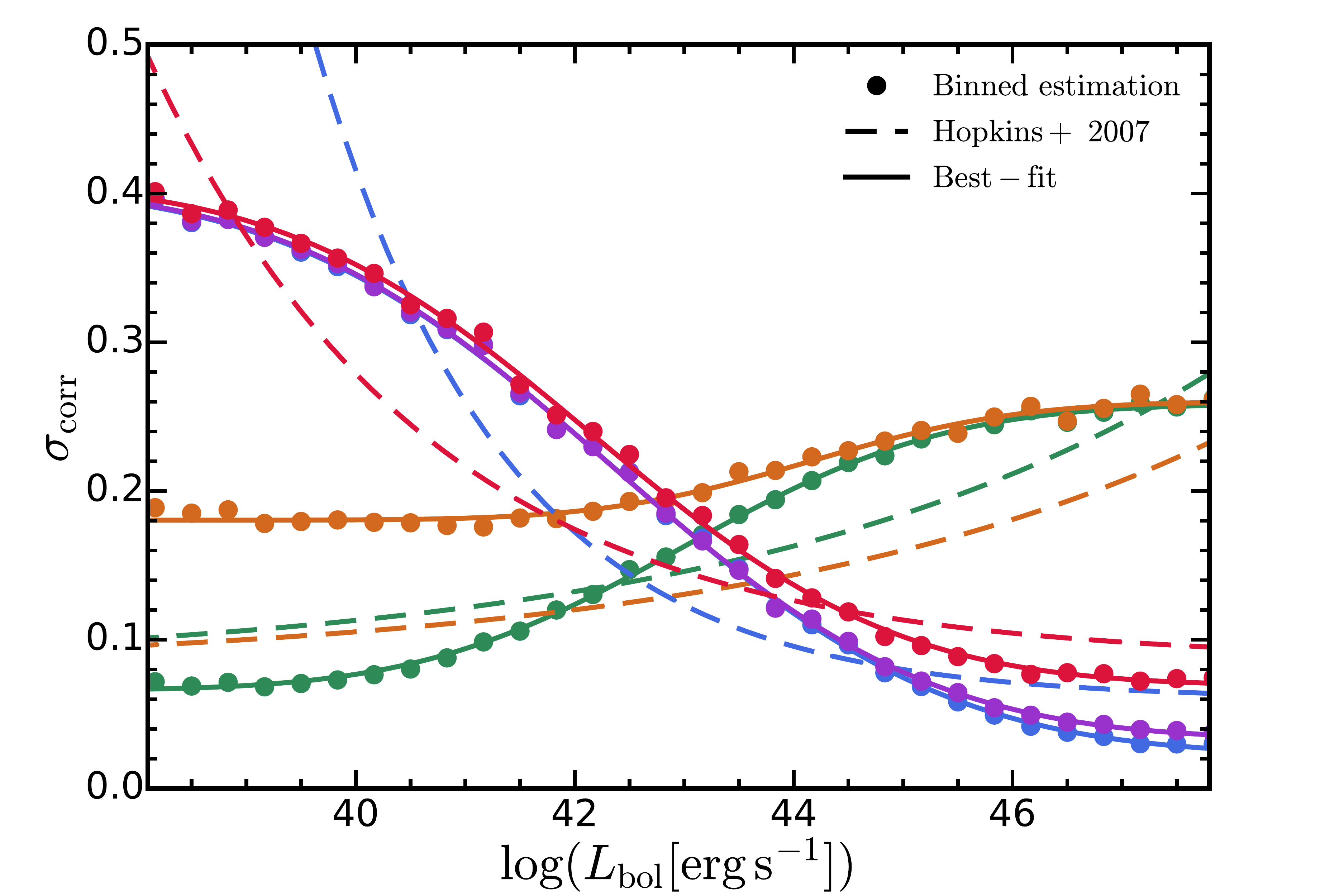}
    \caption{{\it Top:} \textbf{Bolometric corrections as a function of bolometric quasar luminosity.} We show the bolometric corrections in the rest-frame B band, UV, mid-IR, soft and hard X-ray determined by our fiducial quasar mean SED template. $1\sigma$ dispersions are shown with shaded region. The bolometric corrections in \citet{Hopkins2007} are shown in dashed lines. {\it Bottom:} \textbf{Magnitude of the dispersions in bolometric corrections as a function of bolometric quasar luminosity.} Both binned estimations and best-fit relations are presented.}
    \label{fig:corr}
\end{figure}

However, quasars do not have a single universal SED. There are real variations in the spectral shape, which translate to scatters in the bolometric corrections and influence the observed QLFs in the bands. To evaluate this, we first create an ensemble of SEDs. The configuration of these SEDs are similar to our fiducial SED: in the IR, we adopt our fiducial SED; in the optical/UV, for simplicity, we adopt a broken power-law with the break point at $912\text{\AA}$, with a fixed slope $-1.70$ at $\lambda<912\text{\AA}$ and a free slope $\alpha_{\rm opt}$ at $\lambda>912\text{\AA}$; in the X-ray, we adopt our fiducial X-ray SED model but with a free photon index $\Gamma$; the optical/UV and X-ray SEDs are connected with a free $\alpha_{\rm ox}$. We generate an ensemble of $10^5$ SEDs with randomly sampled $L_{\nu}(2500\text{\AA})$, $\alpha_{\rm opt}$, $\Gamma$ and $\alpha_{\rm ox}$. In sampling $\alpha_{\rm opt}$, $\Gamma$ and $\alpha_{\rm ox}$, we adopt a normal distribution around median value with a constant scatter. We adopt $\overline{\Gamma}\pm \sigma_{\Gamma}=1.9\pm0.2$~\citep[e.g.,][]{Ueda2014,Aird2015a}, $\overline{\alpha_{\rm opt}}\pm\sigma_{\rm opt}=-0.44\pm0.125$~\citep{VandenBerk2001,Richards2003}, $\sigma_{\rm ox}\simeq 0.1$~\citep[e.g.,][]{Steffen2006,Lusso2010}. The bolometric luminosity and bolometric corrections for each realization of the SED are calculated. Then we divide the SEDs based on their bolometric luminosities into $30$ uniformly log-spaced bins from $10^{38}$ to $10^{48}\,\erg\,{\rm s}^{-1}$. We evaluate the standard deviation of the bolometric correction of each band in each bolometric luminosity bin, shown in the bottom panel of Figure~\ref{fig:corr}. Double plateaus show up at the bright and faint ends where a certain band is dominant or negligible in the bolometric luminosity. In \citet{Hopkins2007}, the dispersion of the bolometric corrections were fitted with: $\sigma_{\rm corr}(L_{\rm bol})=\sigma_{1}(L_{\rm bol}/10^{9}\Lsun)^{\beta}+\sigma_2$. However, we find this formula no longer appropriate to fit our results, so we fit the dispersion with an error function:
\begin{equation}
    \sigma_{\rm corr}(\log{L_{\rm bol}})=\sigma_{2}+\sigma_{1}\Bigg[\dfrac{1}{2}+\dfrac{1}{2} {\rm erf}\Big(\dfrac{\log{L_{\rm bol}}-\log{L_{0}}}{\sqrt{2}\sigma_{3}}\Big)\Bigg],
\end{equation}
which naturally exhibits a double plateau shape. The best-fit parameters are listed in Table~\ref{tab:bands}. The fitted relations are also shown in the bottom panel of Figure~\ref{fig:corr}. These results indicate a $\sim 0.1\,{\rm dex}$ uncorrelated dispersion in quasar SEDs that is consistent with observations.

In the top panel of Figure~\ref{fig:corr}, we show the bolometric corrections as a function of bolometric luminosity for all bands along with their dispersions shown with shaded regions. The bolometric corrections are generally similar to the \citet{Hopkins2007} model except for the differences at the faint end driven by the updates in the X-ray SED. Following \citet{Hopkins2007}, we fit the dependence of the bolometric corrections on bolometric luminosity with a double power-law:
\begin{equation}
    \dfrac{L_{\rm bol}}{L_{\rm band}}= c_1 \Big(\dfrac{L_{\rm bol}}{10^{10}\Lsun}\Big)^{k_1} + c_2 \Big(\dfrac{L_{\rm bol}}{10^{10}\Lsun}\Big)^{k_2}.
\end{equation}
The best-fit parameters are listed in Table~\ref{tab:bands}.

We note that the derivation of the optical/UV and X-ray luminosities using these bolometric corrections has not considered extinction yet. The observed luminosities will be further affected by extinction, which will be discussed in the following section.

\subsection{Dust and gas extinction}
\label{sec:extinction}

\begin{figure}
    \centering
    \includegraphics[width=0.48\textwidth]{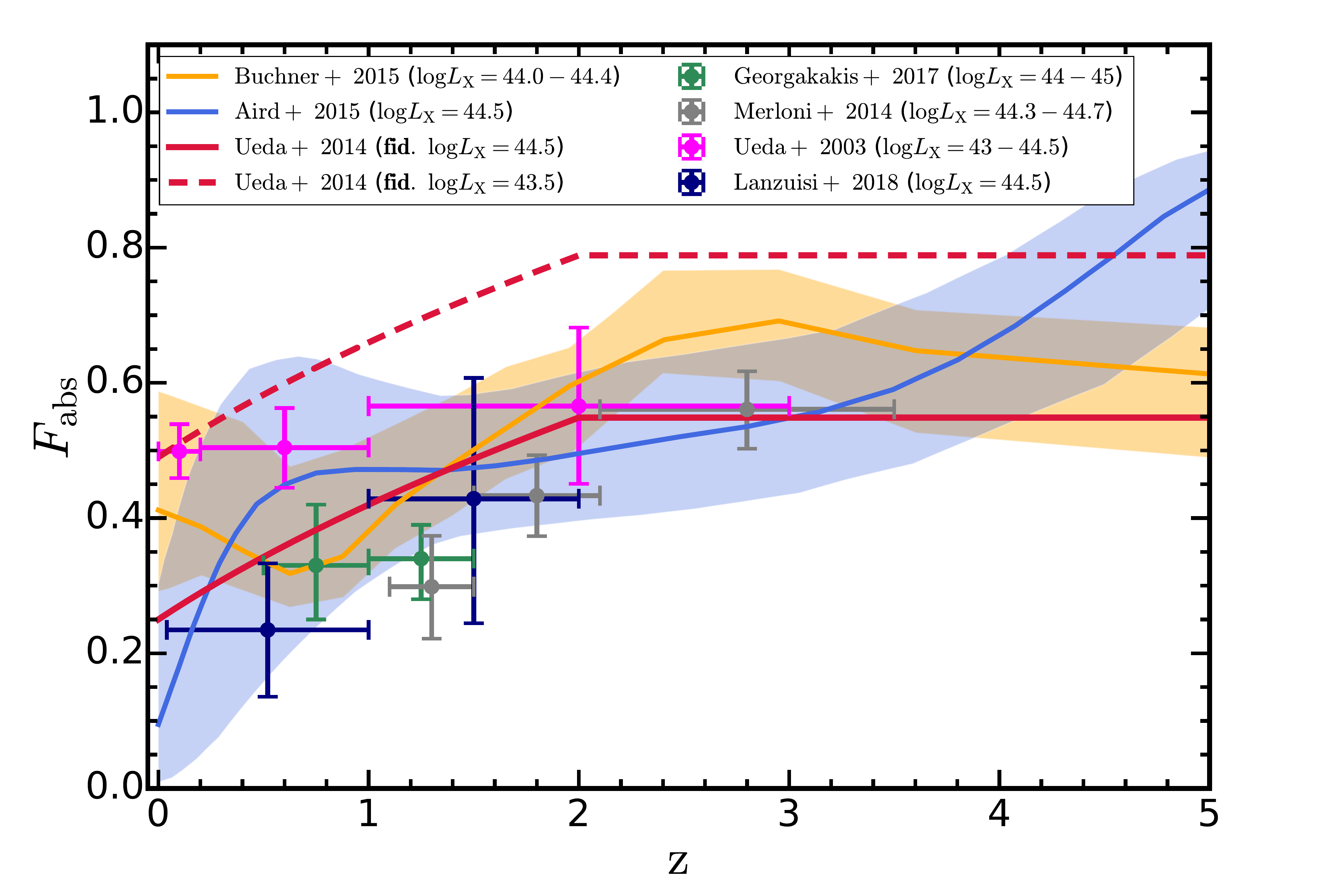}
    \includegraphics[width=0.48\textwidth]{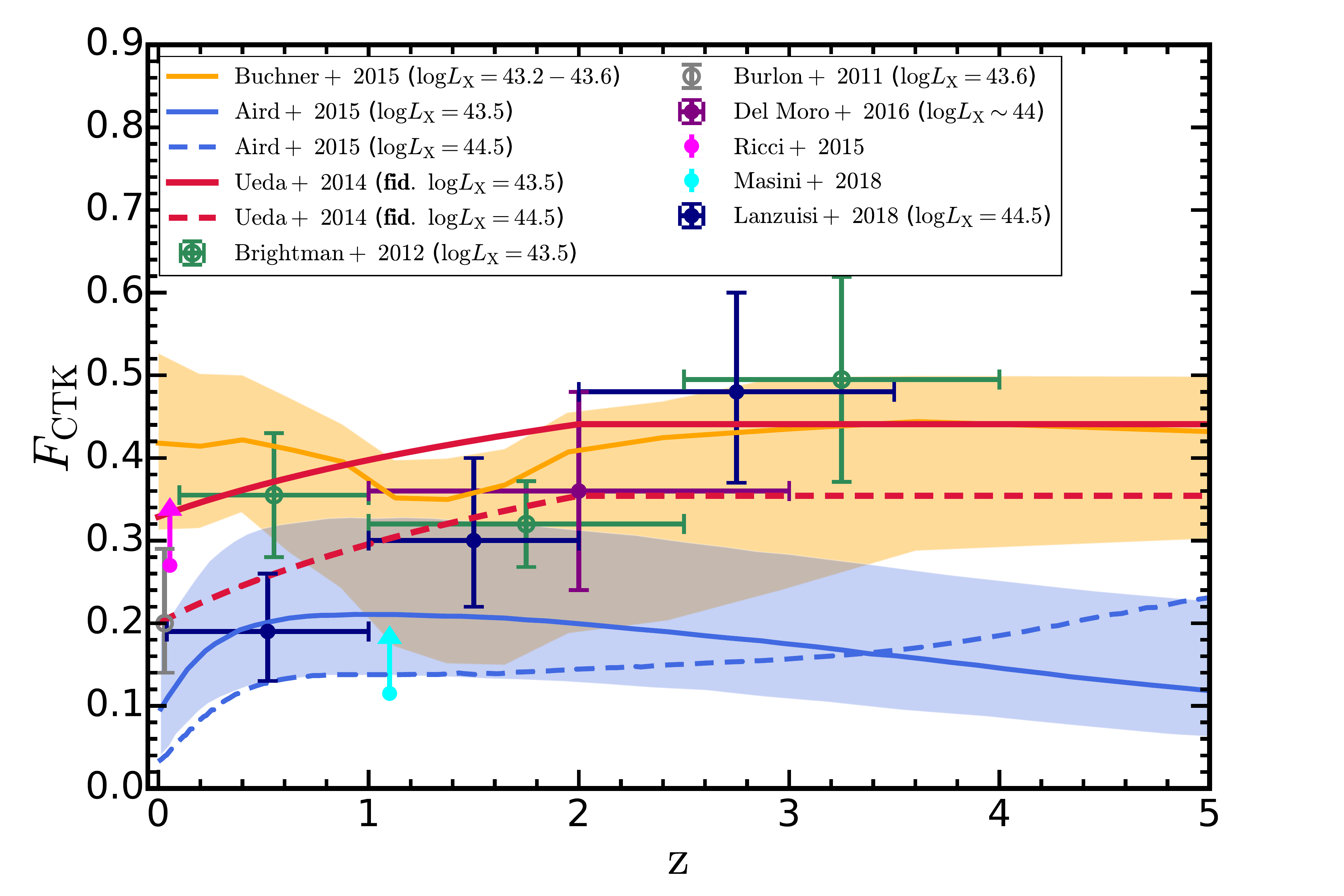}
    \includegraphics[width=0.48\textwidth]{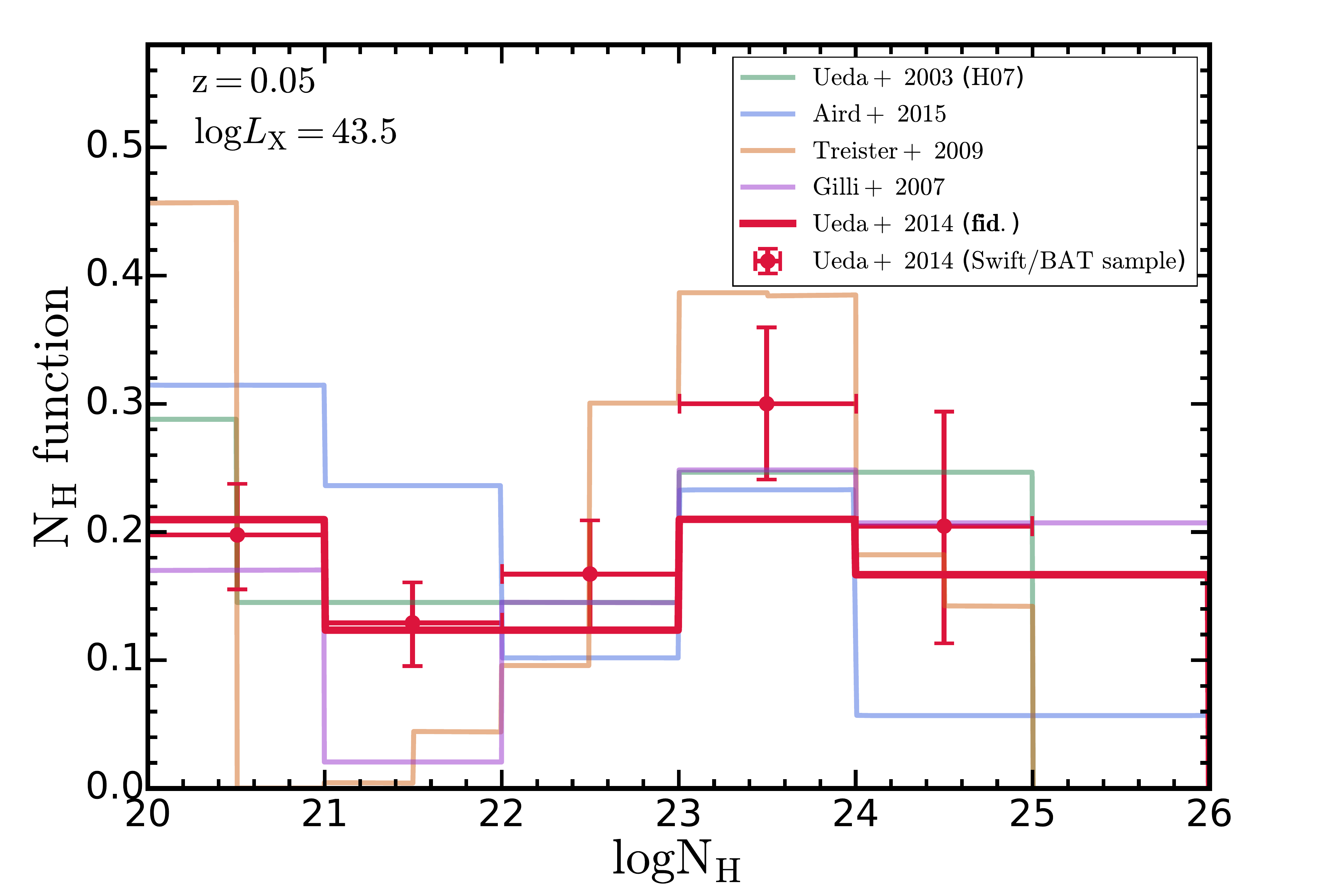}
    \caption{{\it Top:} \textbf{Absorbed quasar fraction at a given hard X-ray luminosity as a function of redshift.} We present the prediction from our fiducial model, the $N_{\rm H}$ model in \citet{Ueda2014}, with red lines. The solid red line is for $\log{L_{\rm X}}=44.5$ while the dashed one is for $\log{L_{\rm X}}=43.5$. We compare the ficucial model with other observations~(labeled). {\it Middle:} \textbf{Compton thick quasar fraction at a given hard X-ray luminosity as a function of redshift.} We compare the fiducial model with other observations~(labeled). Here, the solid red line is for $\log{L_{\rm X}}=43.5$ while the dashed one is for $\log{L_{\rm X}}=44.5$. {\it Bottom:} \textbf{$N_{\rm H}$ distribution at $\log{L_{\rm X}}=43.5$ and $\rm z=0.05$.} We compare our fiducial model with other models~(labeled) and the ${\rm N}_{\rm H}$ distribution of Swift/BAT samples~\citep{Ueda2014}.}
    \label{fig:fabs} 
\end{figure}

The absorption and scattering of surrounding gas and dust further modifies the intrinsic emission of quasars. Neutral hydrogen photoelectric absorption is crucial to the extinction in the X-ray while dust is crucial to the extinction in the optical/UV. Here, we first introduce the neutral hydrogen column density ($N_{\rm H}$) distribution model which determines the extinction in the X-ray. Then, $N_{\rm H}$ is converted to the column density of dust assuming a dust-to-gas ratio. The dust abundance determines the extinction in the optical/UV.

In \citet{Hopkins2007}, where the constant $N_{\rm H}$ model was shown to fail, the $N_{\rm H}$ distribution model from \citet{Ueda2003} was adopted as the fiducial model. Here, we update the $N_{\rm H}$ distribution with the results from \citet{Ueda2014}, which was based on measurements of $N_{\rm H}$ and the intrinsic hard X-ray luminosity for each individual object in their sample. The model provides the probability distribution of $N_{\rm H}$, $f(L_{\rm X},z;N_{\rm H})$, 
at a given intrinsic hard X-ray luminosity~(denoted as $L_{\rm X}$) and a redshift. $f(L_{\rm X},z;N_{\rm H})$ is normalized in the compton thin~(CTN, $\log{N_{\rm H}}\leq24$) regime:
\begin{equation}
    \int_{20}^{24} f(L_{\rm X},z;N_{\rm H})\, {\rm d}\log{N_{\rm H}} = 1,
\end{equation}
where the unit of $N_{\rm H}$ is assumed to be $\cm^{-2}$ and the lower limit of $\log{N_{\rm H}}=20$ is a dummy value introduced for convenience and \citet{Ueda2014} has assigned $\log{N_{\rm H}}=20$ for all the quasars with $\log{N_{\rm H}}<20$.

$f(L_{\rm X},z;N_{\rm H})$ is characterized by three parameters: $\psi(L_{\rm X},z)$, the fraction of absorbed quasars ($22\leq\log{N_{\rm H}}\leq24$) in total CTN quasars; $f_{\rm CTK}$, the fraction of compton thick~(CTK, $\log{N_{\rm H}}\geq24$) quasars relative to the fraction of absorbed CTN quasars; $\epsilon$, the ratio of the quasars with $23\leq\log{N_{\rm H}}\leq24$ to those with $22\leq\log{N_{\rm H}}\leq23$. This $N_{\rm H}$ distribution can then be written as~\citep{Ueda2014}:

\begin{equation}
f(L_{\rm X},z;N_{\rm H}) =\, 
\begin{cases} 
1-\dfrac{2+\epsilon}{1+\epsilon}\psi(L_{\rm X},z) \quad & [20\leq\log{N_{\rm H}}<21] \\
\dfrac{1}{1+\epsilon}\psi(L_{\rm X},z) \quad & [21\leq\log{N_{\rm H}}<22] \\
\dfrac{1}{1+\epsilon}\psi(L_{\rm X},z) \quad & [22\leq\log{N_{\rm H}}<23] \\
\dfrac{\epsilon}{1+\epsilon}\psi(L_{\rm X},z) \quad & [23\leq\log{N_{\rm H}}<24] \\
\dfrac{f_{\rm CTK}}{2}\psi(L_{\rm X},z) \quad & [24\leq\log{N_{\rm H}}<26] \\
\end{cases}
\end{equation}
when $\psi(L_{\rm X},z)<\dfrac{1+\epsilon}{3+\epsilon}$ and:

\begin{equation}
f(L_{\rm X},z;N_{\rm H}) =\, 
\begin{cases} 
\dfrac{2}{3}-\dfrac{3+2\epsilon}{3+3\epsilon}\psi(L_{\rm X},z) \quad & [20\leq\log{N_{\rm H}}<21] \\
\dfrac{1}{3}-\dfrac{\epsilon}{3+3\epsilon}\psi(L_{\rm X},z) \quad & [21\leq\log{N_{\rm H}}<22] \\
\dfrac{1}{1+\epsilon}\psi(L_{\rm X},z) \quad & [22\leq\log{N_{\rm H}}<23] \\
\dfrac{\epsilon}{1+\epsilon}\psi(L_{\rm X},z) \quad & [23\leq\log{N_{\rm H}}<24] \\
\dfrac{f_{\rm CTK}}{2}\psi(L_{\rm X},z) \quad & [24\leq\log{N_{\rm H}}<26] \\
\end{cases}
\end{equation}
when $\psi(L_{\rm X},z)\geq\dfrac{1+\epsilon}{3+\epsilon}$. The model assumes $\epsilon=1.7,\, f_{\rm CTK}=1$ and:
\begin{equation}
    \psi(L_{\rm X},z)={\rm min}[\psi_{\rm max},{\rm max}[\psi_{\rm 43.75}(z)-0.24(\log{L_{\rm X}}-43.75),\psi_{\rm min}]],  
    \label{eq:absorbed_frac}
\end{equation}
where $\psi_{\rm min}=0.2$, $\psi_{\rm max}=0.84$, $\psi_{\rm 43.75}(z)$ depends on redshift as:
\begin{equation}
    \psi_{\rm 43.75}(z)=
    \begin{cases}
    0.43(1+z)^{\rm 0.48} \quad & [z<2]\\
    0.43(1+2)^{\rm 0.48} \quad & [z\geq2]\\
    \end{cases}
\end{equation}
The model describes a negative dependence of the absorbed quasar fraction on the intrinsic quasar hard X-ray luminosity as well as redshift at $z<2$.

Given this $N_{\rm H}$ distribution model, both the absorbed and the CTK quasar fractions decrease at higher hard X-ray luminosities and increase at higher redshift with a plateau at $z\geq2$. The studies of the QLF and extinction properties in the X-ray have many variations in the data used, fitting methods, assumptions of the spectrum form and $N_{\rm H}$ distribution function form, treatments of redshift uncertainties and sources without counterparts. Therefore, it is worth comparing our fiducial extinction model with models determined in other works. In the top and middle panels of Figure~\ref{fig:fabs}, we compare the predictions on the absorbed quasar fraction and the CTK quasar fraction from this model with other observational constraints~\citep{Ueda2003,Burlon2011,Brightman2012,Merloni2014,Aird2015a,Buchner2015,Ricci2015,DelMoro2016,Georgakakis2017,Masini2018,Lanzuisi2018}. In the comparison, we do not show the hard X-ray luminosity from \citet{Ricci2015} and \citet{Masini2018} since these observations were in harder X-ray bands and the $2-10\keV$ X-ray luminosity was not available. The absorbed fraction $F_{\rm abs}$ in the top panel of Figure~\ref{fig:fabs} is defined as the fraction of absorbed quasars relative to total CTN quasars. The compton thick fraction $F_{\rm CTK}$ in the middle panel of Figure~\ref{fig:fabs} is defined as the fraction of CTK quasars relative to all quasars. We find a good agreement with other observations in the absorbed quasar fraction which monotonically increases towards higher redshift. Our fiducial model~(the \citet{Ueda2014} model) is in agreement with the \citet{Buchner2015} and the \citet{Aird2015a} models. Besides, we also find a good consistency in the CTK quasar fraction with most of the observations, except for \citet{Aird2015a} which determined the $N_{\rm H}$ distribution by reconciling the hard X-ray luminosity function of soft X-ray and hard X-ray selected quasars. Compared with the \citet{Buchner2015} model, the \citet{Ueda2014} model is consistent with it except for mild differences at $z<2$. We note that some recent studies~\citep{Masini2018,Georgantopoulos2019} using NuSTAR, which is more sensitive in the hard X-ray, found very small lower bounds of $F_{\rm CTK}$, $\sim 10-20\%$. Assuming that the CTK quasars are completely absent in observations, the uncertainty in $F_{\rm CTK}$ can result in $\log{\big((1-F_{\rm CTK}^{\rm min})/(1-F_{\rm CTK}^{\rm max})\big)}\sim 0.2\,{\rm dex}$ uncertainty in the binned estimations of the bolometric QLFs. In the bottom panel of Figure~\ref{fig:fabs}, we show the $N_{\rm H}$ distribution at $\log{L_{\rm X}}=43.5$, $z=0.05$ comparing different models~\citep{Ueda2003,Gilli2007,Treister2009,Ueda2014,Aird2015a}.

\begin{table*}
\centering
\begin{tabular}{
p{0.11\textwidth}|p{0.15\textwidth}|p{0.20\textwidth}|p{0.21\textwidth}|p{0.20\textwidth}}
\hline 
\hline
Name & Redshift & Data & Fitting function & Parameter fixing\\
\hline
\hline
local "free" fit & each individual redshift & data with its redshift bin covering the target redshift & double power-law luminosity function & all free\\
local "polished" fit & each individual redshift & data with its redshift bin covering the target redshift & double power-law luminosity function but manually reduced to a single power-law at $z\geq 5.8$ & $\phi_{\ast}$ is fixed according to the linear evolutionary trend found in the local "free" fits; other parameters are free\\
\textbf{global fit A} & \textbf{all redshifts simultaneously} & \textbf{all the data compiled} & \textbf{functions of the global evolution model; the faint-end slope has a flexible polynomial evolutionary pattern} & \textbf{all free; uniform priors in the Bayesian inference}\\
\textbf{global fit B} & \textbf{all redshifts simultaneously} & \textbf{all the data compiled} & \textbf{functions of the global evolution model; the faint-end slope is restricted to evolve monotonically with redshift, it has a power-law evolutionary pattern} & \textbf{all free; uniform priors in the Bayesian inference}\\
\hline
\end{tabular}
\caption{Overview of the fits we perform in this paper. The results of the local "free" and "polished" fits are presented in Table~\ref{tab:parameters_at_z}. The results of the global fits A and B are presented in Table~\ref{tab:parameters_global}. Unless otherwise specified, all the predictions and implications presented in this paper are based on the results of the global fits which are highlighted in the table.}
\label{tab:fits}
\end{table*}

Given $N_{\rm H}$, we calculate the extinction using the photoelectric absorption cross section in \citet{Morrison1983} and the non-relativistic Compton scattering cross section. To determine the dust abundance, a dust-to-gas ratio is required. In \citet{Hopkins2007}, a constant dust-to-gas ratio was assumed to convert ${\rm N}_{\rm H}$ to dust column density and a SMC-like extinction curve from \citet{Pei1992} was adopted. However, in this work, we find that these assumptions along with our fiducial ${\rm N}_{\rm H}$ distribution model result in a systematic inconsistency between UV, B band and X-ray observations. The UV and B band luminosities are under-predicted and the phenomenon is more severe in the UV than in the B band in a luminosity and redshift-dependent manner. This indicates that the extinction in the optical/UV is over-predicted by the model with the constant dust-to-gas ratio and the SMC-like extinction curve. Observations have revealed that the mass-metallicity relation of galaxies has a redshift evolution~\citep[e.g.,][]{Zahid2013} with the gas-phase metallicity of typical quasar host galaxies dropping $\sim 0.5\,{\rm dex}$ from $z=0$ to $z=2$. Similar evolution was also seen in numerical simulations~\citep[e.g.,][]{Ma2016}. Assuming that the dust-to-metal ratio remains a constant, the decrement in the gas-phase metallicity of quasar host galaxies will lead to a decrement in the dust-to-gas ratio at higher redshift. In addition, some observations have suggested that the extinction curve of AGN might be shallower than the commonly assumed SMC-like extinction curve~\citep[e.g.,][]{Maiolino2001,Gaskell2004,Gaskell2007,Czerny2004}. Given the observational updates, we choose to adopt a redshift-dependent dust-to-gas ratio which scales as the gas-phase metallicity given by the fit in \citet{Ma2016}. The value of the dust-to-gas ratio in the local Universe still follows \citet{Hopkins2007} with $(A_{\rm B}/N_{\rm H})=8.47 \times10^{-22} \cm^{2}$. We adopt the Milky Way-like extinction curve in \citet{Pei1992} which is shallower than the SMC-like curve. Although the extinction curve of quasars does not exhibit the $2175\text{\AA}$ bump feature as found in the Milky Way, our results are not affected by this since none of the bands we study in this paper are close to $2175\text{\AA}$.

We note that the extinction in the X-ray would also be affected by the decrement of the gas-phase metallicity. In addition, although the metallicities of quasar host galaxies decrease with redshift, the metallicities of broad line regions do not evolve as strongly. The relative contributions of host galaxies and near quasar obscuration are still largely unknown. Here, our choice in the dust-to-metal ratio empirically prefers the scenario that near quasar obscuration is more important in the X-ray and obscuration in host galaxies contributes more to the extinction in the optical/UV. Our choice is motivated by making the X-ray and optical/UV observations more consistent with each other at all redshifts. Similar argument applies to our choice of the extinction curve. The shallow extinction curves found in some studies are still under debate and our choice here is only for empirical needs.

The extinction model and the bolometric corrections introduced in this and previous sections allow us to link the bolometric QLF with the observed QLF in a certain band, and resolve the discrepancies described above. We note that for all the subsequent analysis in the paper, unless otherwise specified, the QLFs presented include both the obscured and unobscured AGN and the observed QLFs presented take account of dust and gas extinction described in this section. 

\begin{figure*}
    \centering
    \includegraphics[width=0.48\textwidth]{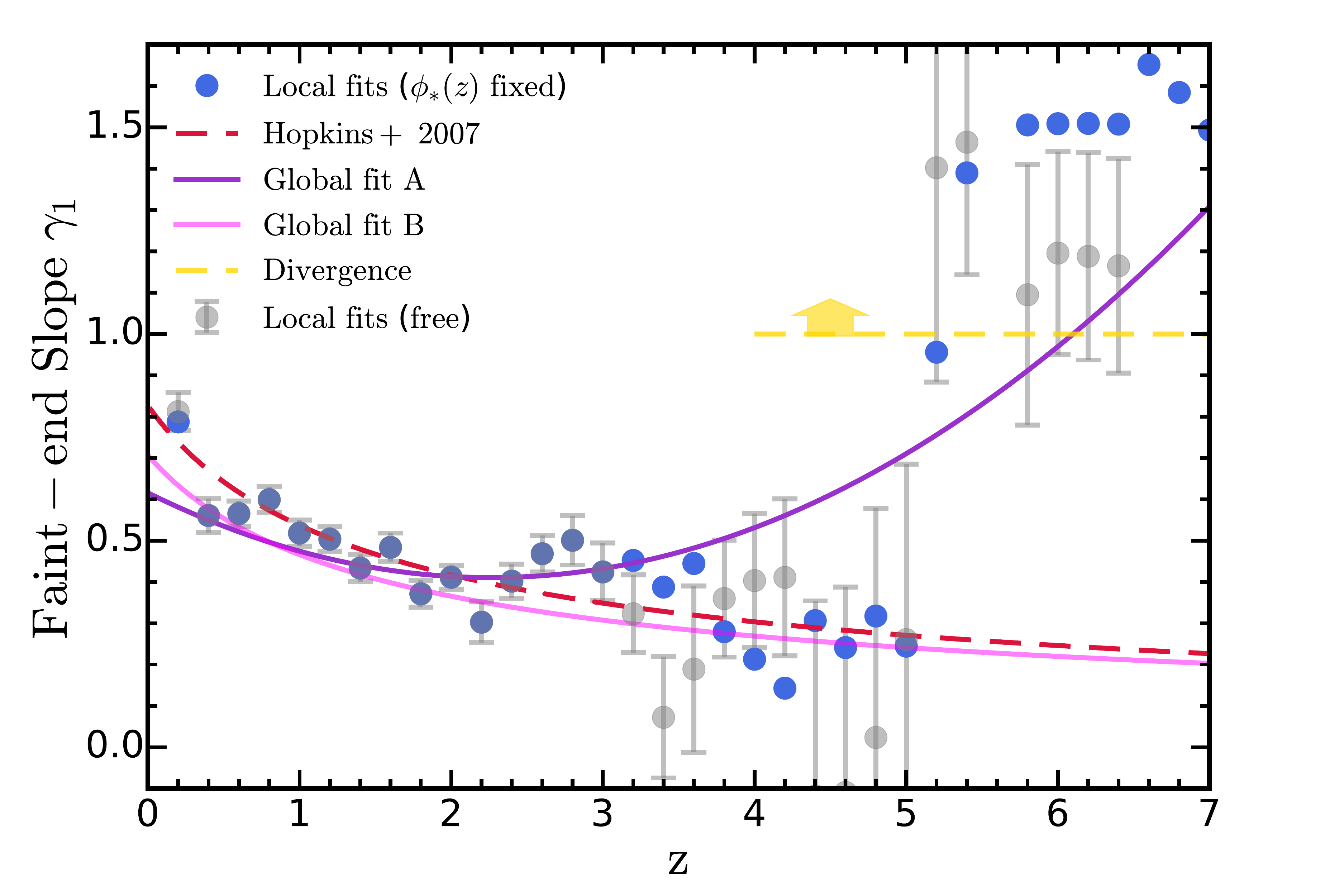}
    \includegraphics[width=0.48\textwidth]{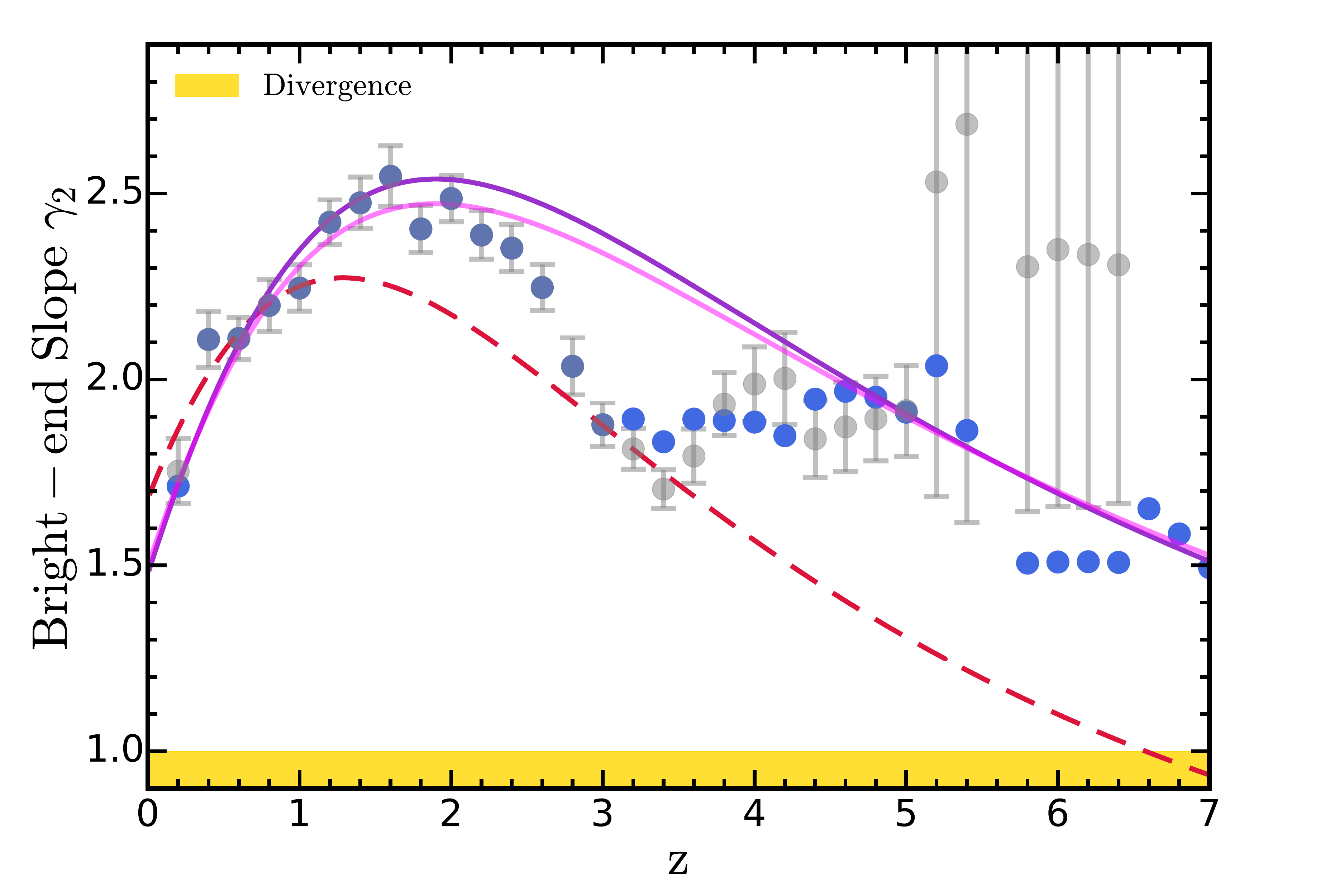}
    \includegraphics[width=0.48\textwidth]{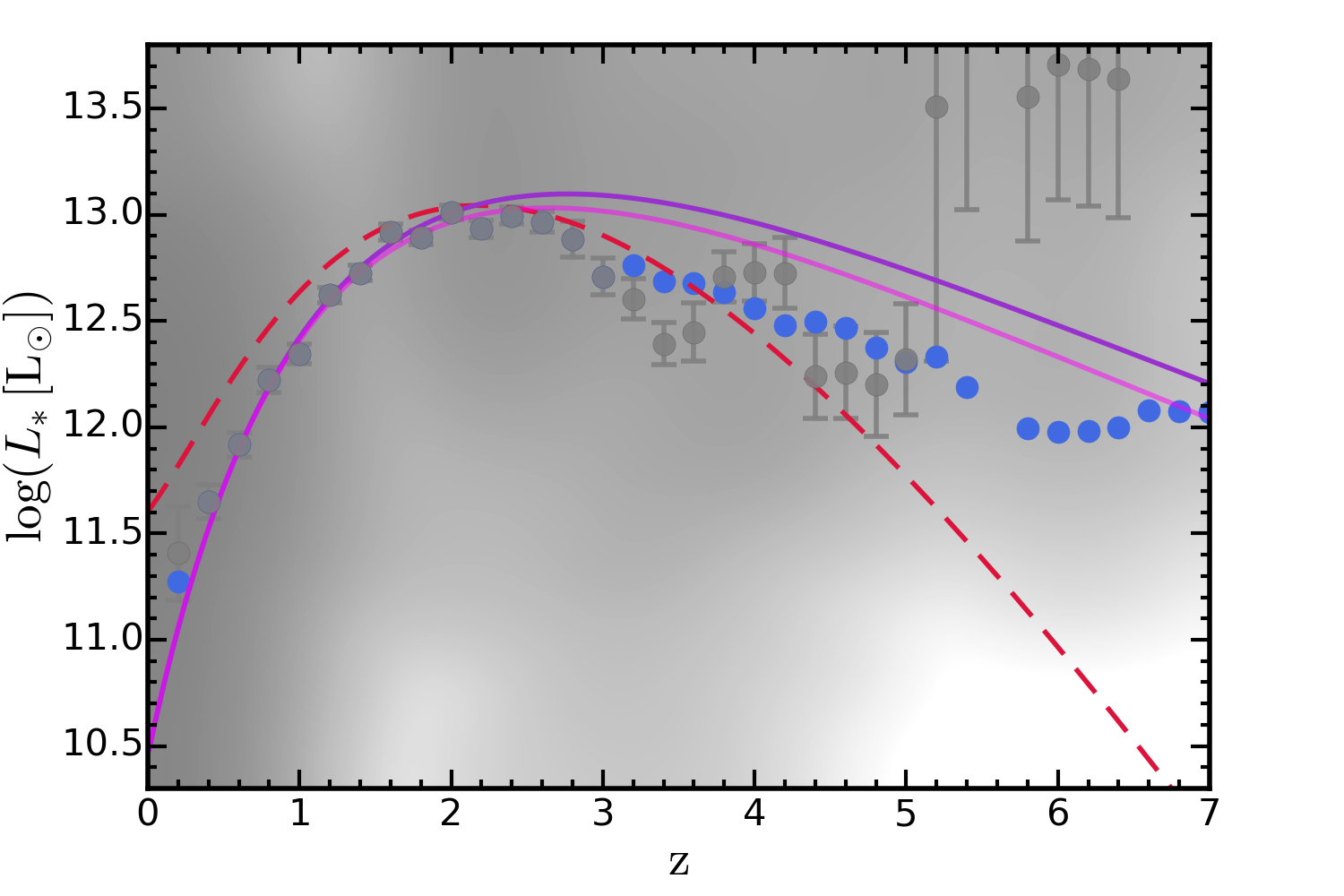}
    \includegraphics[width=0.48\textwidth]{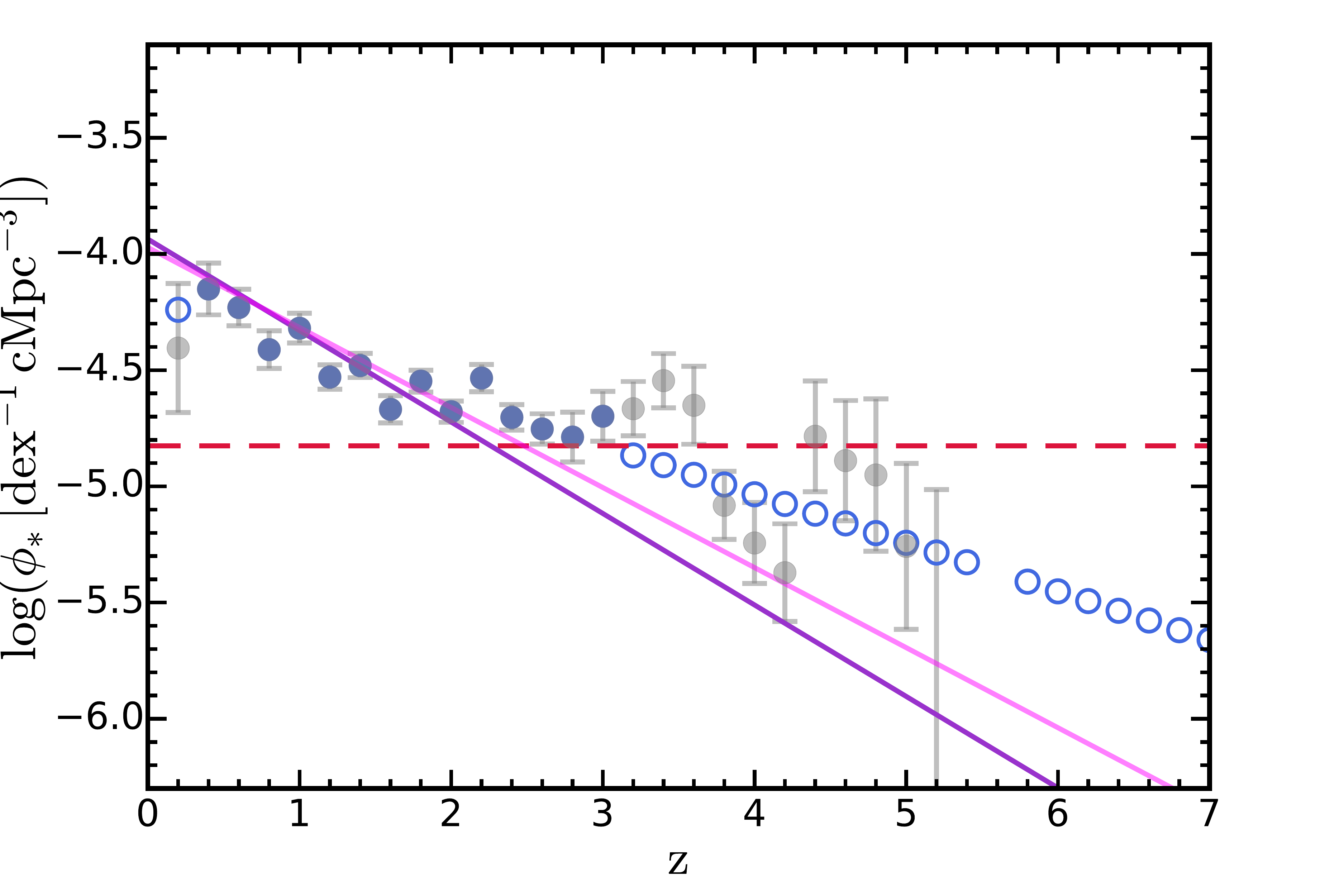}
    \caption{\textbf{Best-fit double power-law parameters of the bolometric QLF at each redshift.} Gray points are the best-fits at individual redshifts (local "free" fits) with error bars indicating $1\sigma$ uncertainties. Blue points are the best-fits when the evolution of $\phi_{\ast}(z)$ is fixed (local "polished" fits). The blue open circles in the bottom right panel indicate where $\phi_{\ast}(z)$ is fixed. For the evolution of the bolometric QLF, the purple (pink) solid lines show the results from the global fit A (B). We compare the evolution of these parameters with that constrained by \citet{Hopkins2007} shown in red dashed lines. In the top left (right) panel, the yellow dashed line (shaded region) indicates where integrated luminosity at the faint (bright) end will diverge. In the bottom left panel, the colormap shows smoothed distribution of the observational data points converted onto the bolometric plane with the bolometric corrections. Darker colors indicate regions with more data points. At $z\gtrsim 5$, the void of data points approaches the break luminosity, indicating that the fits at those redshifts are potentially affected by limited data points at the faint end.}
    \label{fig:fit_at_z_parameters}
\end{figure*}

\section{Bolometric Quasar Luminosity Function}
\label{sec:LF}
\subsection{Bolometric quasar luminosity function at a certain redshift}
\label{sec:LF_at_z}

We first study the bolometric QLF at a certain redshift. Following the standard practice, we parameterize the bolometric QLF with a double power-law:
\begin{equation}
    \phi_{\rm bol}(L)=\dfrac{{\rm d}n}{{\rm d}\log{L}}=\dfrac{\phi_{\ast}}{(L/L_{\ast})^{\gamma_1}+(L/L_{\ast})^{\gamma_2}},
\end{equation}
where $\phi_{\ast}$ is the comoving number density normalization, $L_{\ast}$ is the break luminosity, $\gamma_1$ and $\gamma_2$ are the faint-end and bright-end slopes respectively. We note that the conventions for double power-law are sometimes different. In optical/UV studies, the double power-law is usually defined as:
\begin{equation}
    \dfrac{{\rm d}n}{{\rm d}L}=\dfrac{\phi^{\prime}_{\ast}/L_{\ast}}{(L/L_{\ast})^{-\alpha}+(L/L_{\ast})^{-\beta}},
\end{equation}
or per unit absolute magnitude as:
\begin{equation}
    \dfrac{{\rm d}n}{{\rm d}M}=\dfrac{\phi^{\prime\prime}_{\ast}}{10^{0.4(\alpha+1)(M-M_{\ast})}+10^{0.4(\beta+1)(M-M_{\ast})}},
\end{equation}
where $\phi^{\prime}_{\ast}$ and $\phi^{\prime\prime}_{\ast}$ are the comoving number density normalizations with different units, $M_{\ast}$ is the break magnitude, $\alpha$ and $\beta$ are the faint-end and bright-end slopes respectively. In our notation, it gives $\alpha=-(\gamma_1+1),\, \beta=-(\gamma_2+1),\, \phi^{\prime}_{\ast}=\phi_{\ast}/\ln{10}$ and $\phi^{\prime\prime}_{\ast}=0.4\phi_{\ast}$. 

For a given bolometric QLF, we can convolve it with the bolometric corrections and extinction corrections discussed in Section~\ref{sec:model} to get the predicted observed QLF in a certain band at the redshift we study. We fit the parameters of the bolometric QLF to match the prediction with the observational binned estimations in all bands at the redshift. We select binned estimations of the QLF from our observation compilation listed in Table~\ref{tab:observations}. A data set is selected if the redshift bin of that observation covers the redshift we study. Since the statistical mean redshift of the quasar samples in the binned estimations in observations does not necessarily perfectly match the redshift we study, we correct the binned estimations with a model-dependent method~(referred to as "number density correction" in this paper). To be specific, for each data set in the UV, we first use the UV QLF model constrained by \citet{Kulkarni2018}~(the Model 2 of the paper) to calculate the "expected" number densities at the redshift we study and at the luminosities where the data points are located. Then, we calculate the mean of the logarithm of the "expected" number densities, representing a mean level of quasar number density. Since the observed quasar samples may center on a slightly different redshift, it is likely that the observed data points exhibit a systematic shift from this "expected" mean level of number density. So we rescale the observed data points to have the "expected" mean value at the redshift we study. We also perform this correction to the X-ray data points with the X-ray QLF model constrained by \citet{Miyaji2015} and to the IR data points with the IR QLF models constrained therein. We note that this correction is model-dependent but the models we choose are representative and have the widest redshift coverage in their bands. They are in good agreement with the observations in their bands. In most of the cases, this correction step improves the clustering of data points from different investigations and reduces the potential bias in redshift estimations of observations. Combining all corrected data points, we can derive the best-fit parameters of the bolometric QLF. The best-fit parameters at some selected redshifts are listed in Table~\ref{tab:parameters_at_z}. The best-fits at all selected redshifts are shown in Figure~\ref{fig:fit_at_z_parameters} with gray points. In the following, we will refer to these fits as the local "free" fits~(see Table~\ref{tab:fits} for details), since none of the parameters are fixed during fitting.

\begin{table}
\raggedright
{\bf Local "free" fits:}\\
\centering
\begin{tabular}{
p{0.01\textwidth}|p{0.085\textwidth}|p{0.085\textwidth}|p{0.095\textwidth}|p{0.095\textwidth}}
\hline 
\hline
z & $\gamma_1$ & $\gamma_2$ & $\log{\phi_{\ast}}$ & $\log{L_{\ast}}$ \\
\hline
\hline
0.2 & $0.812 \pm 0.046$ & $1.753 \pm 0.087$ & $-4.405 \pm 0.278$ & $11.407 \pm 0.223$ \\
0.4 & $0.561 \pm 0.041$ & $2.108 \pm 0.075$ & $-4.151 \pm 0.111$ & $11.650 \pm 0.080$ \\
0.8 & $0.599 \pm 0.031$ & $2.199 \pm 0.070$ & $-4.412 \pm 0.080$ & $12.223 \pm 0.059$ \\
1.2 & $0.504 \pm 0.030$ & $2.423 \pm 0.060$ & $-4.530 \pm 0.052$ & $12.622 \pm 0.036$ \\
1.6 & $0.484 \pm 0.034$ & $2.546 \pm 0.082$ & $-4.668 \pm 0.058$ & $12.919 \pm 0.040$ \\
2.0 & $0.411 \pm 0.029$ & $2.487 \pm 0.063$ & $-4.679 \pm 0.046$ & $13.011 \pm 0.032$ \\
3.0 & $0.424 \pm 0.070$ & $1.878 \pm 0.058$ & $-4.698 \pm 0.107$ & $12.708 \pm 0.086$ \\
4.0 & $0.403 \pm 0.162$ & $1.988 \pm 0.099$ & $-5.244 \pm 0.174$ & $12.730 \pm 0.134$ \\
5.0 & $0.260 \pm 0.425$ & $1.916 \pm 0.123$ & $-5.258 \pm 0.357$ & $12.319 \pm 0.261$ \\
6.0 & $1.196 \pm 0.246$ & $2.349 \pm 0.692$ & $-8.019 \pm 1.099$ & $13.709 \pm 0.639$ \\
\hline
\end{tabular}
\\
\raggedright
{\bf Local "polished" fits:}\\
\centering 
\begin{tabular}{
p{0.01\textwidth}|p{0.085\textwidth}|p{0.085\textwidth}|p{0.095\textwidth}|p{0.095\textwidth}}
\hline
\hline
0.2 & $0.787 \pm 0.024$ & $1.713 \pm 0.046$ & $-4.240$           & $11.275 \pm 0.023$ \\ 
0.4 & $0.561 \pm 0.041$ & $2.108 \pm 0.075$ & $-4.151 \pm 0.111$ & $11.650 \pm 0.080$ \\ 
0.8 & $0.599 \pm 0.031$ & $2.199 \pm 0.070$ & $-4.412 \pm 0.080$ & $12.223 \pm 0.059$ \\ 
1.2 & $0.504 \pm 0.030$ & $2.423 \pm 0.060$ & $-4.530 \pm 0.052$ & $12.622 \pm 0.036$ \\ 
1.6 & $0.484 \pm 0.034$ & $2.546 \pm 0.082$ & $-4.668 \pm 0.058$ & $12.919 \pm 0.040$ \\ 
2.0 & $0.411 \pm 0.029$ & $2.487 \pm 0.063$ & $-4.679 \pm 0.046$ & $13.011 \pm 0.032$ \\ 
3.0 & $0.424 \pm 0.070$ & $1.878 \pm 0.058$ & $-4.698 \pm 0.107$ & $12.708 \pm 0.086$ \\ 
4.0 & $0.213 \pm 0.092$ & $1.885 \pm 0.052$ & $-5.034$           & $12.562 \pm 0.027$ \\ 
5.0 & $0.245 \pm 0.211$ & $1.912 \pm 0.086$ & $-5.243$           & $12.308 \pm 0.062$ \\ 
6.0 & $1.509 \pm 0.058$ & $1.509 \pm 0.058$ & $-5.452$           & $11.978 \pm 0.055$ \\ 
\hline
\end{tabular}
\caption{The best-fit double power-law parameters of the bolometric QLF at selected redshifts. We present the results of the local "free" and the local "polished" fits~(see Table~\ref{tab:fits} for details).}
\label{tab:parameters_at_z}
\end{table}

\begin{table}
\centering
\begin{tabular}{
p{0.08\textwidth}|p{0.08\textwidth}|p{0.12\textwidth}|p{0.10\textwidth}}
\hline 
\hline
 & Parameter & Best-fit A & Best-fit B\\
\hline
\hline
$\gamma_1$ & $a_0$ &$\hspace{1.8mm}0.8569^{+0.0247}_{-0.0253}$ &$\hspace{1.8mm}0.3653^{+0.0115}_{-0.0114}$\\
           & $a_1$ &$-0.2614^{+0.0162}_{-0.0164}$              &$-0.6006^{+0.0422}_{-0.0417}$\\
           & $a_2$ &$\hspace{1.8mm}0.0200^{+0.0011}_{-0.0011}$ &\\
$\gamma_2$ & $b_0$ &$\hspace{1.8mm}2.5375^{+0.0177}_{-0.0187}$ &$\hspace{1.8mm}2.4709^{+0.0163}_{-0.0169}$\\
           & $b_1$ &$-1.0425^{+0.0164}_{-0.0182}$              &$-0.9963^{+0.0167}_{-0.0161}$\\
           & $b_2$ &$\hspace{1.8mm}1.1201^{+0.0199}_{-0.0207}$ &$\hspace{1.8mm}1.0716^{+0.0180}_{-0.0181}$\\
$L_{\ast}$ & $c_0$ &$\hspace{0.4mm}13.0088^{+0.0090}_{-0.0091}$&$\hspace{0.4mm}12.9656^{+0.0092}_{-0.0089}$\\
           & $c_1$ &$-0.5759^{+0.0018}_{-0.0020}$              &$-0.5758^{+0.0020}_{-0.0019}$\\
           & $c_2$ &$\hspace{1.8mm}0.4554^{+0.0028}_{-0.0027}$ &$\hspace{1.8mm}0.4698^{+0.0025}_{-0.0026}$\\
$\phi_{\ast}$ & $d_0$ &$-3.5426^{+0.0235}_{-0.0209}$           &$-3.6276^{+0.0209}_{-0.0203}$\\
              & $d_1$ &$-0.3936^{+0.0070}_{-0.0073}$           &$-0.3444^{+0.0063}_{-0.0061}$\\
\hline
\end{tabular}
\caption{The best-fit parameters of the global evolution model of the bolometric QLF~(see Equation~\ref{eq:fitformulae} and Equation~\ref{eq:fitformulae2}). We present the best-fit parameters of the global fits A and B~(see Table~\ref{tab:fits} for details), respectively.}
\label{tab:parameters_global}
\end{table}

\begin{figure*}
    \centering
    \includegraphics[width=0.48\textwidth]{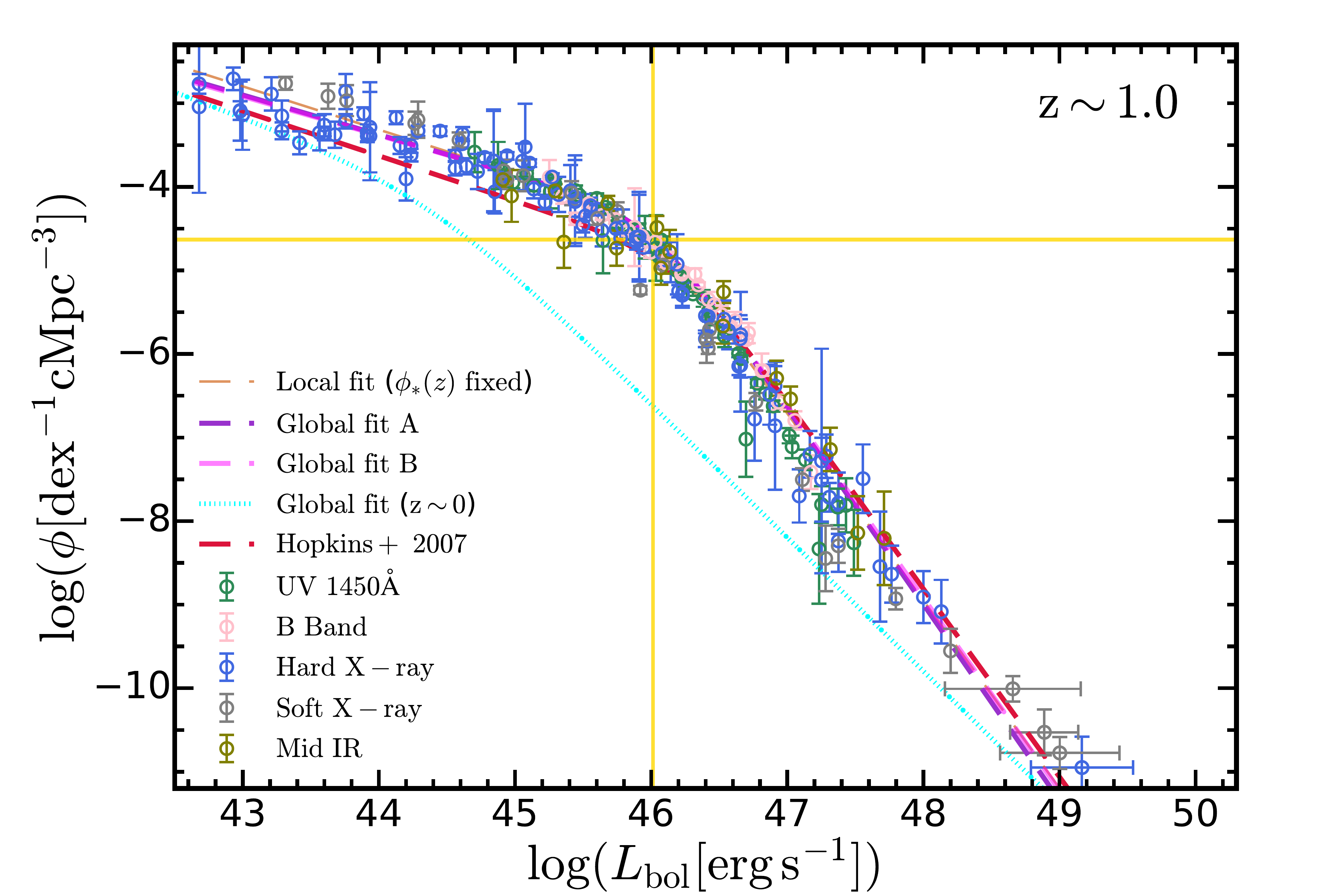}
    \includegraphics[width=0.48\textwidth]{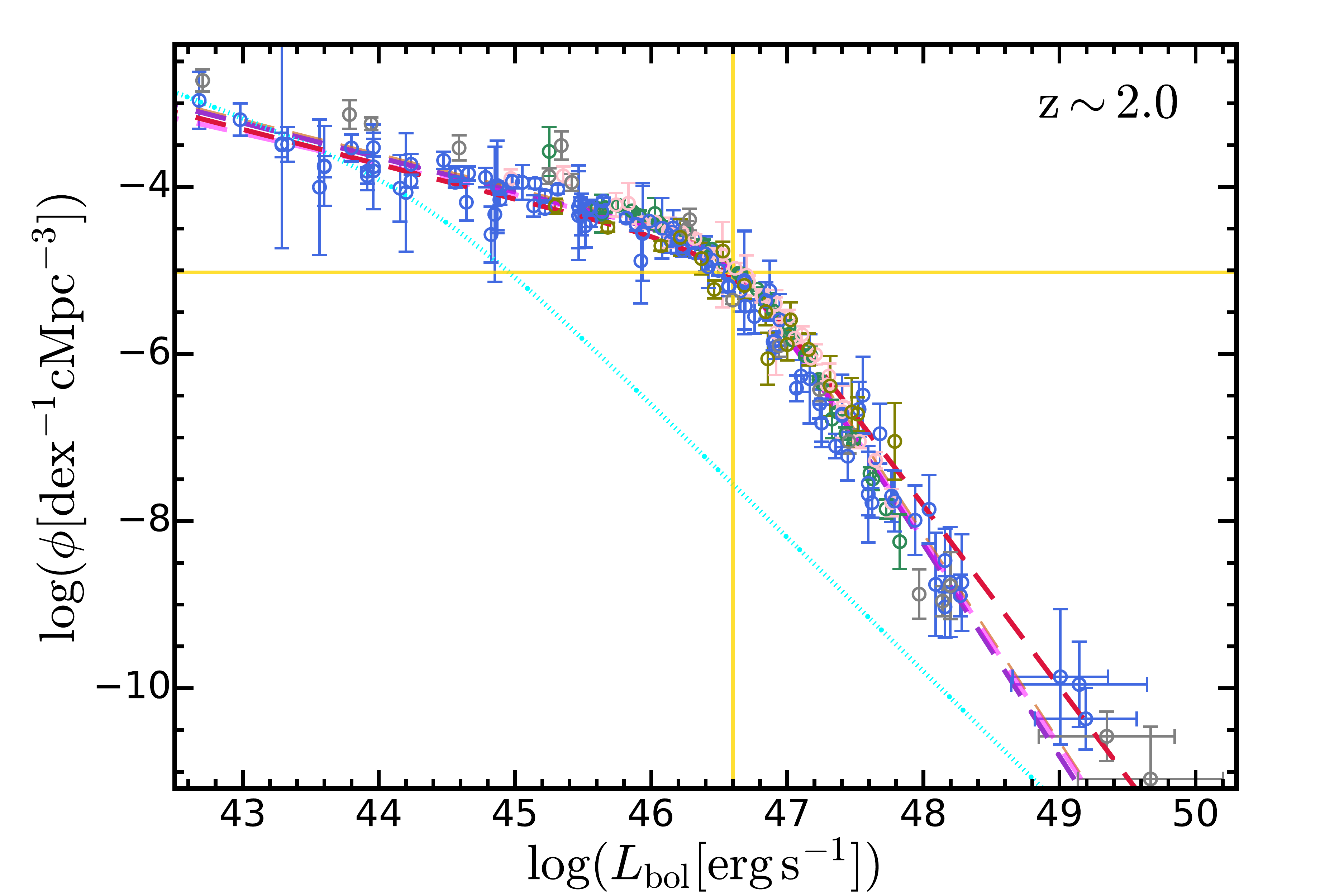}
    \includegraphics[width=0.48\textwidth]{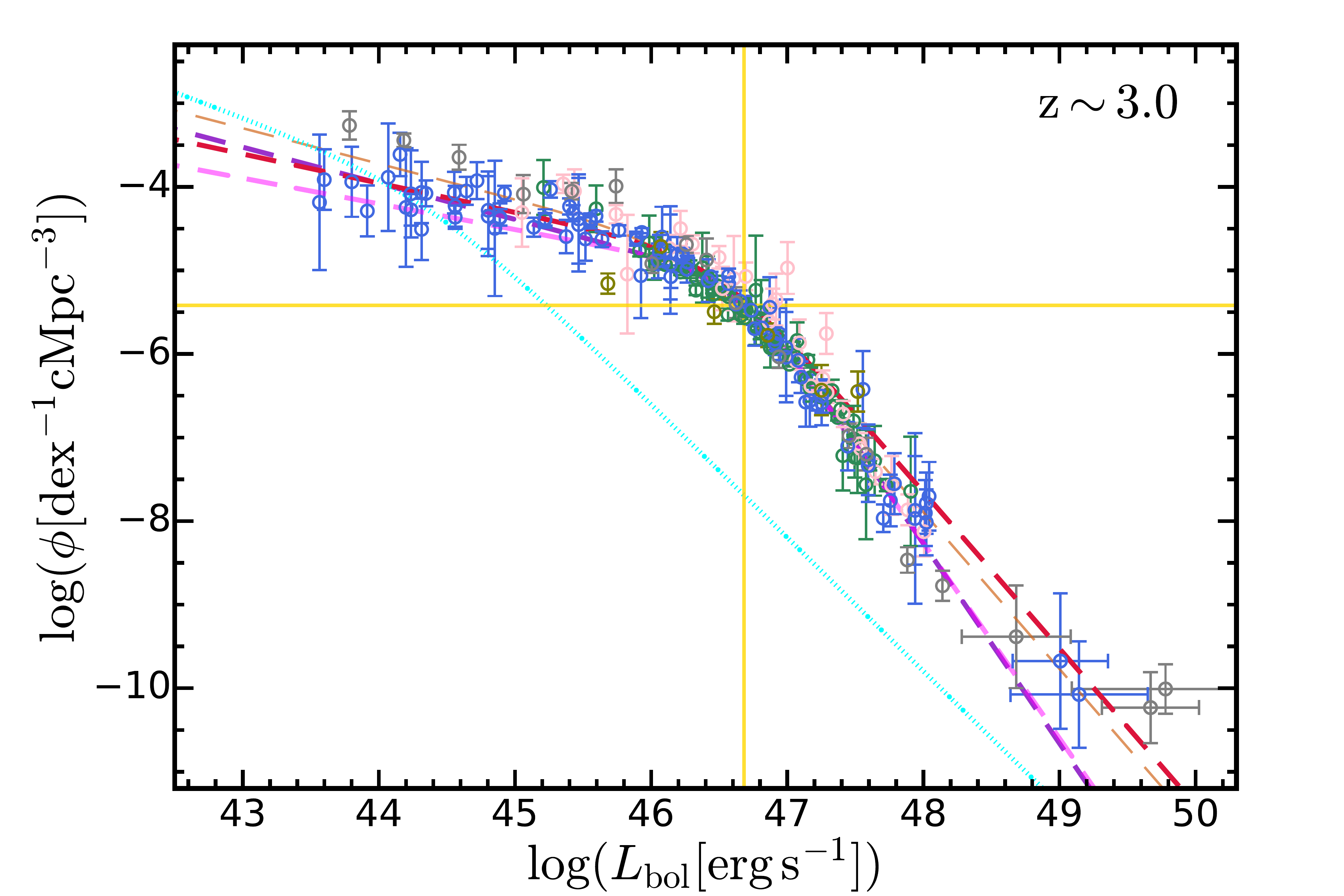}
    \includegraphics[width=0.48\textwidth]{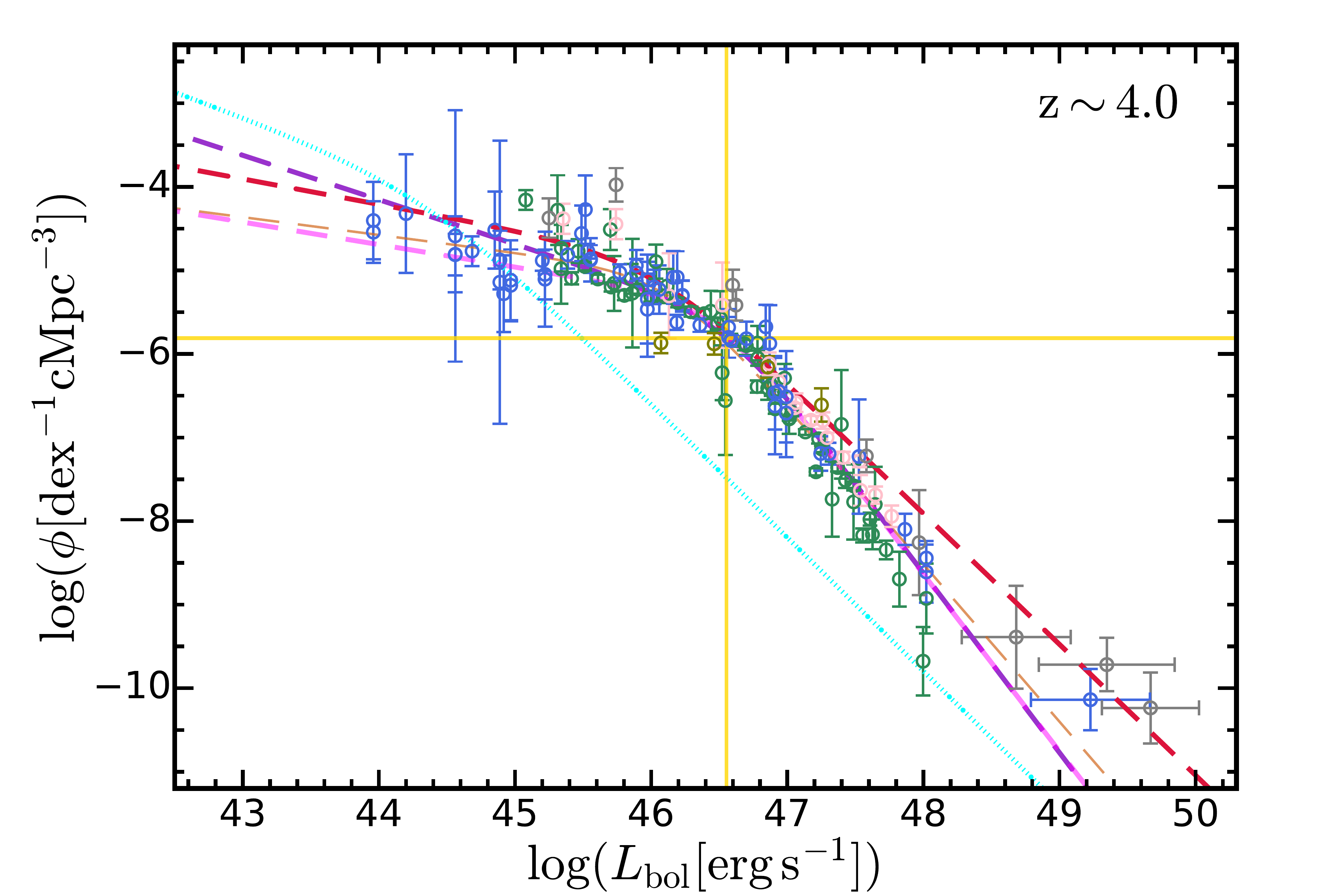}
    \includegraphics[width=0.48\textwidth]{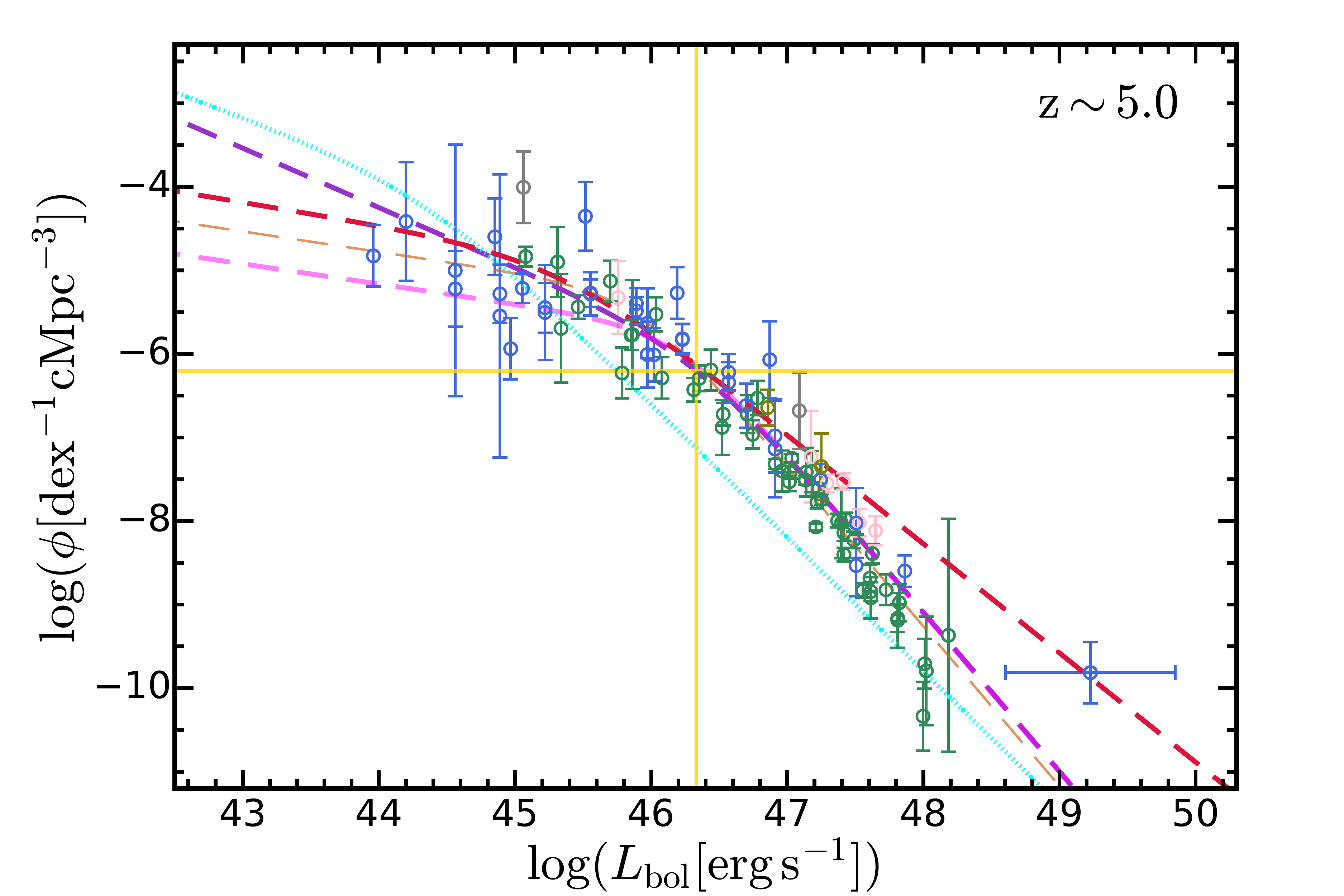}
    \includegraphics[width=0.48\textwidth]{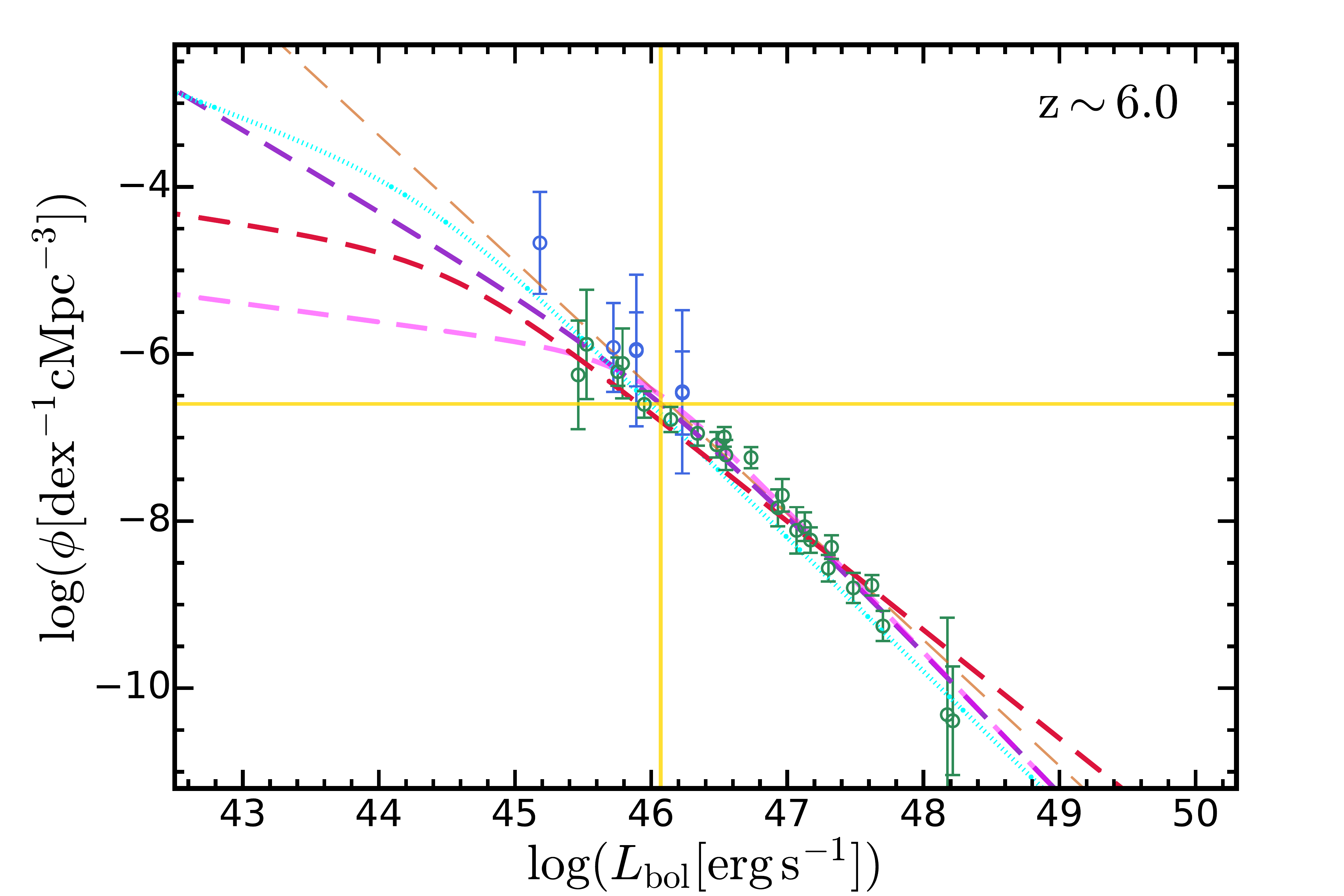}
    \caption{\textbf{Best-fit bolometric QLFs at $6$ selected redshifts.} We compare the predictions with the observational binned estimations converted onto the bolometric plane. We present the best-fit bolometric QLFs at individual redshifts (local "free" fits) in orange dashed lines. The bolometric QLFs constrained by our global fit A (B) are shown in purple (pink) dashed lines. The bolometric QLFs from \citet{Hopkins2007} are also presented in red dashed lines for comparison. The $z=0$ bolometric QLF constrained by our global fit A is shown in cyan dashed lines. The observational data are converted onto the bolometric plane with the bolometric corrections and the $N_{\rm obs}/N_{\rm mod}$ method. The vertical and horizontal yellow lines show the break luminosity and the number density normalization at each redshift.}
    \label{fig:LF_bol}
\end{figure*}

Since the parameters of the double power-law bolometric QLF have significant degeneracy, which manifests as large covariance in fitting, the best-fit parameters exhibit large coherent fluctuations at some redshifts. The degeneracy prevents us from finding the optimal functional form to describe the redshift evolution of the parameters. To improve the fits, we fix the number density normalization to depend linearly on redshift which is quite clear even in the "free" fits. The linear relation is determined by the best-fits at $z=0.4-3.0$. We then redo the fitting at redshifts outside $z=0.4-3.0$ with $\phi_{\ast}(z)$ fixed. Apart from that, we find that the bolometric QLF at $z\geq 5.8$ behaves as a single power-law at least in the regime covered by existing observations. Thus we reduce the fitting formula to a single power-law by restricting the faint and bright-end slope to be the same at these redshifts. The fitting procedure with these updates is referred to as the local "polished" fits~(see Table~\ref{tab:fits} for details). The "polished" best-fits are also shown in Figure~\ref{fig:fit_at_z_parameters} with blue points. Based on the local "polished" fits, the bright-end slope and break luminosity evolution clearly have a double power-law shape, similar to what was seen in \citet{Hopkins2007}, and the faint-end slope has a polynomial-like dependence on redshift.

\subsection{Parameterized evolution model of the bolometric QLF}
\label{sec:global_fit}

In this section, we aim to describe the evolution of the bolometric QLF with simple formulae and to perform a global fit on all the observational data at all redshifts. Following the discussion in the previous section, we describe the QLF as a double power-law with parameters that evolve with redshift as:
\begin{align}
    & \gamma_1(z) = a_0 T_0(1+z) + a_1 T_1(1+z) + a_2 T_2(1+z); \nonumber \\
    & \gamma_2(z) = \dfrac{2\,b_0}{ \Big(\dfrac{1+z}{1+z_{\rm ref}}\Big)^{b_1} + \Big(\dfrac{1+z}{1+z_{\rm ref}}\Big)^{b_2} };\nonumber \\
    & \log{L_{\ast}}(z) = \dfrac{2\,c_0}{ \Big(\dfrac{1+z}{1+z_{\rm ref}}\Big)^{c_1} + \Big(\dfrac{1+z}{1+z_{\rm ref}}\Big)^{c_2} }; \nonumber \\
    & \log{\phi_{\ast}}(1+z) = d_0 T_0(1+z)+ d_1 T_1(1+z), \label{eq:fitformulae} \\
    & \Big(T_0(x) = 1,\, T_1(x) = x,\, T_2(x) = 2x^2-1\Big) \nonumber
\end{align}
where $T_{\rm n}$ is the n-th order Chebyshev polynomial and $z_{\rm ref}$ is chosen to be $2$. The evolution of the bolometric QLF is therefore controlled by 11 parameters: \{$a_0$, $a_1$, $a_2$\}; \{$b_0$, $b_1$, $b_2$\}; \{$c_0$, $c_1$, $c_2$\}; \{$d_0$, $d_1$\}. This parameterization is adequate to describe the evolution of the bolometric QLF parameters. We have tried to extend the parameterization with higher order polynomials and find their contributions are negligible.

In the next step, we perform a global fit~(referred to as the global fit A, see Table~\ref{tab:fits} for details) on all the observational data from the compilation at all redshifts simultaneously. To do this, we adopt a Monte Carlo Markov Chain (MCMC) method using the {\bf emcee}~\footnote{\href{https://emcee.readthedocs.io/en/stable/}{https://emcee.readthedocs.io/en/stable/}} package~\citep{emcee}. Given a proposed parameter set of the evolution model, we calculate the resulting observed QLF in bands and compare that with observational data. For a redshift bin of a given data set, the predicted observed QLF is calculated at the center of the redshift bin. The observational data points are also rescaled to the center of the redshift bin with the number density correction discussed in the previous section. The likelihood function is then calculated in a standard way:
\begin{equation}
    \ln{\mathcal{L}} = -\dfrac{1}{2}\sum\limits_{n} \Bigg[ W(z_{\rm n})\dfrac{(\log{\phi_{\rm mod}} - \log{\phi_{\rm obs})^{2}} }{\sigma^{2}_{\rm n}} + \ln(2\pi\sigma^{2}_{\rm n})\Bigg],
\end{equation}
where $\log{\phi_{\rm mod}}$ and $\log{\phi_{\rm obs}}$ are the predicted and observed number density respectively, $\sigma_{\rm n}$ is the uncertainty of the measurement and $W(z)$ is a weighting function introduced to balance the statistical power of high and low redshift data. (Otherwise, the fact that there is more data at low redshifts would skew the fits, sacrificing large discrepancies at high redshifts for marginal improvements at low redshifts.) The summation is taken over all the observational data points at all redshifts. We choose $W(z)=1$ when $z<3$, $W(z)=\Big(\dfrac{1+z}{1+3}\Big)^2$ when $3\leq z <4$ and $W(z)=\Big(\dfrac{1+z}{1+4}\Big)^3 \Big(\dfrac{1+4}{1+3}\Big)^2$ when $z\geq4$. This weighting function makes the weights of data points roughly the same at $z=2-6$ and helps achieve a converged and decent fit on high redshift data. We adopt uniform priors for all the parameters involved, so that the posterior probability function is the same as the likelihood function given above. The global best-fit parameters of this evolution model are listed in Table~\ref{tab:parameters_global} and this best-fit model will be referred to as the global fit A. In Figure~\ref{fig:LF_bol}, we show the best-fit bolometric QLFs at $6$ selected redshifts compared with the observational data converted onto the bolometric plane with the bolometric corrections and the $N_{\rm obs}/N_{\rm mod}$ method~(moving data points across different QLF planes by fixing the ratio between observed and model-predicted number densities). In general, the global fit A does comparably well to the local best-fit at each redshift in matching the observational data. The best-fit bolometric QLFs of the global fit A are qualitatively different from the \citet{Hopkins2007} model. The bright end of the QLF is steeper at $z\gtrsim 2$. The faint end of the QLF is steeper at $z\gtrsim3$ and becomes progressively steeper at higher redshifts. We achieve a better agreement with observations than the \citet{Hopkins2007} model at $z\gtrsim 3$. The evolution of the double power-law parameters of the bolometric QLF determined by the global fit A is also shown in Figure~\ref{fig:fit_at_z_parameters} with purple lines. In the top left (right) panel of Figure~\ref{fig:fit_at_z_parameters}, we indicate with yellow dashed line (shaded region) the regime where integrated luminosity towards infinite low (high) luminosity will diverge. Compared with \citet{Hopkins2007}, extrapolating our new model to $z>7$ will not lead to any divergence at the bright end. However, the integrated luminosity at the faint end will diverge at $z\gtrsim 6$ due to the steep faint-end slope constrained in the global fit A. In the bottom left panel of Figure~\ref{fig:fit_at_z_parameters}, the colormap shows smoothed distribution of the observational data points converted on to the bolometric plane with the bolometric corrections. Darker colors indicate regions with more data points. At $z\gtrsim5$, the void of data points approaches the break luminosity, indicating that the fits at those redshifts are affected by limited data points at the faint end. For example, in the extreme case that there are no data points fainter than the break luminosity, the fitted faint-end slope will simply be equal to the bright-end slope. The steepening of the faint-end slope at high redshift we find in both the local fits and the global fit A may be seriously affected by this. So, in parallel to the global fit A introduced above, we perform another independent global fit~(referred to as the global fit B, see Table~\ref{tab:fits} for details) assuming a different evolution model for the faint-end slope. We adopt the function form used in \citet{Hopkins2007}:
\begin{equation}
    \gamma_1(z) = a_0 \Big(\dfrac{1+z}{1+z_{\rm ref}}\Big)^{a_1},
    \label{eq:fitformulae2}
\end{equation}
where $z_{\rm ref}$ is again chosen to be $2$. Different from the function form used in the global fit A, the faint-end slope here is restricted to evolve monotonically as a function of redshift. This will by construction prevent the steepening of the faint-end slope at high redshift. Other double-power-law parameters have the same evolution model as the global fit A. We perform exactly the same fitting procedure for this model and the results are also shown in Figure~\ref{fig:fit_at_z_parameters} with pink lines. The best-fit bright-end slopes, break luminosities and number density normalizations of the global fit B are similar to those of the global fit A. However, the faint-end slope in this model remains shallow at $z\gtrsim 5$ on contrary to the global fit A. The best-fit bolometric QLFs of the global fit B are also presented in Figure~\ref{fig:LF_bol} compared with observational binned estimations. The global fit B is consistent with observations equally well as the global fit A while behaves qualitatively differently at the faint end at $z\gtrsim 5$. Future observations are required to test these two models.

The evolution models of the bolometric QLF we described above are constrained by observational data at $0<z<7$. Making predictions beyond the redshift frontier certainly requires extrapolations of the model. For the global fit A, the best-fit faint-end slope becomes the same as the bright-end slope at $z\sim 7$ and the double power-law bolometric QLF tends to behave like a single power-law approaching $z\sim 6-7$. Therefore, extrapolating to $z>7$, we postulate that the bolometric QLF simply has a single power-law shape. The evolution of the single power-law slope follows the extrapolation of the evolution of the bright-end slope at $z<7$. On the other hand, for the global fit B, the faint-end slope remains shallow at $z>7$ and we can simply extrapolate the evolution of the parameters to get luminosity functions at $z>7$. We note all these extrapolations involve assumptions on the shape of the QLF in the regime where no observational evidence is available. There are serious uncertainties there.

\begin{figure}
    \centering
    \includegraphics[width=0.48\textwidth]{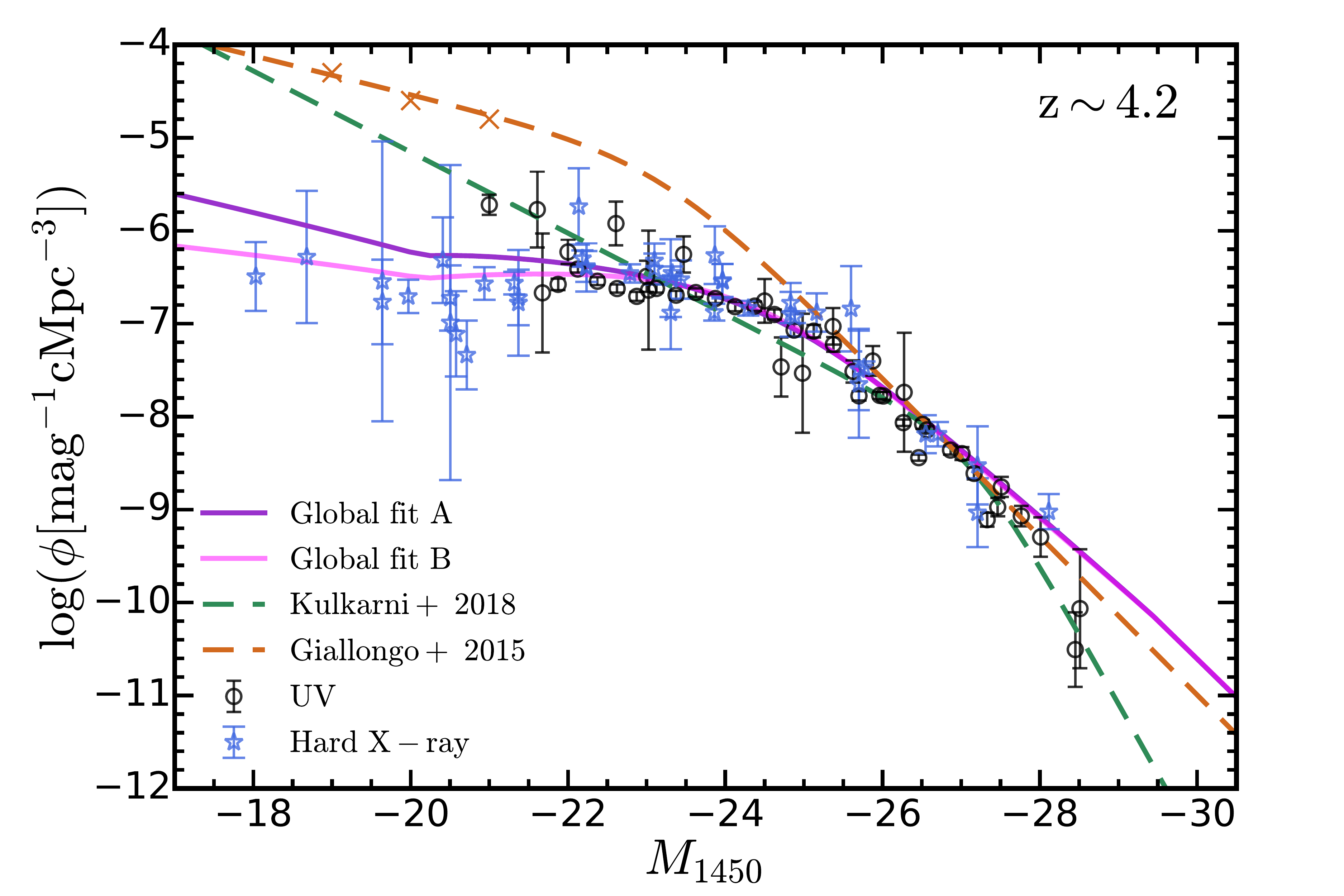}
    \includegraphics[width=0.48\textwidth]{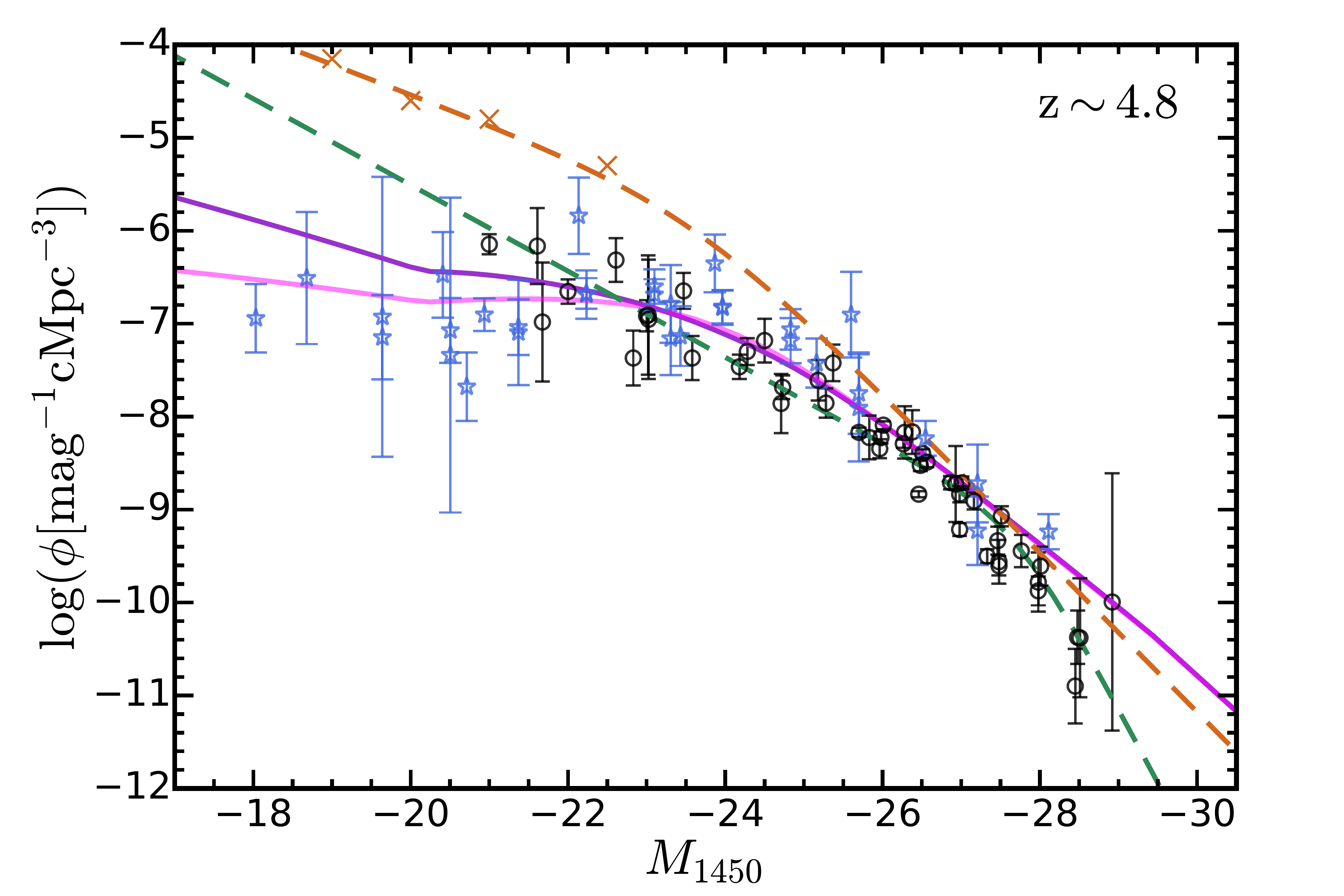}
    \includegraphics[width=0.48\textwidth]{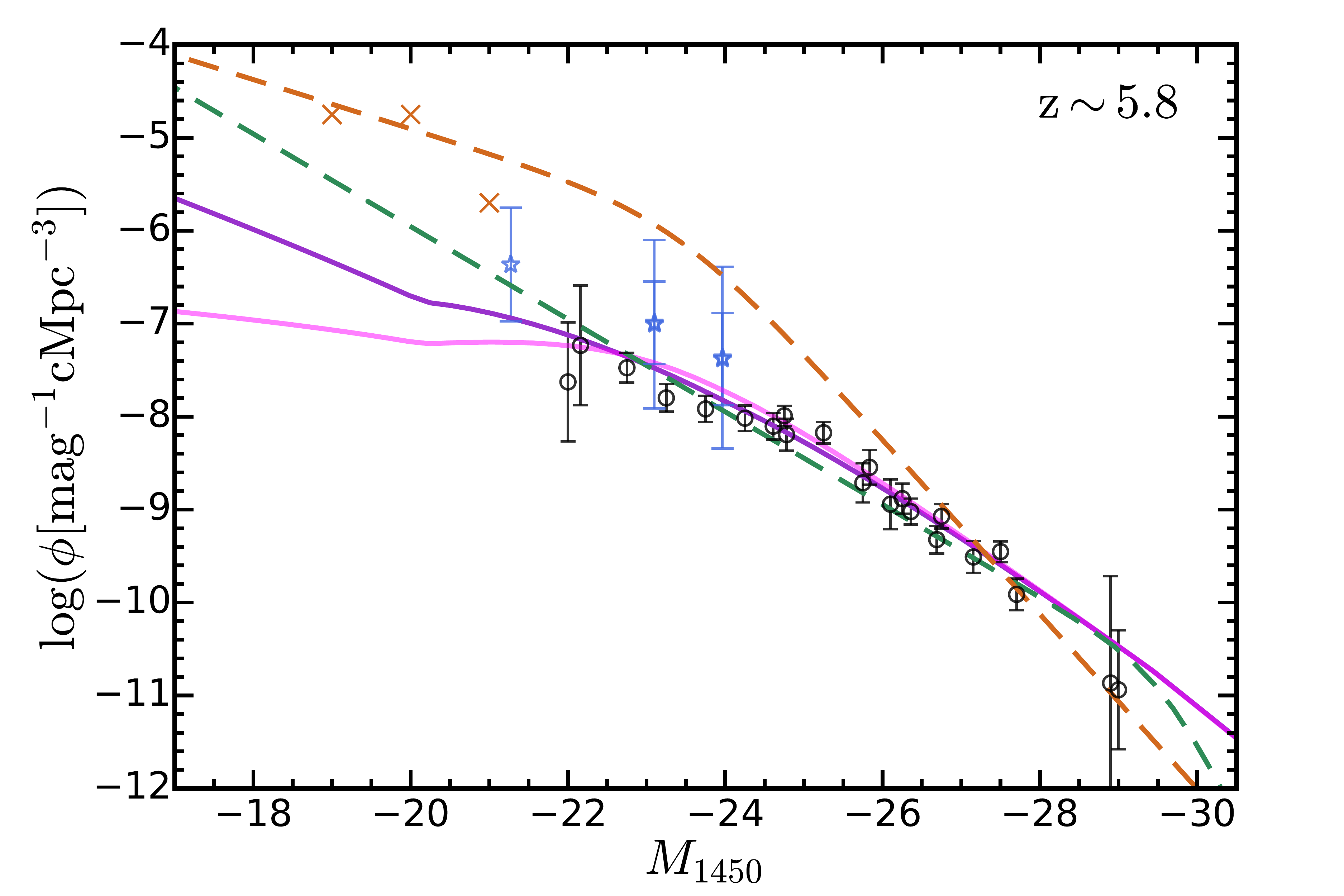}
    \caption{\textbf{UV QLFs at $z=4.2,4.8,5.8$.} The observational data points in the UV are shown in black circles. The observational data points in the X-ray are moved onto the UV QLF plane with the $N_{\rm obs}/N_{\rm mod}$ method and are shown in blue stars. The predicted UV QLFs from the global fits A and B are shown with purple and pink lines, respectively. The UV QLFs constrained in \citet{Kulkarni2018} is shown with green dashed lines. The UV QLF fitted by \citet{Giallongo2015} and their binned estimations are shown in orange dashed lines with orange crosses. The inferred high number density of quasars at the faint end is disfavored by X-ray observations.
    }
    \label{fig:tension}
\end{figure}

\subsection{Tensions in the UV QLF at $z = 4-6$}
\label{sec:tension}

The measurements of the UV QLF presented in \citet{Giallongo2015}, followed by the updates in \citet{Giallongo2019}, indicated a high number density of faint AGN at $z\sim 4-6$. This has motivated conjectures on whether quasars alone can be responsible for the reionization of hydrogen at $z>6$. However, other recent observations~\citep[e.g.,][]{Akiyama2018,Matsuoka2018,McGreer2018} have presented measurements that are in conflict with the \citet{Giallongo2015} results (as illustrated in Figure~4 in \citet{Giallongo2019}). These tensions serve as a reminder that the potential uncertainties associated with the selection of quasars and host galaxy contamination at high redshift are still substantial. 

In the fiducial analysis of this paper, we do not include the \citet{Giallongo2015} data in our fits. In order to check the robustness of our QLF constraints in the UV, we investigate the tensions in the UV QLF at $z\sim 4-6$ in Figure~\ref{fig:tension}. We show the UV QLF determinations with various approaches at $z=4.2,4.8,5.8$ including the \citet{Giallongo2015} measurement, the compiled observational binned estimations in the UV and X-ray and the prediction from the global fits A and B. The redshifts are chosen to be close to the centers of the redshift bins in \citet{Giallongo2015}. The \citet{Giallongo2015} data points (the orange crosses at the faint end) are clearly in tension with other observations in the intermediate luminosity range. The X-ray data points are moved onto the UV QLF plane with the $N_{\rm obs}/N_{\rm mod}$ method. They also disfavor the high number density of faint quasars measured by some UV observations. We show the UV QLFs constrained by \citet{Kulkarni2018} in green dashed lines. The overall normalization of the \citet{Kulkarni2018} QLFs is consistent with that of the X-ray data, despite a somewhat steeper evolved faint-end slope. Both of the global fits A and B achieve a better agreement with multi-band observational data than the \citet{Kulkarni2018} model, provided that the observational data themselves are internally consistent. At the faint end where observation data is limited, the global fits A and B behave differently. The global fit B predicts shallow faint-end slope at $z\gtrsim5$ while the global fit A predicts progressively steeper faint-end slope. Given current available observational constraints, we are not able to tell which model is more accurate.

\section{Evolution of the bolometric QLF}
\label{sec:LFE}
In this section, we explore the evolution of the bolometric QLF in detail and investigate the physical interpretation of the evolution based on our global fits discussed in Section~\ref{sec:LF}. 

\begin{figure}
    \centering
    \includegraphics[width=0.48\textwidth]{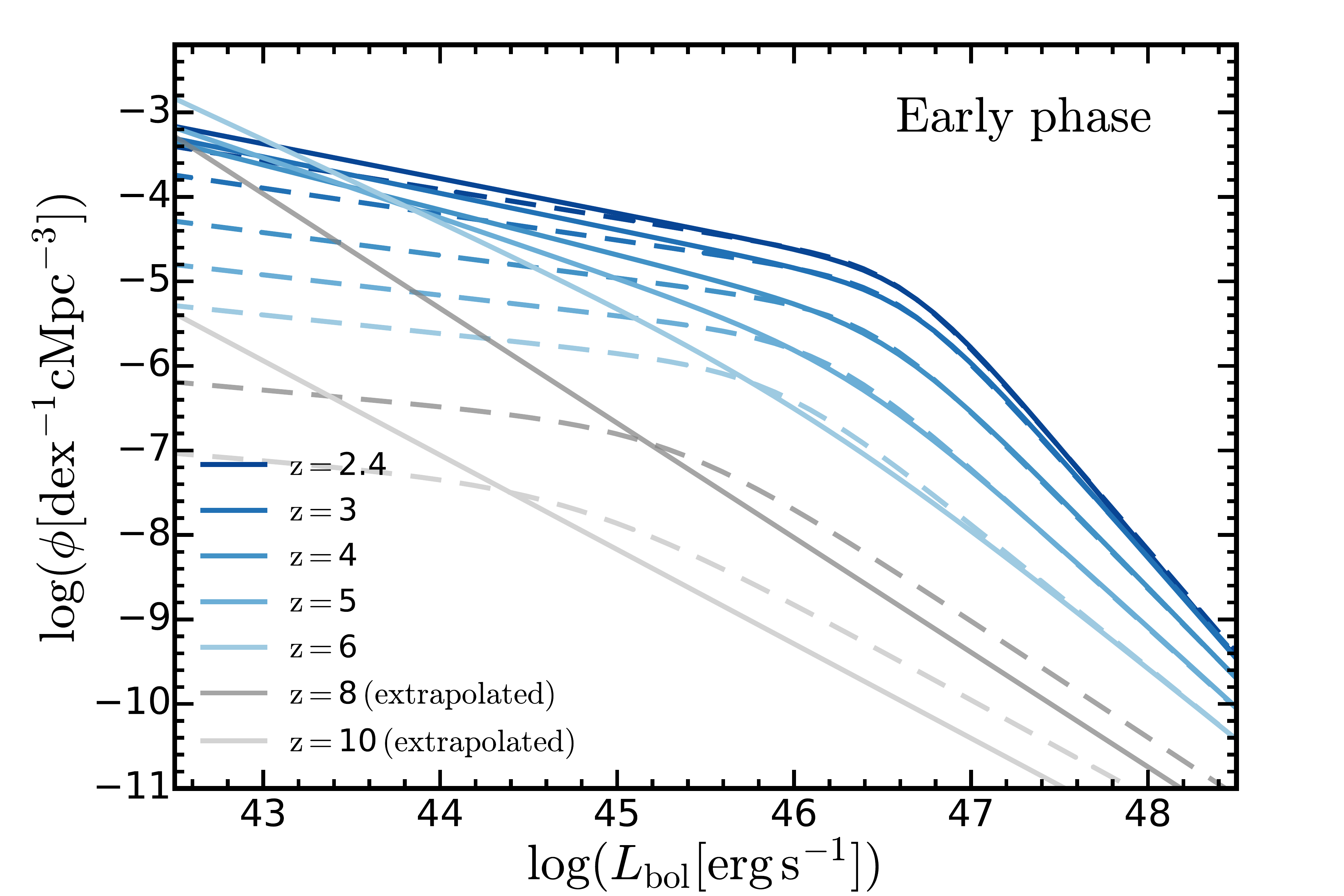}
    \includegraphics[width=0.48\textwidth]{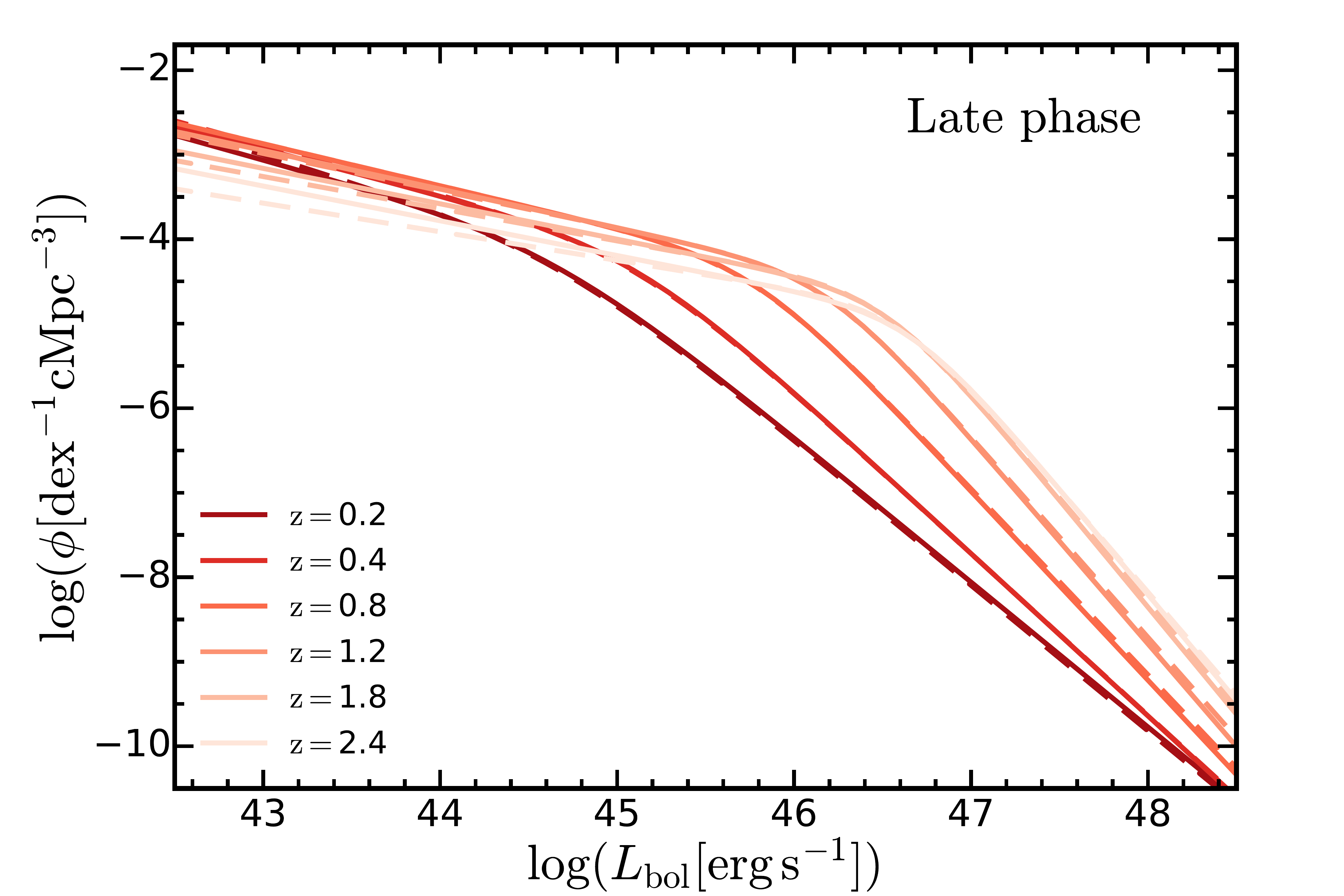}
    \caption{\textbf{Evolution of the bolometric QLF divided into two phases.} The solid (dashed) lines show predictions from the global fit A (B). The lines are color coded to indicate redshifts. The bolometric QLFs at higher redshifts are represented by lighter colors~(as labeled). At high redshift~("early phase"), the QLF rises up likely following the hierarchical build-up of structures in the Universe, similar to the halo mass functions. At low redshift~("late phase"), the characteristic luminosity of quasars declines rapidly and the QLF gets systematically shifted to the low luminosity regime, indicating "quenching". The two phases are separated at $z\simeq 2-3$. We also show the bolometric QLFs extrapolated to $z=8,10$~(see text in Section~\ref{sec:global_fit}).}
    \label{fig:LF_evolve}
\end{figure}

In Figure~\ref{fig:LF_evolve}, we compare the bolometric QLFs at different redshifts predicted by the global fits A (solid lines) and B (dashed lines). We divide the evolution of the bolometric QLF into two phases, the early phase at $z \gtrsim 2-3$ and late phase at $z\lesssim 2-3$. In the early phase, the bolometric QLF rises up monotonically following the hierarchical build-up of structures in the Universe. For the global fit A, approaching lower redshift, the relative abundance of faint quasars decreases accompanied by the increased abundance of brighter populations, forming a sharper "break" in the QLF. As a consequence of this change in the relative abundance, the faint-end slope becomes shallower and the bright-end slope becomes steeper. For the global fit B, the relative abundance of faint and bright quasars remains stable towards lower redshift, accompanied by the growth of the break luminosity. In both fits, the evolution at the bright end ($L_{\rm bol}\gtrsim 48$) is milder than that in the intermediate luminosity range. In the late phase, the bolometric QLF stops rising up. Instead, the bolometric QLF shows a systematic and continuous horizontal shift towards the low luminosity regime. The faint end has almost no evolution in this phase. This indicates processes other than the hierarchical build-up of structures dominating the evolution of the quasar population at late times. AGN feedback is potentially responsible for this evolutionary pattern. AGN feedback is believed to shut down the supply of cold gas to galaxy centers and thus could systematically decrease the bolometric quasar luminosities. Surprisingly, at $z\lesssim 0.5$, the bright end stops evolving and the bright-end slope becomes slightly shallower again. We note that the global fits A and B give similar evolutionary pattern in the late phase. Across the entire evolution history of the QLF, the evolution at the bright end of the bolometric QLF is apparently milder compared to other luminosity regimes. This suggests potential regulation on the abundance of the most luminous quasars. In Figure~\ref{fig:LF_evolve}, we also present the bolometric QLF extrapolated to $z=8,10$. The extrapolations are done as introduced in Section~\ref{sec:global_fit}. The rapidly dropping number density normalization makes the detection of quasars progressively difficult at these redshifts.

\begin{figure}
    \centering
    \includegraphics[width=0.48\textwidth]{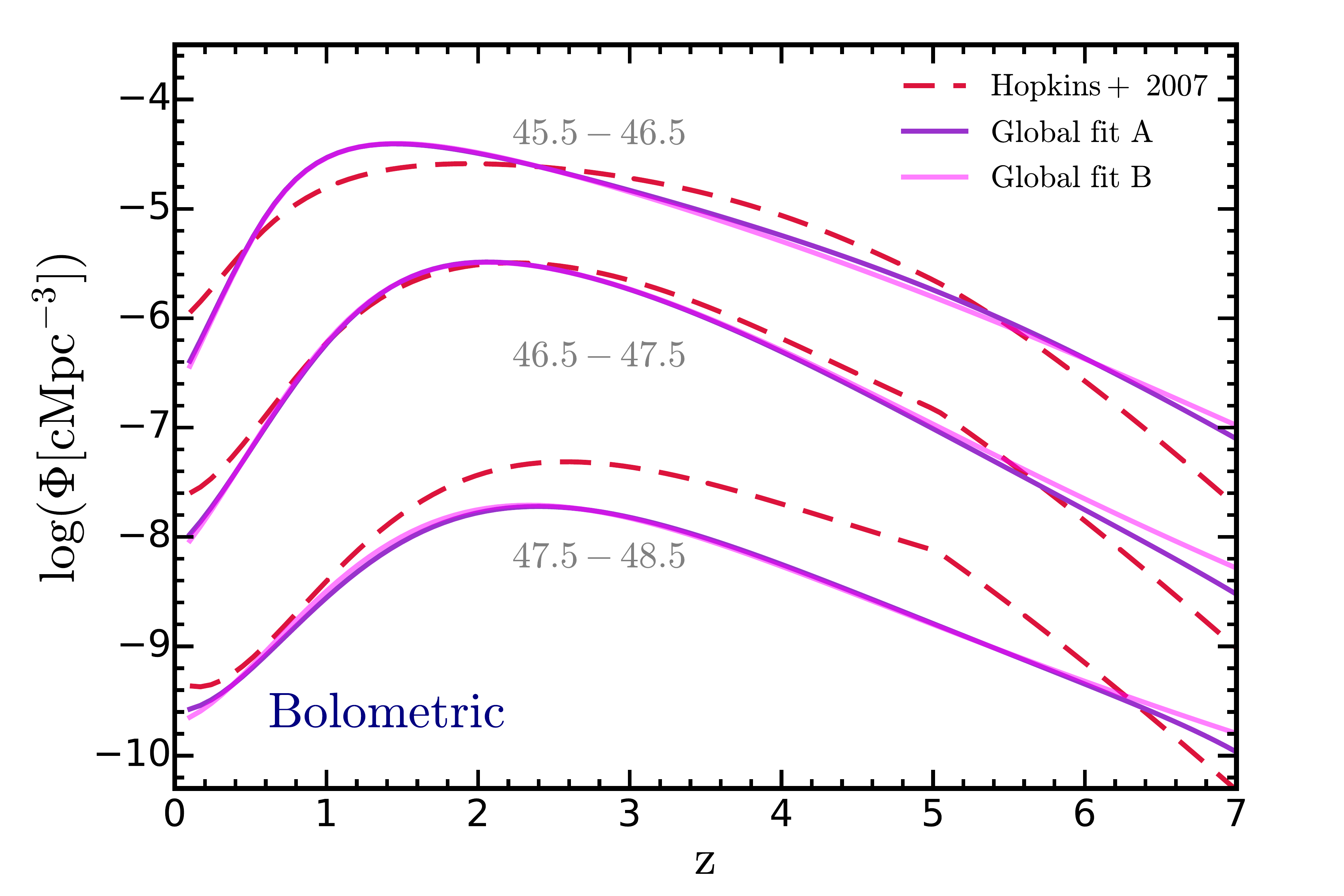}
    \includegraphics[width=0.48\textwidth]{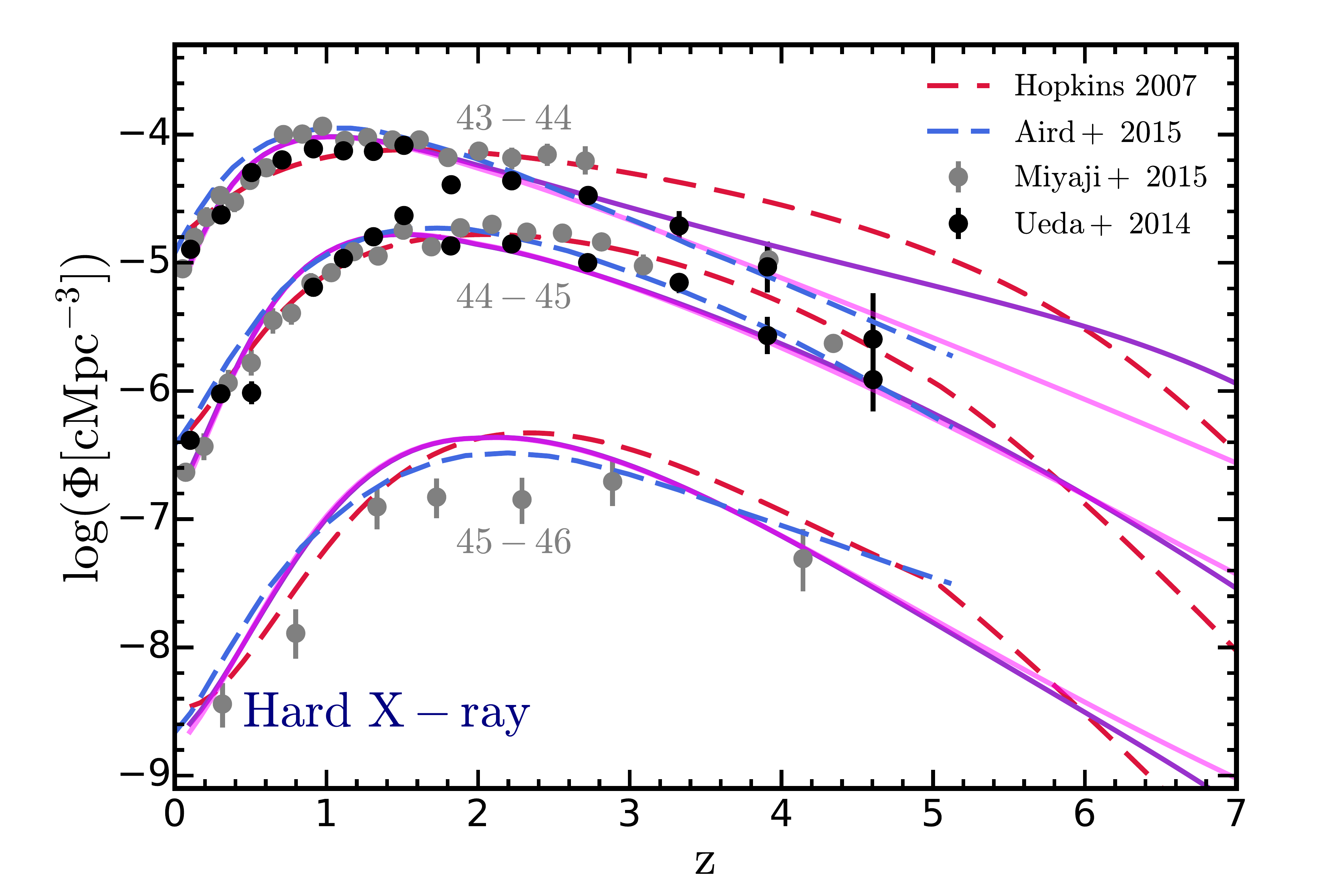}
    \includegraphics[width=0.48\textwidth]{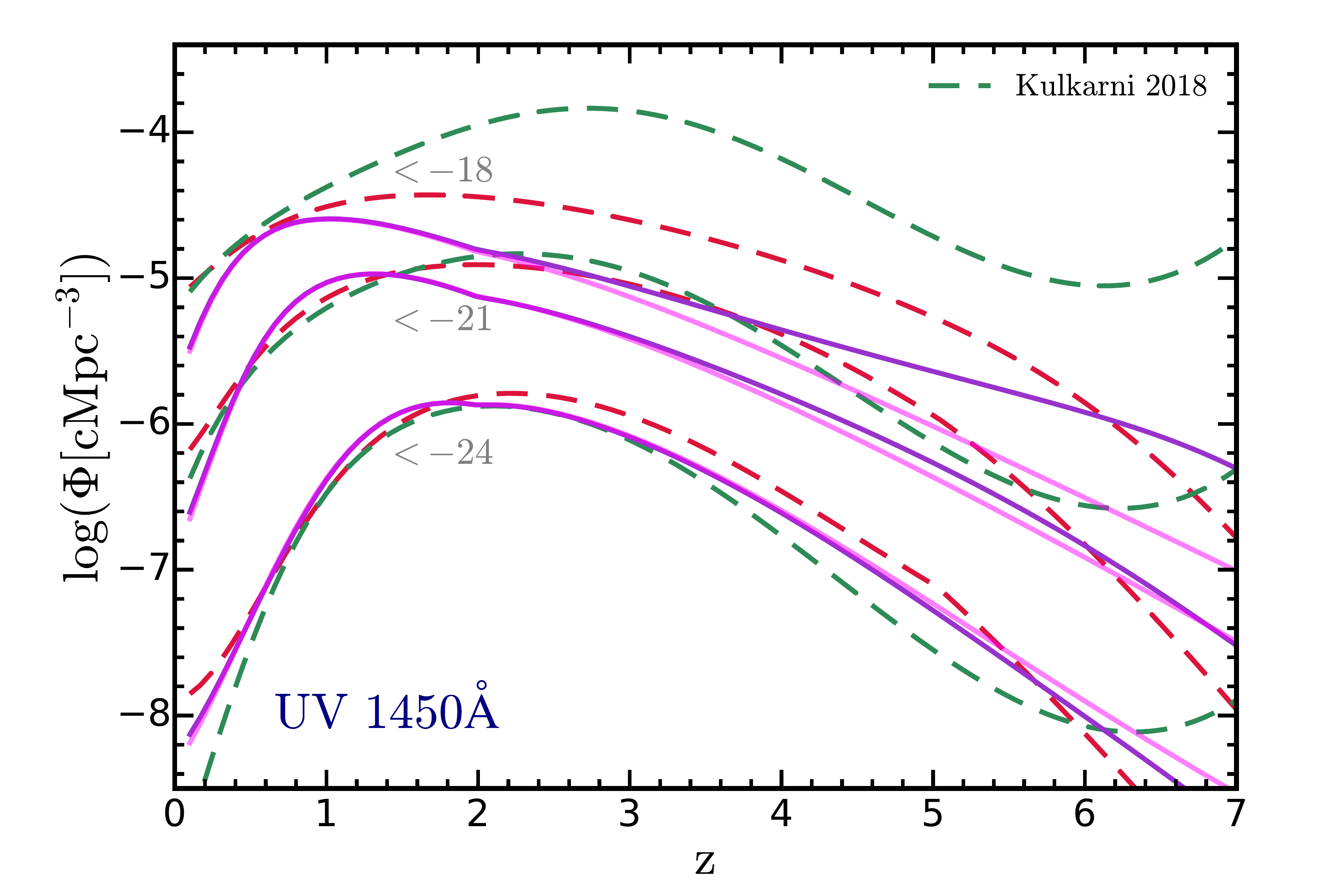}
    \caption{\textbf{Cumulative number density of quasars in a certain luminosity ($L_{\rm band}[\erg\,{\rm s}^{-1}]$) or magnitude ($M_{\rm 1450}$) interval as a function of redshift.} The predictions from the global fits A and B are shown in purple and pink lines, respectively. The constraints from \citet{Hopkins2007} and \citet{Kulkarni2018} are shown in red and green dashed lines respectively. In the X-ray, we compare our prediction with observations~\citep{Ueda2014,Aird2015a,Miyaji2015}.}
    \label{fig:cumu_num}
\end{figure}

In Figure~\ref{fig:cumu_num}, we show the evolution of the cumulative number density of quasars in different luminosity bins in different bands. We show the predictions from the global fits A and B with purple and pink lines, respectively. The cumulative number densities predicted from the two models overlap in the bright luminosity bins. The global fit B predicts lower number density than the global fit A in the faintest UV/X-ray luminosity bin at $z\gtrsim 3$. Apparently, the number density of faint quasars peaks at lower redshift than that of bright ones, consistent with the observed "cosmic downsizing" trend~\citep[e.g.,][]{Cowie1996,Barger2005,Hasinger2005} of AGN at $z\lesssim 2-3$. Compared with the \citet{Hopkins2007} model which is shown in red dashed lines, our models agree well at $z \lesssim 2$ but differences show up at high redshift where new data from the past decade modifies the predictions. Since we predict steeper bright-end slopes than the \citet{Hopkins2007} model at $z\gtrsim 2$, it is not surprising that we predict lower number density at $2<z<6$ in the most luminous bin of the bolometric luminosity. The lower number density normalization we predict at high redshift gives rise to the lower cumulative number density in the UV, when integrated down to the faint end, compared with the \citet{Hopkins2007} model. In the faintest bin of the UV luminosity, at $z\gtrsim6$, the prediction of the global fit A in the cumulative number density does not drop as fast as the \citet{Hopkins2007} model and the global fit B primarily because it predicts steeper faint-end slopes at those redshifts. In the UV, we also compare our prediction with the results in \citet{Kulkarni2018} which is an optical/UV-only study. In the bright luminosity bins, we are consistent with their estimations. However, at the faint end, we predict much lower number density of quasars at $z\gtrsim 2$ primarily driven by the much less steep faint-end slope we constrain. We note that the estimations of the number density in \citet{Kulkarni2018} did not reach $M_{\rm UV} \sim -21/-18$, so their predictions on the cumulative number density depends on the extrapolation of their measurements at brighter parts ($M_{\rm UV}\sim -23$) of the QLF. The steeper faint-end slope they constrained results in the higher cumulative number density in their prediction at $z\gtrsim2$. The steep faint-end slope of UV QLF constrained in \citet{Kulkarni2018} is potentially affected by the paucity of X-ray observations in their study, which provide better constraints at the faint end than present UV observations. Crucially, our models do not have the unphysical upturn at $z>6$ in the \citet{Kulkarni2018} model.

\begin{figure*}
    \centering
    \includegraphics[width=0.48\textwidth]{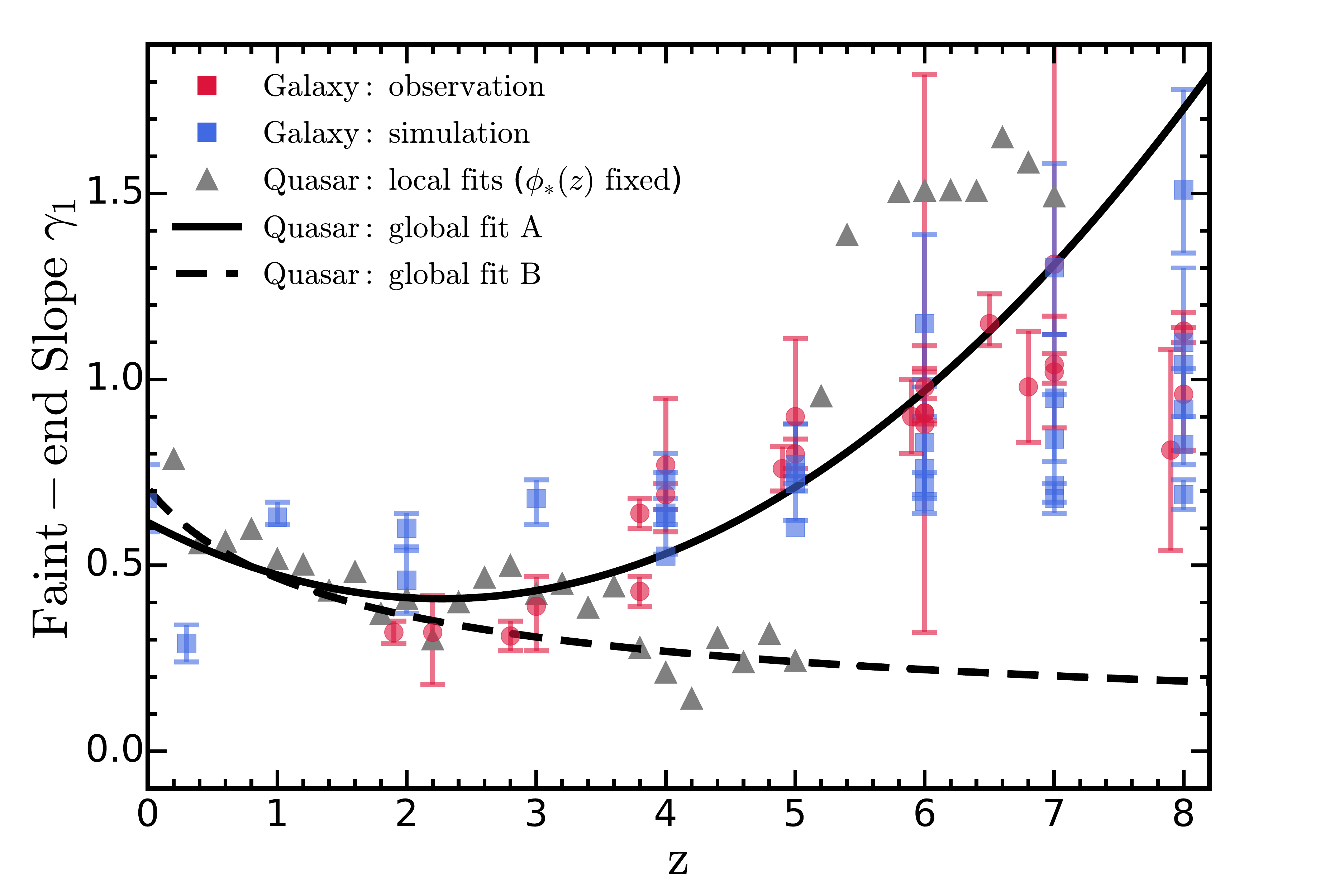}
    \includegraphics[width=0.48\textwidth]{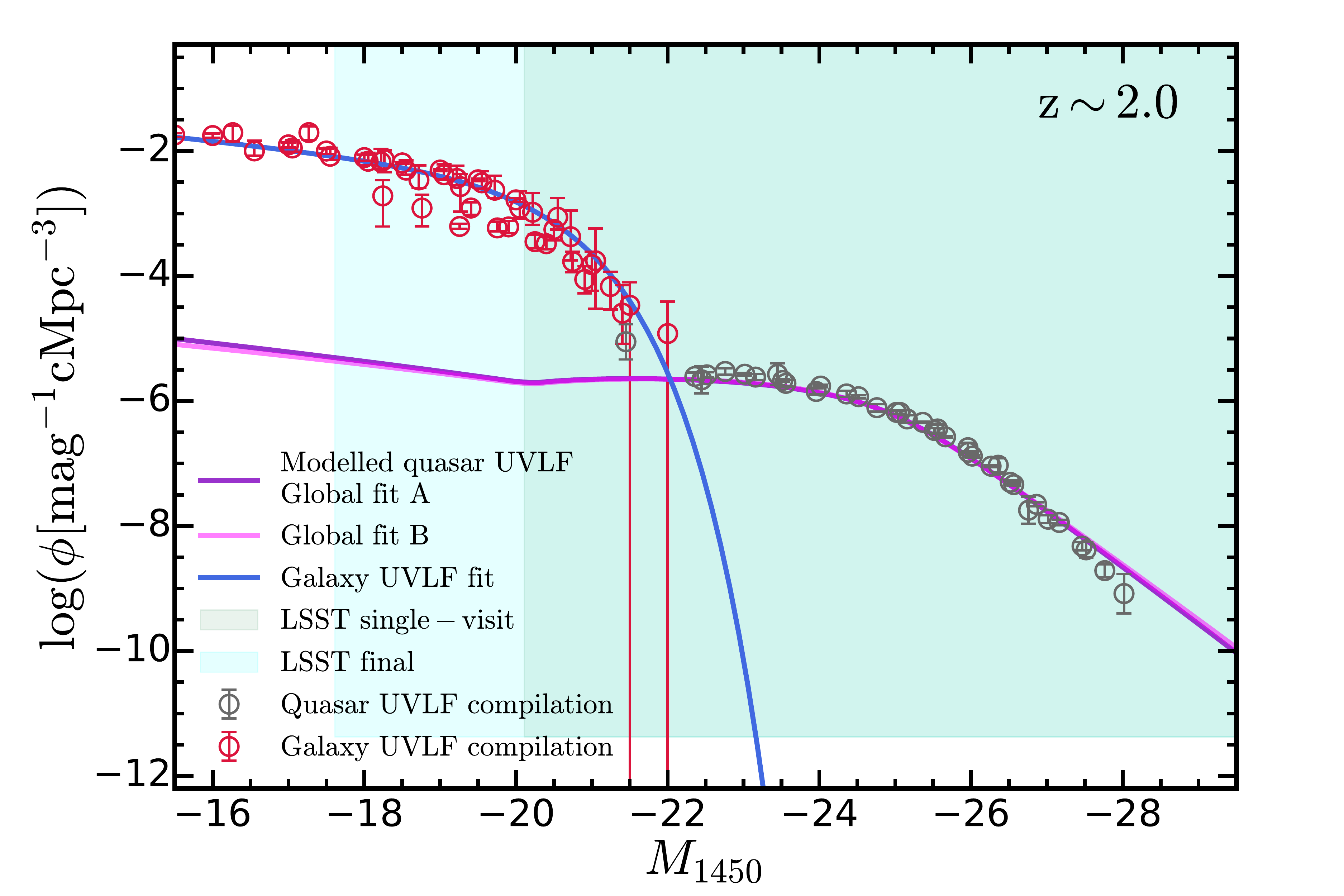}
    \includegraphics[width=0.48\textwidth]{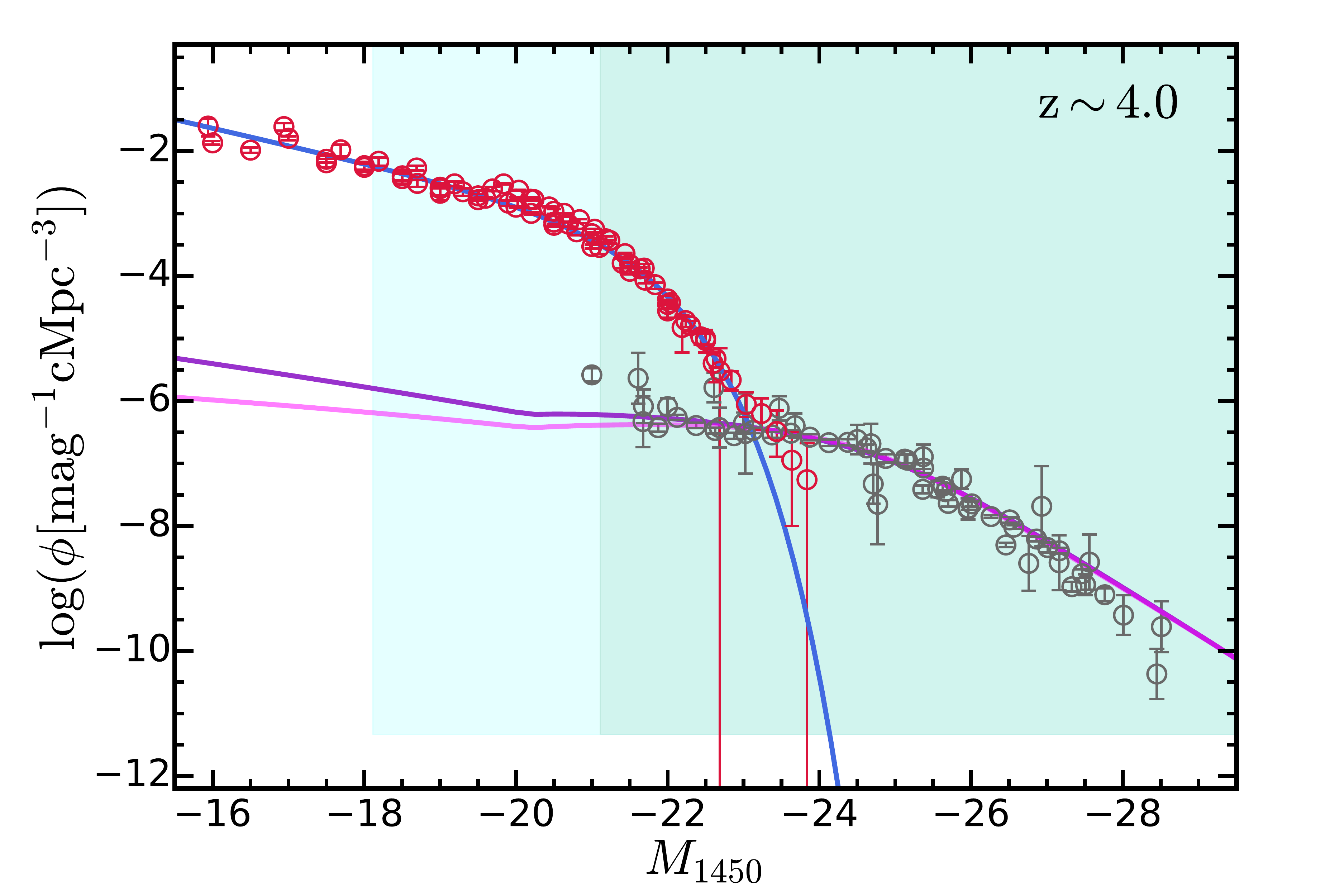}
    \includegraphics[width=0.48\textwidth]{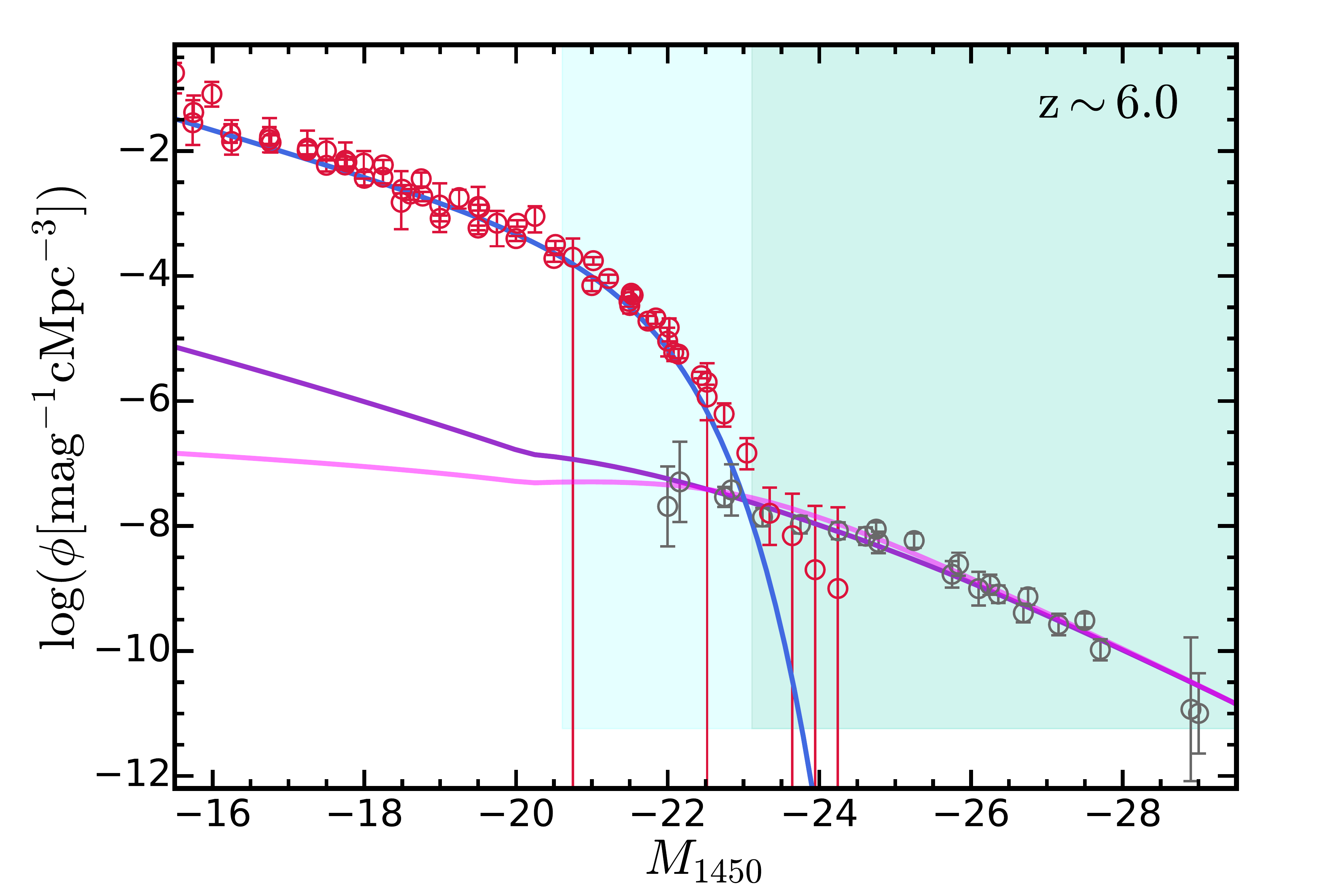}
    \caption{{\it Top left:} \textbf{Comparison between the faint-end slope of the bolometric QLF and the galaxy UV LF~(GUVLF).} Red (blue) squares represent observational (theoretical) constraints on the GUVLF. Gray triangles represent the best-fit faint-end slopes of the bolometric QLF at individual redshifts. The black solid (dashed) line represents the prediction from the global fit A (B). \textbf{In the other three panels, we show a detailed comparison between the UV QLF and the GUVLF at $z=2,4,6$.} Binned estimations from observations are shown with points~(see text in Section~\ref{sec:obs} (Section~\ref{sec:LFE}) for the sources of the UV QLF~(GUVLF) data). The GUVLF always strongly dominates the faint UV population below $M_{\rm 1450}\gtrsim -23$. For the shaded regions, the two vertical boundaries show the single-visit and final detection limit of the Legacy Survey of Space and Time~\citep[LSST, ][]{LSST2009}. The horizontal boundary shows a reference number density corresponding to one object in the field-of-view of LSST ($\sim 20000\, {\rm deg}^2$) with a survey depth $\Delta z=1$.}
    \label{fig:vs_galaxy}
\end{figure*}

In the top left panel of Figure~\ref{fig:vs_galaxy}, we compare the faint-end slope of our best-fit bolometric QLF with the faint-end slope of the rest-frame UV luminosity function of galaxies observed at $z=0-8$. The predictions from the global fits A and B are shown in purple and pink lines, respectively. For the galaxy UV luminosity function~(GUVLF), constraints on the faint-end slope come from: observations~\citep{Duncan2014,Bowler2015,Bouwens2015,Parsa2016,Finkelstein2016,Mehta2017,Atek2015,Atek2018,Ishigaki2018} and theoretical studies~\citep{Jaacks2012,Tacchella2013,Mason2015,Wilkins2017,Tacchella2018,Yung2018}. The faint-end slopes of the bolometric QLF and the GUVLF are roughly the same at $z\lesssim2$. Towards higher redshift, the faint end of the GUVLF starts to become steeper at $z=2-3$ where the QLF is still flat. For the global fit A, the faint-end slope of the QLF soon catches up that of the GUVLF and become even steeper at $z\gtrsim5$. For the global fit B, the faint-end slope of the QLF remains shallow at high redshift. Again, these differences are caused by the paucity of observations at the faint end of the QLF.

In the other three panels of Figure~\ref{fig:vs_galaxy}, we compare the UV QLF with the GUVLF at $z=2,4,6$. Both the binned estimations and the best-fit luminosity function models are shown. The binned estimations of the GUVLF include: the compilation from \cite{Finkelstein2016} at $z=4-10$, \citet{Alavi2014,Mehta2017} at $z=2$, \citet{Parsa2016} at $z=2,4$, \citet{VanderBurg2010} at $z=4$, \citet{Bouwens2017,Atek2018} at $z=6$. We use the best-fit Schechter function in \citet{Finkelstein2016} for the blue curves in the figure. The point where the UV QLF and the GUVLF cross each other becomes progressively higher from $z=6$ to $z=2$ which indicates enhanced significance of quasars at late times. But at all redshifts, the GUVLF appears to strongly dominate over the faint quasar UVLF at $M_{\rm 1450}\gtrsim -23$, wherever data exists. This is true even in models predicting high number density of faint quasars. It is also clear that the global fits A and B only show discrepancies in the regime where no observation is available. For the shaded regions, the two vertical boundaries show the single-visit and final detection limits of the Legacy Survey of Space and Time~\citep[LSST, ][]{LSST2009} which will be conducted with the Simonyi Survey Telescope at the Vera Rubin Observatory. The horizontal boundary shows a reference number density corresponding to one object in the field-of-view of LSST ($\sim 20000\, {\rm deg}^2$) with a survey depth $\Delta z=1$. As illustrated in the figure, LSST will expand our knowledge by observing faint quasars at high redshift. This will be particularly important in resolving the knee of the QLF at $z\gtrsim 6$, if it exists, and more reliably determine the faint end slope of the QLF at high redshift. Meanwhile, LSST will boost the statistics of both galaxies and quasars at the bright end. 

\section{Implications and Predictions}
\label{sec:pred}

In this section, we will make predictions based on our global best-fit bolometric QLF models. The predictions involve quasars' cumulative emissivity in the UV and their contribution to hydrogen ionization, the cosmic X-ray radiation background spectrum, the evolution of the cosmic SMBH mass density and the local SMBH mass functions. Comparing the predictions with observations in these independent channels tests the validity of our bolometric QLF model. We note that, for all the predictions made in this section, the global fits A and B give indistinguishable predictions. The major difference between the global fits A and B is the faint-end slope which does not have strong impact on cumulative luminosities of all quasars, unless the faint-end slope is extremely steep. Therefore, in the following sections, we will only show the result of the global fit A and refer to it as "the global fit" for simplicity.

\subsection{Contribution to hydrogen ionization}
\label{sec:ion}
Faint galaxies have long been considered the dominant source of ionizing photons for the reionization of hydrogen in the Universe \citep[e.g.,][]{Kuhlen2012, Robertson2015}. However, some observations of high-redshift quasars~\citep[e.g.,][]{Giallongo2015,Giallongo2019} have inferred much higher number density of quasars at the faint end than other measurements. This suggets the idea that faint quasars could potentially account for the reionization photons~\citep{Haardt2015}. In this section, we quantify the quasar contribution to the photoionization of intergalactic hydrogen using the bolometric QLF derived in this paper.

Following standard modeling of UV background~\citep[UVB; e.g.,][]{Haardt1996,Haardt2012,FG2009,FG2020,Khaire2019}, the $\rm H \Rmnum{1}$ photoionization rate is:
\begin{equation}
    \Gamma_{\rm H \Rmnum{1}}(z)=\int_{\nu_{\rm 912}}^{\infty}{\rm d}\nu\, \sigma_{\rm H  \Rmnum{1}}(\nu)\, c\, n_{\nu}(\nu,z),
    \label{eq:gamma_HI}
\end{equation}
where $\sigma_{\rm H \Rmnum{1}}(\nu)$ is the $\rm H \Rmnum{1}$ photoionization cross section 
%~\citep{Hui1997,Draine2011} 
and $n_{\nu}(\nu,z)$ is the number density of ionizing photons per unit frequency at redshift $z$. In principle, ionizing photons emitted at all $z^{\prime}>z$ should contribute to the ionizing background $n_{\nu}(\nu,z)$:
\begin{align}
    n_{\nu}(\nu,z) & =\dfrac{(1+z)^{3}}{h\nu}\int_{z}^{\infty}{\rm d}z^{\prime} \dfrac{{\rm d}t}{{\rm d}z^{\prime}}\epsilon_{\nu}(\nu_{\rm em},z^{\prime})\, e^{-\tau_{\rm eff}(z,z^{\prime},\nu)} \nonumber \\
    & =\dfrac{(1+z)^{3}}{h\nu}\int_{z}^{\infty}{\rm d}z^{\prime} \dfrac{1}{H(z^{\prime})(1+z^{\prime})}\epsilon_{\nu}(\nu_{\rm em},z^{\prime})\, e^{-\tau_{\rm eff}(z,z^{\prime},\nu)},
    \label{eq:n_ion}
\end{align}
where $\epsilon_{\nu}(\nu_{\rm em},z^{\prime})$ is the comoving emissivity of $\rm H \Rmnum{1}$ Lyman continuum sources at redshift $z^{\prime}>z$ at emitting frequency $\nu_{\rm em}=\nu(1+z^{\prime})/(1+z)$ and $\tau_{\rm eff}(z,z^{\prime},\nu)$ is the effective optical depth of photons at $z$ emitted at $z^{\prime}$. First, to simplify the calculation, we adopt the "local source" approximation~\citep[e.g.,][]{Schirber2003,FG2008,Hopkins2007}, which assumes that only ionizing sources with optical depth $\tau_{\rm eff}\leq 1$ contribute to the  ionizing background~(we will relax this assumption below). Then approximately, Equation~\ref{eq:n_ion} is reduced to:
\begin{equation}
    n_{\nu}(\nu,z)\simeq\dfrac{(1+z)^{3}}{h\nu} \dfrac{\Delta l(\nu, z)}{c}\epsilon_{\nu}(\nu,z),
    \label{eq:n_ion_local}
\end{equation}
where $\Delta l(\nu, z)$ is the mean free path of ionizing photons defined by $\tau_{\rm eff}(\Delta l)=1$. Based on the results in \citet{FG2008}, the frequency dependence of the mean free path can be described as $\Delta l(\nu, z) = \Delta l(\nu_{912}, z)\, (\nu/\nu_{912})^{3(\beta -1)}$, where the $\beta$ is the power-law index of the intergalactic HI column density distribution. For our local source approximation, we assume that the HI column distribution can be approximated by a single power-law index $\beta \approx 1.5$~\citep[e.g.,][]{Madau1999}. Assuming a power-law shape for the extreme UV quasar continuum, we have:
\begin{equation}
    \epsilon_{\nu}(\nu,z)=\epsilon_{\rm 912}(z)\Big(\dfrac{\nu}{\nu_{\rm 912}}\Big)^{-\alpha_{\rm UV}}.
\end{equation}
Since $\sigma_{\rm H \Rmnum{1}}(\nu) \propto \nu^{-3}$, the $\sigma_{\rm H  \Rmnum{1}}(\nu)\, c\, n_{\nu}(\nu,z)$ term in Equation~\ref{eq:gamma_HI} will be proportional to $\nu^{-(4+\alpha_{\rm UV}-3(\beta-1))}$. Then integrating Equation~\ref{eq:gamma_HI} gives $\Gamma_{\rm H  \Rmnum{1}}(z) = \dfrac{\sigma_{\rm H  \Rmnum{1}}(\nu_{912})\, c\, n_{\nu}(\nu_{912},z)\,\nu_{912}}{3+\alpha_{\rm UV}-3(\beta-1)}$. Plugging in Equation~\ref{eq:n_ion_local}, 
 we finally obtain:
\begin{align}
    \dfrac{\Gamma_{\rm H  \Rmnum{1}}(z)}{10^{-12}} \simeq \dfrac{0.46}{3+\alpha_{\rm UV}-3(\beta-1)} \Big(\dfrac{1+z}{4.5}\Big)^{3-\eta} \Big(\dfrac{\Delta l^{912}_{z=3.5}}{50\Mpc}\Big) 
    \ \ \ \ \ \ \ \ \ \ \ \ \ \ \ \ \ \ \ \ \ \ & \nonumber \\
    \Big(\dfrac{\epsilon_{\rm 912}(z)}{10^{24}\erg {\rm s}^{-1}\Hz^{-1}\cMpc^{-3}}\Big). &
\end{align}
where we adopt $\alpha_{\rm UV}$ = 1.7~\citep{Lusso2015} and $\Delta l^{912}_{z=3.5} = 50\Mpc$ with a power-law index $\eta=4.44$ for the redshift dependence of $\Delta l$~\citep{Songaila2010}. Here, we only consider the contribution from quasars. The emissivity at Lyman limit $\epsilon_{912}(z)$ can be linked with the UV emissivity $\epsilon_{1450}(z)$ of quasars as: 
\begin{equation}
    \epsilon_{912}(z)=\epsilon_{1450}(z) \Big(\dfrac{\nu_{912}}{\nu_{1450}}\Big)^{-0.61}
\end{equation}
assuming a power-law shape of the UV continuum with index $-0.61$~\citep{Lusso2015}, which is in good agreement with our SED model. We note that here we have assumed the escape fraction $f_{\rm esc}=100\%$ for the ionizing photon produced by quasars. It is common to adopt $100\%$ escape fractions for optically-bright quasars. However, some fraction of quasars have known dust and gas obscuration that would severely limit the escape of ionizing photons. So, the results we derive here should be interpreted as an upper limit of quasars' contribution to ionization. To derive the comoving UV emissivity of quasars, we integrate luminosity over the UV QLF predicted by our global best-fit models:
\begin{align}
    \epsilon_{1450}(z) &=\int_{L_{\rm min}}^{L_{\rm max}}L_{\nu}\,\phi(L_{\nu},z)\,{\rm d}\log{L_{\nu}} \nonumber\\
    &=\int_{M_{\rm min}}^{M_{\rm max}}L^{0}_{\nu}\,10^{-0.4M_{1450}}\,\phi^{\rm (M)}(M_{1450},z)\,{\rm d}M_{1450},
\end{align}
where $L^{0}_{\nu}$ is the zero-point luminosity of the AB magnitude system, $M_{\rm min}$ and $M_{\rm max}$ are the magnitude bounds for integration. We adopt $M_{\rm min}=-18$ and $M_{\rm max}=-35$. 

\begin{figure*}
    \centering
    \includegraphics[width=0.48\textwidth]{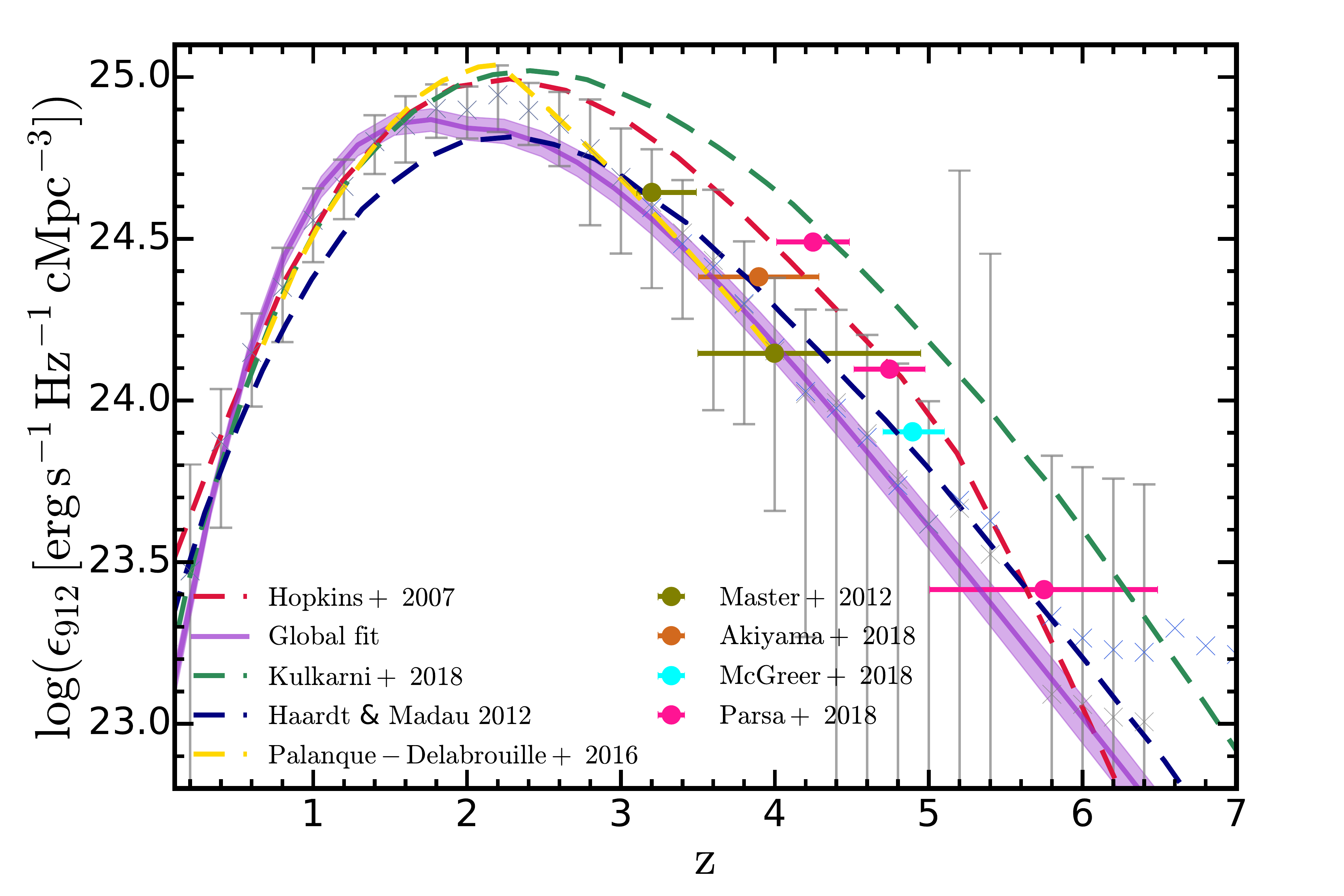}
    \includegraphics[width=0.48\textwidth]{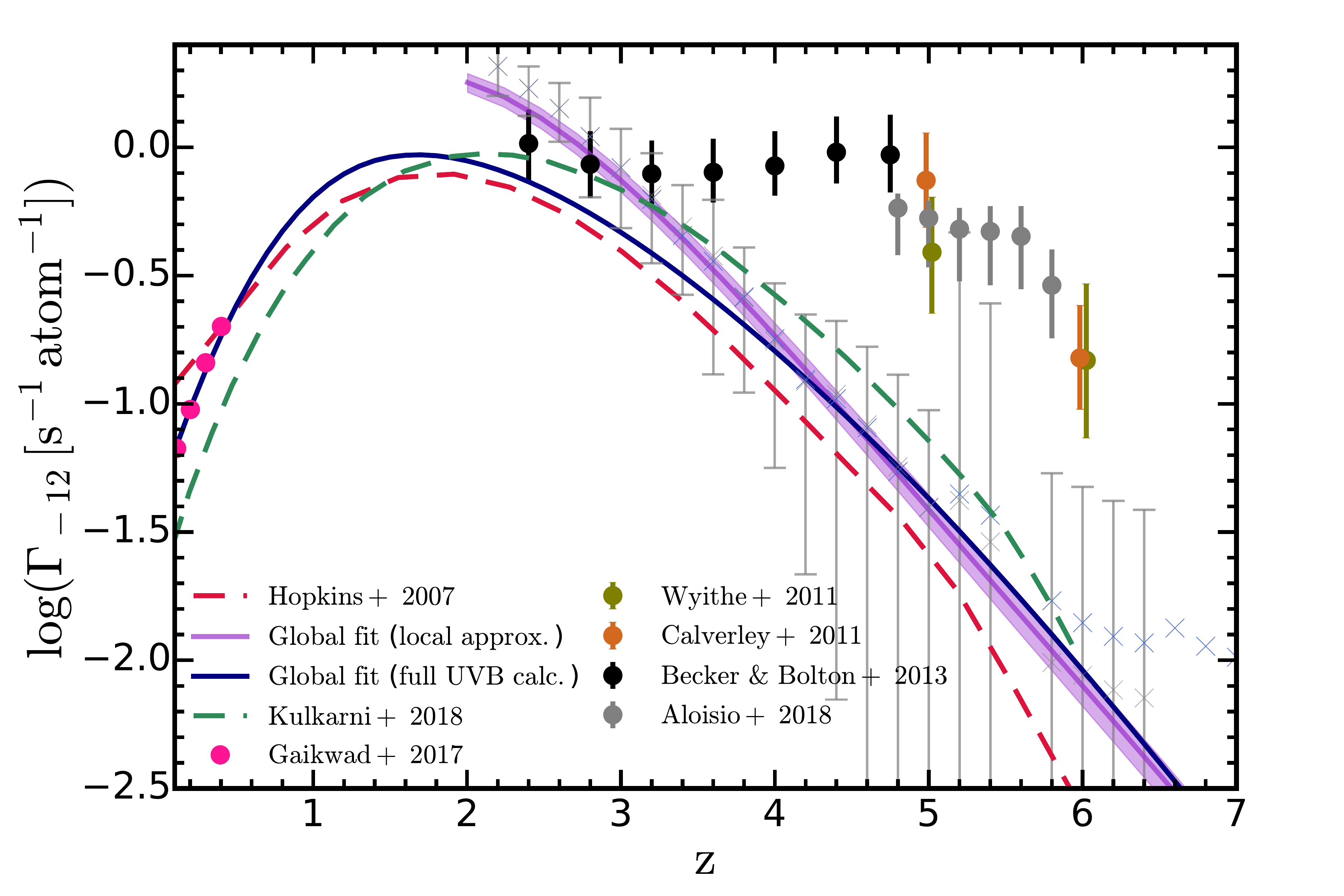}
    \caption{{\it Left:} \textbf{Predicted Lyman limit comoving emissivity of quasars versus redshift.} The prediction from our global best-fit model is shown in the purple line with $1\sigma$ confidence interval shown with the shaded region. The predictions from the fits at individual redshifts are shown in the blue (with $\phi_{\ast}(z)$ fixed) and gray (leaving $\phi_{\ast}(z)$ free) crosses with error bars indicating $1\sigma$ uncertainties. The predictions from other models are shown in: \citet{Hopkins2007}, the red dashed line; \citet{Haardt2012}, the dark blue dashed line; \citet{Kulkarni2018}, the green dashed line. We compare these results with the estimations from observations~(labeled). {\it Right:} \textbf{Predicted hydrogen photoionization rate from quasars versus redshift.} The prediction from our global best-fit model, assuming a local source approximation, is shown in the purple line with $1\sigma$ confidence interval shown with the shaded region. The prediction from a full UV background calculation \citep[using the code from][]{FG2020} is shown with the dark blue line. The predictions from the fits at individual redshifts are shown in the blue (with $\phi_{\ast}(z)$ fixed) and gray (leaving $\phi_{\ast}(z)$ free) crosses with error bars indicating $1\sigma$ uncertainties. The prediction from the \citet{Hopkins2007} model is shown in the red dashed line. The prediction from \citet{Kulkarni2018} is shown in the green dashed line. We compare these results with the measurements of the total hydrogen photoionization rate from observations (labeled). The predicted photoionization rate contributed by quasars is an order of magnitude lower than the measured total rate at $z\sim6$ and becomes close to the total rate at $z\lesssim3$. The results indicate that quasars are subdominant to the hydrogen reionization at $z\gtrsim6$.
    }
    \label{fig:ion}
\end{figure*}

In the left panel of Figure~\ref{fig:ion}, we present the predicted Lyman limit comoving emissivity $\epsilon_{\rm 912}$ versus redshift. At low redshifts, our prediction is close to the results of \citet{Hopkins2007} and \citet{Kulkarni2018}. At high redshifts, our prediction agrees well with the \citet{Haardt2012} model and is much lower than the \citet{Kulkarni2018} prediction due to the less steep faint-end slope we constrain. The prediction is in agreement with observational estimations in narrow redshift bins from \citet{Masters2012,Palanque2016,Akiyama2018}. We predict lower emissivity compared to the estimations of \citet{McGreer2018,Parsa2018}. We fit the redshift dependence of the emissivity with a five-parameter functional form~\citep{Haardt2012}:
\begin{equation}
    \epsilon_{\rm 912}=\epsilon_{\rm 0}\,(1+z)^{a}\,\dfrac{\exp{(-bz)}}{\exp{(cz)}+d},
\end{equation}
and we obtain the best-fit as:
\begin{align}
    \epsilon_{\rm 912} = (10^{24.108}\erg\,{\rm s}^{-1}{\rm Hz}^{-1}&{\cMpc}^{-3})\,(1+z)^{5.865} \nonumber \\
    & \times \dfrac{\exp{(0.731 z)}}{\exp{(3.055z)}+15.60}.
\end{align}

In the right panel of Figure~\ref{fig:ion}, we present our prediction for the hydrogen photoionization rate contributed by quasars. We find that the prediction using the local source approximation severely over-predicts the hydrogen photoionization rate at $z\lesssim 2$ where the mean free path of ionizing photons grows comparable to (and eventually larger than) the Hubble radius, so that the local source approximation fails significantly. Therefore, we perform a full UVB calculation using the method described in \citet{FG2020}. The result is also shown in the right panel of Figure~\ref{fig:ion}. The prediction from this full UVB calculation almost overlaps with the prediction with the local source approximation at $z\gtrsim 4$, despite slight differences. The slight differences are due to more physics incorporated in the full UVB calculation that make the UVB spectrum (filtered by IGM absorption and including recombination emission) different from the simple power-law that we have assumed above \citep[see][]{FG2020}. We compare our predicted hydrogen photoionization rates (from quasars only) with observational inferences of the total rates from \citet{Wyithe2011,Calverley2011,Becker2013,Gaikwad2017,DAloisio2018}. The predicted hydrogen photoionization rate contributed by quasars is an order of magnitude lower than the measured total rate at $z\sim6$ and only becomes close to the total rate at $z\lesssim3$. The results indicate that quasars are subdominant to the hydrogen reionization at $z\gtrsim 6$, but they start to dominate the ionization budget at $z\lesssim3$. Interestingly, that the hydrogen photoionization rates predicted using our new bolometric QLF are quite similar to the results of \citet{Hopkins2007}, which used a different bolometric QLF and adopted a different mean free path model. We have assumed that all ionizing photons produced by the quasar can escape the host galaxy even for faintest quasars. Given this favorable assumption, the predicted contribution of quasars to the hydrogen reionization is still subdominant. Similar conclusion has been reached in \citet{Ricci2017} who adopted a different approach. We have tested that, even including the \citet{Giallongo2015} data in the fit and neglecting all the data points that are incompatible with it, quasars can only have a maximum of $\sim 50\%$ contribution to the ionization budget at $z\sim 5.8$, under the assumption that the escape fraction $f_{\rm esc}=100\%$ even for quasars much fainter than typical star-forming or Seyfert galaxies.

\subsection{Cosmic X-ray background}

Since quasars dominate the radiation budget in the X-ray in the Universe, the cosmic X-ray radiation background (CXB) serves as an important channel to cross check our model of the bolometric QLF. The observation of the CXB does not require spatially resolving and identifying quasars and thus can even probe the contribution from faint-end quasars at any redshift.

In general, to get the cosmic radiation background contributed by quasars, we integrate the spectrum of quasars at $z=0-7$ as:
\begin{align}
    I_{\rm RB}(\nu) & = \int_0^{7}{\rm d}z \dfrac{\epsilon_{\nu}(\nu_{\rm em},z)}{4\pi d_{\rm L}^{2}(z)} \dfrac{{\rm d}V}{{\rm d}\Omega{\rm d}z}(z) \nonumber \\
    & = \int_0^{7}{\rm d}z \dfrac{\epsilon_{\nu}(\nu_{\rm em},z)}{4\pi d_{\rm L}^{2}(z)} \dfrac{{\rm d}V}{{\rm d}\Omega{\rm d}z}(z),
    \label{eq:XRB}
\end{align}
where $\nu_{\rm em}=(1+z)\nu$ and $\dfrac{{\rm d}V}{{\rm d}\Omega{\rm d}z}(z)$ is the differential comoving volume element at $z$. $\epsilon_{\nu}(\nu_{\rm em},z)$ is derived by integrating over the luminosity function of the emission at $\nu_{\rm em}$ predicted by our best-fit model.

\begin{figure}
    \centering
    \includegraphics[width=0.48\textwidth]{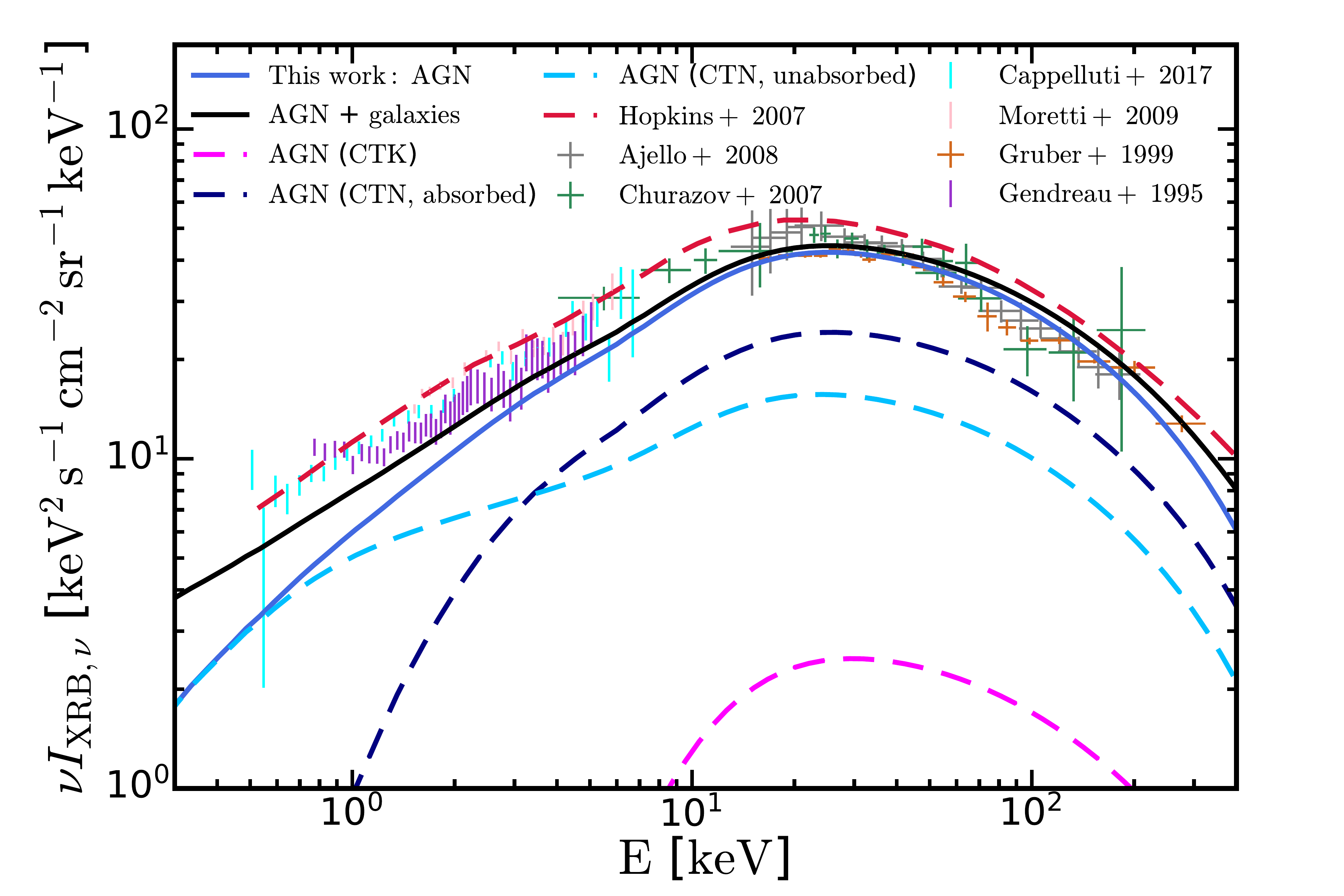}
    \caption{\textbf{Predicted CXB spectrum.} The prediction from our global best-fit model, which only includes the contribution from quasars, is shown with the blue solid line. The prediction that accounts for a simplified constant $2\keV^{2}{\rm s}^{-1}{\rm sr}^{-1}\keV^{-1}$ contribution from galaxies is shown with the black solid line. The predictions that only include CTK~($\log{N_{\rm H}}\geq24$) or absorbed CTN~($22\leq\log{N_{\rm H}}\leq24$) or unabsorbed CTN AGN~($\log{N_{\rm H}}\leq22$) are shown in dashed lines. We compare the predictions to the measurements from \citet{Gendreau1995,Gruber1999,Churazov2007,Ajello2008,Moretti2009,Cappelluti2017}. We also show the prediction from the \citet{Hopkins2007} model with the red dashed line. The prediction from this work is generally in agreement with the observations despite a $\sim 0.05\,{\rm dex}$ lower at $E\lesssim 20\keV$.}
    \label{fig:cxb}
\end{figure}

In practice, we have found that simply adopting the X-ray SED template with the median photon index $\Gamma=1.9$ leads to an under-prediction for the CXB. Considering that the photon index has a significant scatter, $\sim 0.2$, the stacked SED of quasars should have a very different shape from a simple cut-off power-law. Therefore, in making predictions on the CXB, we adopt the stacked SED of $1000$ sampled SEDs with a normal distribution of photon indexes with median value $1.9$ and scatter $0.2$. In Figure~\ref{fig:cxb}, we show the predicted CXB spectrum and compare it with the measurements from \citet{Gendreau1995,Gruber1999,Churazov2007,Ajello2008,Moretti2009,Cappelluti2017}. For simplicity, we have assumed the galaxies' contribution to the CXB to be a constant $2\keV^{2}{\rm s}^{-1}{\rm sr}^{-1}\keV^{-1}$. We find our predicted CXB spectrum agrees well with observations at high energy end while it is roughly $\sim 0.05\,{\rm dex}$ lower at $E\lesssim 20\keV$. Imperfectness in the extinction model may be responsible for this though it is hard to argue the source of this level of inconsistency. The \citet{Hopkins2007} model systematically over-predicts the CXB spectrum. We also show separately the contribution to CXB from CTK, absorbed CTN and unabsorbed CTN AGN. The absorbed CTN AGN are the major sources of the CXB in the high energy regime while the unabsorbed CTN AGN overtake at $E\lesssim 3\keV$. The CTK AGN are subdominant to the CXB.

\subsection{Growth history of SMBHs}
The bolometric quasar luminosity is connected with the accretion of the SMBH that powers quasar activities. Thus, based on our bolometric QLF model, constraints can be put on the growth history of SMBHs in the Universe. Here we focus on the evolution of the cosmic SMBH mass density and the SMBH mass function.

\begin{figure}
    \centering
    \includegraphics[width=0.48\textwidth]{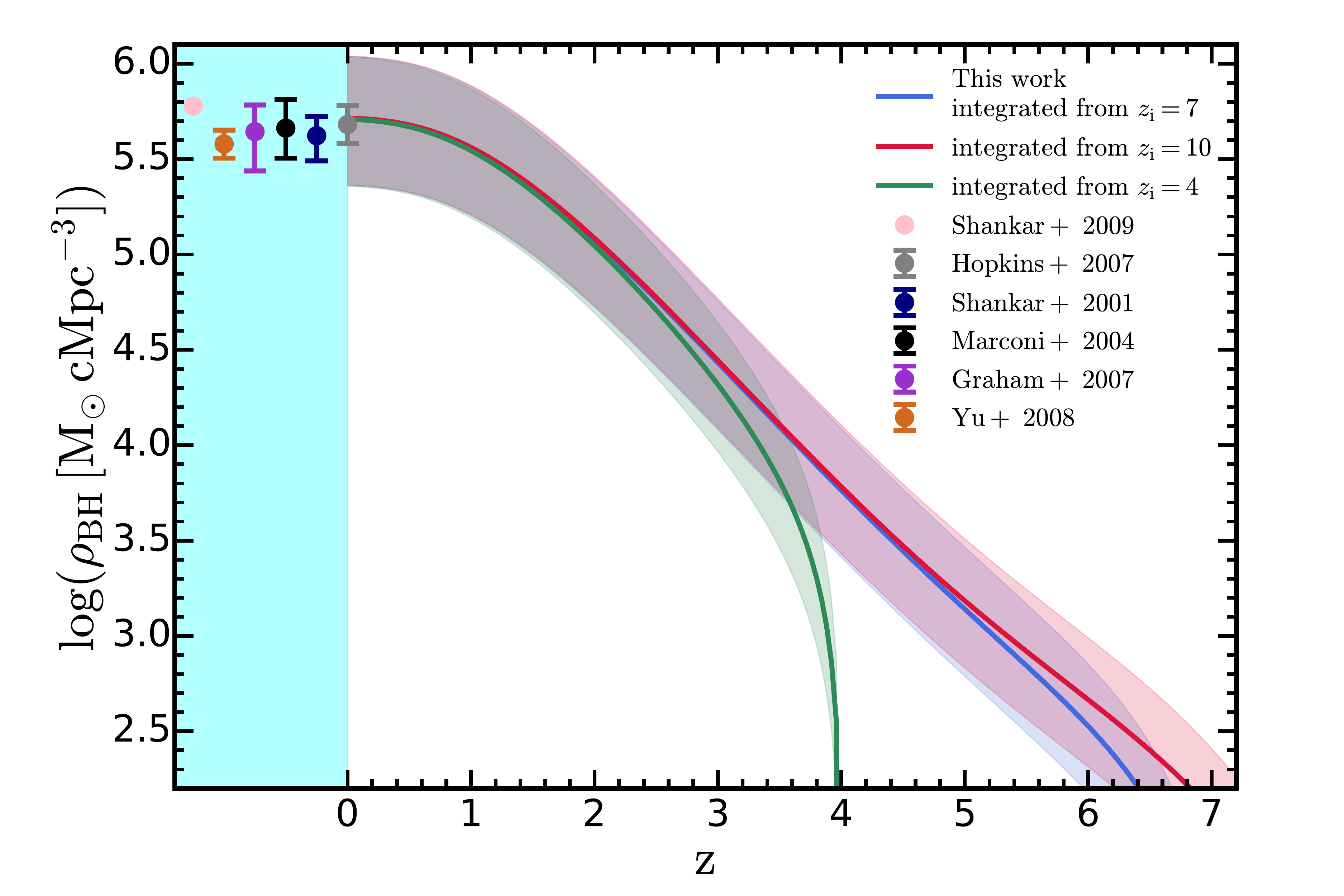}
    \caption{\textbf{Predicted evolution of the cosmic SMBH mass density at $z=0-7$.} The red, blue and green lines represent the predictions with starting redshift of integration $z_{\rm i}=10,7,4$ respectively. We assume the averaged radiative efficiency $\epsilon_{\rm r}=0.1$. Shaded regions show the uncertainties when increasing or decreaing $\epsilon_{\rm r}$ by $2$ times. The data points show the estimated SMBH mass density in the local Universe from \citet{Shankar2004,Marconi2004,Graham2007,Yu2008,Shankar2009}. The local SMBH mass density is mainly dominated by the SMBH growth at $z<4$.}
    \label{fig:bh_density}
\end{figure}

\subsubsection{Cosmic SMBH mass density}

Assuming a constant averaged radiative efficiency $\epsilon_{\rm r}\simeq 0.1$ for the SMBH accretion, the bolometric quasar luminosity can be related to the accretion rate of the SMBH as:
\begin{equation}
    L_{\rm bol}=\epsilon_{\rm r}\dot{M}c^{2}.
\end{equation}
Therefore, the integrated luminosity density can be translated to the rate of change in the total SMBH mass density as:
\begin{equation}
    \dfrac{{\rm d}\rho_{\rm BH}}{{\rm d}z}=\dfrac{1-\epsilon_{\rm r}}{\epsilon_{\rm r}c^{2}H(z)(1+z)} \int_{L_{\rm min}}^{L_{\rm max}}L_{\rm bol}\,\phi(L_{\rm bol},z)\,{\rm d}\log{L_{\rm bol}}.
\end{equation}
where we adopt $\log{L_{\rm min}}=43,\log{L_{\rm max}}=48$ here. Starting from an initial redshift for SMBH growth $z_{\rm i}$ and integrating over redshift, we derive the evolution of $\rho_{\rm BH}$. In Figure~\ref{fig:bh_density}, we show the redshift evolution of the SMBH mass density with $z_{\rm i}=10,7,4$ in the red, blue and green lines. Shaded regions show the uncertainties when increasing or decreasing $\epsilon_{\rm r}$ by $2$ times. The build-up of the SMBH mass density is completely dominated by the accretion at $z<4$. Compared with local constraints~\citep{Shankar2004,Marconi2004,Graham2007,Yu2008,Shankar2009}, we predict slightly higher SMBH mass density at $z=0$. There are several uncertainties that could impact the comparison made here. These local constraints were calculated by translating galaxy central spheroid properties to the mass of SMBH. New calibrations of the scaling relations between the mass of SMBH and galaxy spheroid properties~\citep[e.g.,][]{Kormendy2013,McConnell2013} have generally found higher intercepts and steeper slopes than the old calibrations. Besides, as discussed in Section~\ref{sec:sed}, variations on the definition of the bolometric luminosity could also lead to systematic shift in the estimated radiation energy budget of SMBHs. Both of these two factors could drive the local constraints and our predictions to be more consistent with each other. However, on the other hand, the selection biases in observed scaling relations could result in an over-estimation of the local SMBH mass density~\citep[e.g.,][]{Shankar2016}. In that case, the discrepancy of our result with local estimations indicates a higher averaged radiative efficiency than the assumed value $0.1$.

\subsubsection{SMBH mass function}
\label{sec:SMBH_MF}
The mass function is one of the most important statistical properties of the SMBH population. In the local Universe, SMBH mass can be determined by various properties of galaxy spheroids, e.g. the velocity dispersion, the bulge mass. Both quiescent and active SMBHs' masses can be estimated in this way. At high redshift, SMBH masses are measured based on direct radiation from the vicinity of active SMBHs. Alternatively, the SMBH mass can be related to the bolometric quasar luminosity with the Eddington ratio. Assuming an Eddington ratio distribution, one can convert the bolometric QLF to the SMBH mass function. Technically, there are two ways to achieve this: 
\begin{itemize}
    \item convolve the bolometric QLF with the measured relation between Eddington ratio and bolometric quasar luminosity. This method is referred to as "convolution". 
    \item assuming an Eddington ratio distribution, fit the parameterized SMBH mass function based on the bolometric QLF. This method is referred to as "deconvolution".
\end{itemize}
    
\begin{figure}
    \centering
    \includegraphics[width=0.48\textwidth]{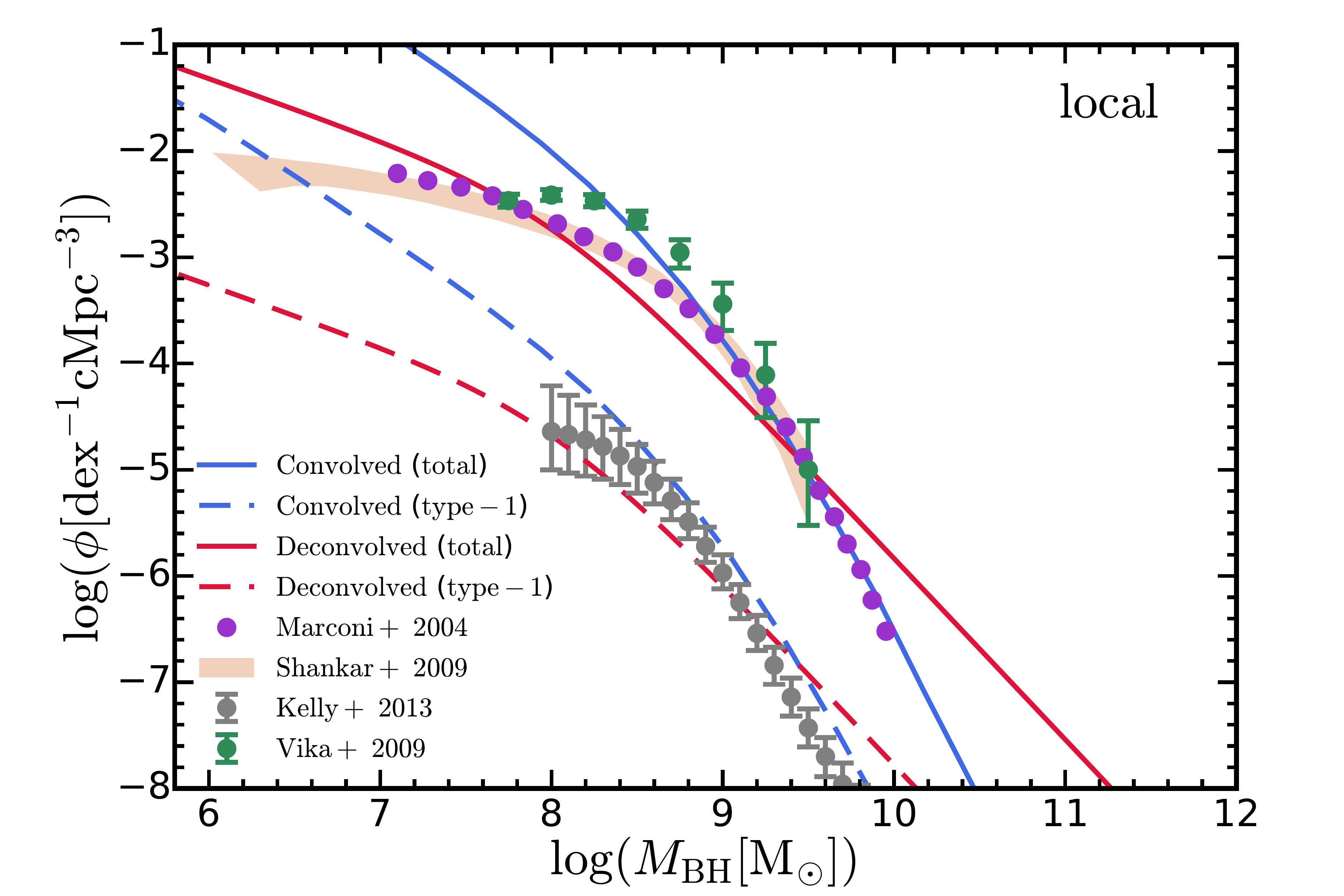}
    \caption{\textbf{Total SMBH mass function and Type-1 AGN mass function in the local Universe.}  We show the predictions "convolved" ("deconvolved") from the bolometric QLF in blue (red) lines~(see text in Section~\ref{sec:SMBH_MF} for details of the two methods). The total SMBH mass functions are shown in solid lines and the Type-1 AGN mass functions are shown in dashed lines. We compare the predictions for the total SMBH mass function with estimations from \citet{Marconi2004,Shankar2009,Vika2009} and compare the predictions on the Type-1 AGN mass function with the estimation from \citet{Kelly2013}.}
    \label{fig:BHMF}
\end{figure}

For the first approach, we adopt the scaling relation~\citep{Nobuta2012}:
\begin{equation}
   \log{\lambda_{\rm Edd}}=0.469\times \log{L_{\rm bol}}-22.46, 
\end{equation}
where $\lambda_{\rm Edd}$ is the Eddington ratio. The relation was measured based on X-ray selected AGN at $z\sim1.4$ and was demonstrated~\citep{Nobuta2012} to be consistent with what had been found in the SDSS DR5 broad-line AGN~\citep{Shen2009}. We also consider the $\sim 0.4\,{\rm dex}$ scatter of this relation~\citep{Nobuta2012}. Convolving the bolometric QLF with this relation, we can derive the SMBH mass function for active SMBHs. We further multiply the fraction of unabsorbed CTN AGN $F\sim 0.38$, estimated at the knee of the local X-ray QLF with our fiducial extinction model, to get the SMBH mass function of Type-1 AGN. We present the predicted SMBH mass function of Type-1 AGN in Figure~\ref{fig:BHMF} with the blue dashed line which is in good agreement with the observation~\citep{Kelly2013}. In order to further deduce the total SMBH (including quiescent ones) mass function, we need to correct for the fraction of AGN that are in the active phase, $f_{\rm duty}$. We find that in order to match the observational constrained total SMBH mass function in the local Universe~\citep{Vika2009,Shankar2009,Marconi2004}, $f_{\rm duty}$ should take the value $\sim 0.03$. After multiplying $1/f_{\rm duty}$ to the predicted SMBH mass function of active SMBHs, we derive the total SMBH mass function shown with the blue solid line in Figure~\ref{fig:BHMF}.

For the second approach, we assume a two component Eddington ratio distribution function (ERDF) for AGN~\citep{Tucci2017}:
\begin{equation}
    P(\log{\lambda})=\Big[(1-F) A \lambda^{1+\alpha}e^{-\lambda/\lambda_1}+\dfrac{F }{\sqrt{2\pi\sigma^{2}}}e^{-(\log{\lambda}-\log{\lambda_2})^{2}/2\sigma^{2}}\Big].
\end{equation}
The first component takes a Schechter function format and describes the ERDF of Type-2 AGN. The prefactor $A$ is set to normalize the total probability of this component to be $1-F$. We choose $\lambda_1=1.5$ and $\alpha=-0.6$ which were found in agreement with observations on low redshift Type-2 AGN~\citep{Hopkins2009,Kauffmann2009,Aird2012}. The second component takes a log-normal format and describes the ERDF of Type-1 AGN of which the parameters were determined by fitting the shape of the ERDFs from \citet{Kelly2013} in different redshift bins and interpolating the results with a linear function~\citep{Tucci2017}:
\begin{align}
    \log{\lambda_{2}} = \max[-1.9+0.45z, \log{0.03}], \nonumber \\ 
    \sigma = \max[1.03-0.15z, 0.6)]/\ln{10}.
\end{align}
We note that a consensus on the shape of the ERDF has not been reached. However, the potential influence of the ERDF assumptions should be limited~(see the Appendix of \citet{Weigel2017}) for our purpose here. We parameterize the total SMBH mass function as a double power-law function. For a proposed total SMBH mass function, multiplying $f_{\rm duty}=0.03$ where we found through the other method, we can derive the SMBH mass function of the active SMBHs with parameters left for fitting. We can convolve this active SMBH mass function with the assumed ERDF to derive the resulting bolometric QLF. By comparing the result with our bolometric QLF model, we derive the best-fit parameter choice for the SMBH mass function. In Figure~\ref{fig:BHMF}, we present constraints on the local SMBH mass function from two different methods and compare it with observations of the total SMBH mass function~\citep{Marconi2004,Shankar2009,Vika2009} and observations of the Type-1 AGN mass function~\citep{Kelly2013}. The constraints from this work are in decent agreement with all the observations in the range $10^{7}$ to $10^{9.5}\msun$. The "convolution" method does better at the massive end while the "deconvolution" method does better at the low mass end. 

We limit our prediction to the local SMBH mass function, since the uncertainties in the ERDF, the active fraction and the absorbed fraction grow much larger at high redshift. A more comprehensive model of the SMBH population and constraints on the evolution of the SMBH mass function will be explored in future works.

\section{Summary and Conclusions}
In this paper, we update the constraints on the bolometric QLF at $z=0-7$ and make various predictions based on this model. Our technique follows the method of \citet{Hopkins2007} but with an updated quasar mean SED model and bolometric and extinction corrections. We have also extended the observational compilation in \citet{Hopkins2007} with new binned estimations of the QLF from the recent decade. These new observations allow more robust determination of the bolometric QLF at $z\gtrsim 3$. Our findings on the bolometric QLF can be summarized as:
\begin{itemize}
    \item We obtain two global best-fit models A and B with different assumptions on the evolution of the faint-end slope at high redshift. As shown in Figure~\ref{fig:fit_at_z_parameters} and Figure~\ref{fig:LF_bol}, comparing with the \citet{Hopkins2007} model, we find the bright-end slope steeper at $z\gtrsim 2$ in both the global fits A and B. In the global fit A, the faint-end slope is steeper than the \citet{Hopkins2007} model at $z\gtrsim 3$ and becomes progressively steeper at higher redshift. In the global fit B, where we adopt a monotonically evolved faint-end slope, the faint-end slope remains shallow at high redshift and is close to the prediction of the \citet{Hopkins2007} model. The uncertainties on the faint-end slope arise from the paucity of measurements of the faint-end QLF at high redshift. Apart from that, we have fixed some extrapolation problems of the \citet{Hopkins2007} model. The integrated luminosity of bright-end quasars would not blow up at $z\gtrsim 7$ and the number density normalization exhibits a more natural evolution towards higher redshift.
    
    \item We investigate the current tension in the UV QLF at $z\simeq 4-6$ shown in Figure~\ref{fig:tension}. We find that the high number density of faint quasars found in \citet{Giallongo2015} is disfavored when compared with current available X-ray observations. Our QLF models achieve a better agreement with the X-ray data at the faint end than the previous QLF models based on optical/UV observations only.
    
    \item The evolution of the bolometric luminosity function can be interpreted as two phases separted at $z\simeq 2-3$, illustrated in Figure~\ref{fig:LF_evolve}. In the early phase, the bolometric QLF rises up monotonically following the hierarchical build-up of structures in the Universe. In the late phase, the bolometric QLF shows a systematic and continuous horizontal shift towards the low luminosity regime. AGN feedback is potentially responsible for this evolutionary pattern. Surprisingly, in both phases, the evolution at the bright end ($L_{\rm bol}\gtrsim 48$) of the bolometric QLF is apparently milder compared to other luminosity regimes. This suggests potential regulation on the abundance of the most luminous quasars.
\end{itemize}

We have made predictions with this new model on the hydrogen photoionization rate contributed by quasars, the CXB spectrum, the evolution of the cosmic SMBH mass density and the local SMBH mass function. We find a general consistency with observations in these channels and our findings can be summarized as:

\begin{itemize}
    \item We find that quasars are subdominant to the hydrogen photoionization rate during the epoch of reionization at $z\gtrsim 6$. They start to dominate the UV background at $z\lesssim3$.
    \item The predicted CXB spectrum shown in Figure~\ref{fig:cxb} agrees well with observations in the high energy regime while lies slightly lower than observations at $E\lesssim 20\keV$.
    \item We predict the evolution of the SMBH mass density at $z=0-7$ shown in Figure~\ref{fig:bh_density}. We find that the prediction is consistent with local observations and the evolution is dominated by the growth of SMBHs at $z<4$.
    \item We make predictions on the local total SMBH mass function and the Type-1 AGN mass function shown in Figure~\ref{fig:BHMF}. We explore two different methods, a "convolution" method and a "deconvolution" method. Both of them can generate consistent results with observations.
\end{itemize}

The new bolometric QLF model constrained in this paper can simultaneously match the multi-band observations on QLF over a wide redshift range up to $z\sim 7$. The model reveals an evolutionary pattern of the bolometric QLF at high redshift that is qualitatively different from the \citet{Hopkins2007} model. The predictions from the new model is in consistent with observations in various channels. We demonstrate the new bolometric QLF model as a solid basis for future studies of high redshift quasar populations and their cosmological impacts. 

\section*{Acknowledgements}
Support for PFH was provided by an Alfred P. Sloan Research
Fellowship, NSF Collaborative Research grant \#1715847 and CAREER grant \#1455342. CAFG was supported by NSF through grants AST-1517491, AST-1715216, and CAREER award AST-1652522; by NASA through grant 17-ATP17-0067; and by a Cottrell Scholar Award and Scialog Award \#26968 from the Research Corporation for Science Advancement. NPR acknowledges support from the STFC and the Ernest Rutherford Fellowship scheme. GTR was supported in part by NASA-ADAP grant NNX17AF04G. DMA thanks the Science and Technology Facilities Council (STFC) for support through grant code ST/P000541/1. Numerical calculations were run on the Caltech computer cluster 'Wheeler'.

%%%%%%%%%%%%%%%%%%%%%%%%%%%%%%%%%%%%%%%%%%%%%%%%%%



%\bibliographystyle{mnras}
%\bibliography{reference} % if your bibtex file is called example.bib

%%%%%%%%%%%%%%%%%%%%%%%%%%%%%%%%%%%%%%%%%%%%%%%%%%

%%%%%%%%%%%%%%%%% APPENDICES %%%%%%%%%%%%%%%%%%%%%

\appendix

\section{Compiled observations}

In Table~\ref{tab:observations}, we list the observational papers compiled in this work along with the details of their observations, including the survey/fields, the band, the luminosity/redshift coverage and the number of quasar samples adopted. 

\begin{table*}
\centering
\begin{tabular}{
p{0.23\textwidth}|p{0.22\textwidth}|p{0.10\textwidth}|p{0.07\textwidth}|p{0.20\textwidth}|p{0.05\textwidth}}
\hline 
\hline
Reference & Survey/Field & Rest-frame & Redshift & Luminosity Range$^{\,\rm a}$ & $N_{\rm AGN}$\\
 &  & Wavelength/Band & Range & [AB mag or \erg/{\rm s}] & \\
\hline
\hline
\bf{optical/UV}\\
\hline
\citet{Kennefick1995} & POSS & B & 4.0-4.5 & $-28.50<M_{\rm B}<-26.50$ & 10\\
\citet{Schimidt1995} & PTGS & B & $\sim$3.5-4.5 & $-27.50<M_{\rm B}<-25.50$ & 8\\
\citet{Fan2001a} & SDSS (equatorial stripe) & 1450\AA & 3.6-5.0 & $-27.50<M_{\rm 1450}<-25.50$ & 39\\
\citet{Fan2001b,Fan2003,Fan2004}$^{\rm b}$ & SDSS (Main \& Southern Survey) & 1450\AA & $\sim$5.7-6.4 & $-28.00<M_{\rm 1450}<-26.50$ & 9\\
\citet{Wolf2003} & COMBO-17 & 1450\AA & 1.2-4.8 & $-28.50<M_{\rm 1450}<-23.50$ & 192\\
\citet{Cristiani2004} & GOODS & 1450\AA & $\sim$4-5.2 & $-23.50<M_{\rm 1450}<-21.00$ & 1-4\\
\citet{Croom2004} & 2QZ/6QZ & B & 0.4-2.1 & $-28.50<M_{\rm B}<-20.50$ &  20905\\
\citet{Hunt2004} & LBG survey & 1450\AA & $\sim$2-4 & $-27.00<M_{\rm 1450}<-21.00$ & 11\\
\citet{Richards2005} & 2dF-SDSS & g & 0.3-2.2 & $-27.00<M_{\rm g}<-21.00$ & 5645\\
\citet{Richards2006b} & SDSS DR3 & i(z=2) & 0.3-5.0 & $-29.00<M_{\rm i}<-22.50$ &  15343\\
\citet{Siana2006} & SWIRE & 1450\AA & $\sim$2.8-3.4 & $-26.50<M_{\rm 1450}<-23.50$ & $\sim$100\\
\citet{Bongiorno2007} & VVDS & B$\rightarrow$1450\AA & 1-4 & $-25.69<M_{\rm 1450}<-20.69$ &  130\\
\citet{Fontanot2007} & SDSS DR3 \& GOODS/ACS & 1450\AA & 3.5-5.2 & $-28.00<M_{\rm 1450}<-21.00$ &  13 \\
\citet{Siana2008} & SWIRE & r$\rightarrow$1450\AA & 2.83-3.44 & $-26.11<M_{\rm 1450}<-23.61$ & 100\\
\citet{Croom2009} & 2SLAQ \& SDSS DR3 & g(z=2) & 0.4-2.6 & $-29.75<M_{\rm g(z=2)}<-20.25$ & 10637\\
\citet{Jiang2009}$^{\rm b}$ & SDSS Main \& Deep & 1450\AA & 5.7-6.6 & $-27.63<M_{\rm 1450}<-25.10$ & 6\\
\citet{Willott2010} & CFHQS & 1450\AA & 5.75-6.45 & $-26.05<M_{\rm 1450}<-22.15$ &  19\\
\citet{Glikman2011} & NDWFS \& DLS \& SDSS DR3 & 1450\AA & 3.8-5.2 & $-28.45<M_{\rm 1450}<-21.61$ &  24+314\\
\citet{Ikeda2012} & COSMOS & 1450\AA & 5.07 & $-23.52< M_{\rm 1450}<-22.52$ & 1\\
\citet{Masters2012} & COSMOS & 1450\AA & 3.1-5.0 & $-24.50<M_{\rm 1450}<-21.00$ & 128\\
\citet{Shen2012} & SDSS DR7 & i(z=2) & $\sim$0.3-5 & $-29.25<M_{\rm i(z=2)}<-22.65$ & $\sim$58000\\
\citet{McGreer2013} & SDSS DR7 \& Stripe 82 & 1450\AA & 4.7-5.1 & $-27.98<M_{\rm 1450}<-24.18$ &  103+59\\
\citet{Palanque2013} & BOSS \& MMT & g(z=2) & 0.68-4.0 & $-28.80<M_{\rm g}<-21.60$ & 1367 \\
\citet{Ross2013} & BOSS DR9 & i(z=2)$\rightarrow$1450\AA   & 2.2-3.5   & $-27.53<M_{\rm 1450}<-23.00$ &  22301\\
 & BOSS Stripe82 & i(z=2)$\rightarrow$1450\AA   & 2.2-3.5   & $-28.42<M_{\rm 1450}<-23.59$ &  5476\\
\citet{Giallongo2015}$^{\rm d}$ & CANDELS GOODS-S & 1450\AA & 4.0-6.5 & $-22.50<M_{\rm 1450}<-19.00$ & 22\\
\citet{Kashikawa2015} & UKIDSS-DXS & 1450\AA & 5.85-6.45 & $M_{\rm 1450}\sim-22.84$ & 2\\
\citet{Jiang2016} & SDSS & 1450\AA & 5.7-6.4 & $-29.00<M_{\rm 1450}<-24.50$ & 52\\
\citet{Palanque2016} & SDSS-\Rmnum{4}/eBOSS & g(z=2) & 0.68-4.0 & $-28.80<M_{\rm g(z=2)}<-22.00$ & 13876\\
\citet{Yang2016} & SDSS \& WISE & 1450\AA & 4.7-5.4 & $-29.00<M_{\rm 1450}<-26.80$ & 99\\
\citet{Akiyama2018} & HSC-SSP & 1450\AA & 3.6-4.3 & $-25.88<M_{\rm 1450}<-21.88$ & 1666\\
\citet{Matsuoka2018} & SDSS \& CFHQS \& SHELLQs & 1450\AA & 5.7-6.5 & $-30.00<M_{\rm 1450}<-22.00$ & 110\\
\citet{McGreer2018} & CFHTLS & 1450\AA & 4.7-5.4 & $-26.35<M_{\rm 1450}<-22.90$ & 25\\
\citet{Wang2018} & DELS \& UHS \& WISE & 1450\AA & 6.45-7.05 & $-27.60<M_{\rm 1450}<-25.50$ & 17 \\
\citet{Yang2018} & Deep CFHT Y-band \& SDSS \& VVDS & 1450\AA & 0.5-4.5 & $-27.00<M_{\rm 1450}<-20.50$ & 109\\
\hline
\hline
\bf{Soft X-ray}\\
\hline
\citet{Miyaji2000,Miyaji2001} & ROSAT & 0.5-2\keV & 0.015-4.8 & $10^{41}<L_{\rm 0.5-2}<10^{47}$ & 691\\
\citet{Hasinger2005} & ROSAT \& CDF-N/S & 0.5-2\keV & 0.015-4.8 & $10^{42}<L_{\rm 0.5-2}<10^{48}$ & 2566\\
\citet{Silverman2005} & CHAMP \& ROSAT & 0.5-2\keV & 0.1-5 & $10^{44.5}<L_{\rm 0.5-2}<10^{46}$ & 217\\
\citet{Ebrero2009} & XMS \& RBS \& RIXOS8 \& RIXOS3 \& UDS \& CDF-S & 0.5-2\keV & 0.01-3 & $10^{40.50}<L_{\rm 0.5-2}<10^{46.81}$ & 1009\\
\hline
\hline
\bf{Hard X-ray}\\
\hline
\citet{Ueda2003} & HEAO-1 \& AMSS-n/s \& ALSS \& ASCA \& CDF-N & 2-10\keV & 0.015-3.0 & $10^{41.5}<L_{\rm 2-10}<10^{46.5}$ & 247\\
\citet{Sazonov2004} & RXTE & 3-20\keV & 0.0-0.1 & $10^{41}<L_{\rm 3-20}<10^{46}$ & 77\\
\citet{Barger2005} & CDF-N/S+CLASXS+ASCA & 2-8\keV & $\sim$0.1-1.2 & $10^{42}<L_{\rm 2-8}<10^{46}$ & 601\\
\citet{LaFranca2005} & HELLAS2XMM & 2-10\keV & 0.0-4.0 & $10^{42}<L_{\rm 2-10}<10^{46.5}$ & 508\\
\citet{Nandra2005}$^{\rm b}$ & GWS \& HDF-N & 2-10\keV & 2.7-3.2 & $10^{43}<L_{\rm 2-10}<10^{44.5}$ & 15\\
\citet{Silverman2005} & CHaMP & 0.3-8\keV & 0.2-4.0 & $10^{42}<L_{\rm 0.3-8}<10^{45.5}$ & 368\\
\citet{Aird2008} & GWS \& HDF-N \& Lynx \& LALA CETUS \& EGS1 & 2-10\keV & 2.5-3.5 & $10^{42.5}<L_{\rm 2-10}<10^{48.0}$ & $\sim$1000\\
%\citet{Silverman2008} & ChaMP \& CLASXS \& CDF-N \& CDF-S \& Lockman Hole \& AMSSn & 2-8\keV & $\sim$0-5 & & \\
\hline
\end{tabular}
\raggedright
\\
$^{\rm a}$ The minimum and maximum luminosity that binned luminosity function data ever reach. One should not expect that this luminosity range holds for all redshift bins.\\
$^{\rm b}$ Old observations that are not included in constraining the QLF. There are more recent works using exactly the same or more extended quasar samples.\\
$^{\rm c}$ Data sets presented in a way that an apple-to-apple comparison cannot be made. But we still list them here for references.\\
$^{\rm d}$ \citet{Giallongo2015} data is not included in our fiducial analysis. 
\caption{Observations compiled.}
\label{tab:observations}
\end{table*}

\begin{table*}
\contcaption{ }
\label{tab:observations_continued}
\centering
\begin{tabular}{
p{0.23\textwidth}|p{0.22\textwidth}|p{0.10\textwidth}|p{0.07\textwidth}|p{0.20\textwidth}|p{0.05\textwidth}}
\citet{Ebrero2009} & XMS \& AMSS \& CDF-S & 2-10\keV & 0.01-3 & $10^{41.83}<L_{\rm 2-10}<10^{45.87}$ & 435 \\
\citet{Aird2010} & CDF-S \& CDF-N \& AEGIS \& ALSS \& AMSS & 2-10\keV & 0-3.5 & $10^{41.3}<L_{\rm 2-10}<10^{45.8}$ & 130\\
\citet{Fiore2012} & CDF-S & 2-10\keV & 3-7.5 & $10^{42.75}<L_{\rm 2-10}<10^{44.5}$ & 54\\
\citet{Ueda2014} & BAT9 \& MAXI7 \& AMSS \& ALSS \& SXDS \& LH/XMM \& H2X \& XBS \& CLASXS \& CLANS \& CDF-N \& CDF-S \& ROSAT surveys & 2-10\keV & 0.002-5 & $10^{41.8}<L_{\rm 2-10}<10^{46.5}$ & 4039\\
\citet{Aird2015a} & CDF-S \& CDF-N \& EGS \& COSMOS \& Bo\"{o}tes field \& AMSS \& ALSS \& ROSAT surveys & 2-10\keV & 0-7 & $10^{38.25}<L_{\rm 2-10}<10^{47.5}$ & 2957+4351\\
\citet{Aird2015b} & NuSTAR & 10-40\keV & 0.1-3 & $10^{42.75}<L_{\rm 10-40}<10^{45.75}$ & 94\\
\citet{Miyaji2015} & Swift BAT \& CDF-S & 2-10\keV & 0.015-5.8 & $10^{42}<L_{\rm 2-10}<10^{46}$ & $\sim$3200\\
\citet{Khorunzhev2018} & XMM-NEWTON Serendipitous & 2-10\keV & 3.0-5.1 & $10^{45}<L_{\rm 2-10}<7.5\times10^{45}$ & 101\\
\hline
\hline
\bf{Near-IR \& Mid-IR}\\
\hline
\citet{Brown2006} & NDWFS Bo\"{o}tes field & $8\micron$ & $\sim$1-5 & $10^{45}<\nu L_{\rm 8\micron}<10^{47}$ & 183\\
\citet{Matute2006} & RMS \& ELIAS \& HDF-N/S & $15\micron$ & $\sim$0.1-1.2 & $10^{42}<\nu L_{\rm 15\micron}<10^{47}$ & 148\\
\citet{Assef2011} & NDWFS Bo\"{o}tes field & J & 0-5.85 & $-28.5<M_{\rm J}<-18.5$ & 1838\\
\citet{Lacy2015} & SWIRE \& XFLS & $5\micron$ & 0.05-3.8 & $10^{43.5}<\nu L_{\rm 5\micron}<10^{46.5}$ & 479\\
\citet{Singal2016}$^{\rm c}$ & SDSS DR7 \& WISE & $22\micron$ & 0.08-4.97 &  & >20000\\
\hline
\hline
\bf{Emission Lines}\\
\hline
\citet{Hao2005} & SDSS (main galaxy sample) & H$\alpha$ & 0-0.33 & $10^{5}{\rm L}_{\odot}<L_{\rm H\alpha}<10^{9}{\rm L}_{\odot}$ & $\sim$3000\\
 & . . . & $\rm [O_{\,\,\Rmnum{2}}]$ & . . . & $10^{5}{\rm L}_{\odot}<L_{\rm [O_{\,\,\Rmnum{2}}]}<10^{8}{\rm L}_{\odot}$ & . . .\\
 & . . . & $\rm [O_{\,\,\Rmnum{3}}]$ & . . . & $10^{5}{\rm L}_{\odot}<L_{\rm [O_{\,\,\Rmnum{3}}]}<10^{9}{\rm L}_{\odot}$ & . . .\\
\hline
\hline
\end{tabular}
\end{table*}

\section{Posterior distribution in the global fit}

\begin{figure*}
    \centering
    \includegraphics[width=0.99\textwidth]{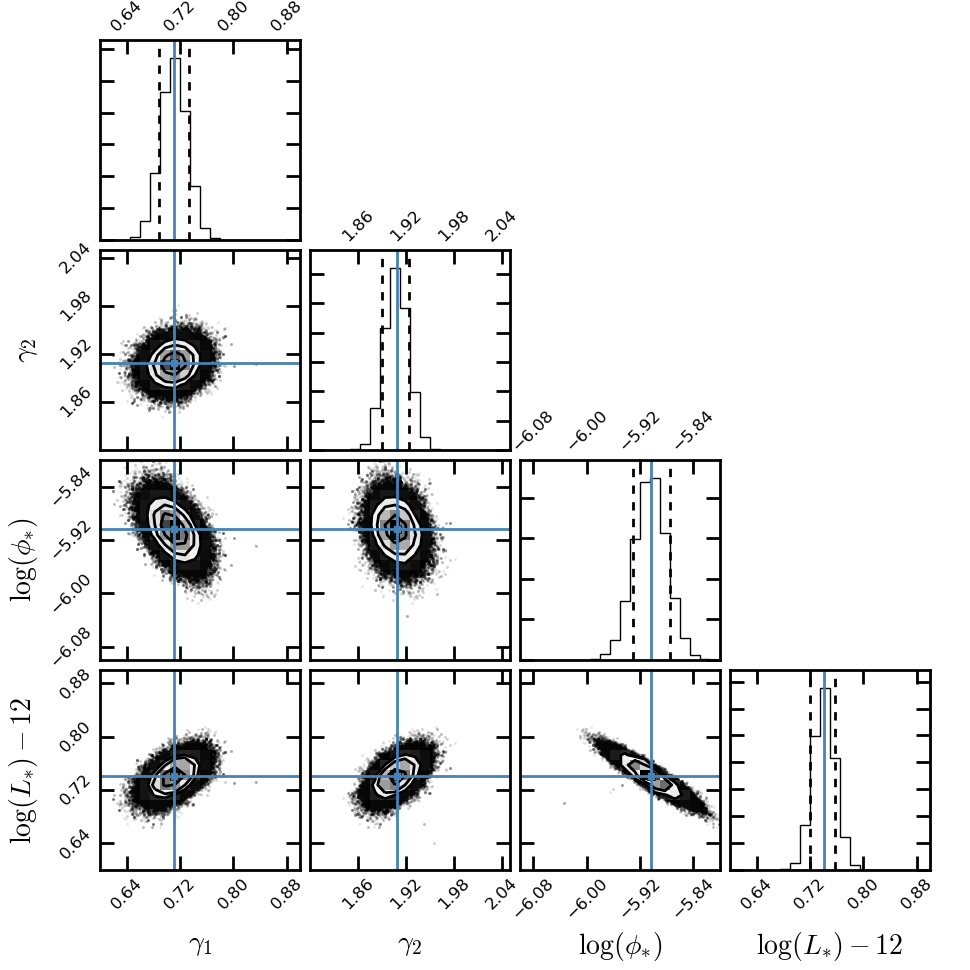}
    \caption{\textbf{Posterior distribution of the four double-power-law parameters at $z=5$ in our global fit A~(see Table~\ref{tab:fits} for details).} The global fit A is originally done in a $11$ dimension parameter space of the QLF evolution model. Here, we project the posterior distribution onto the $4$ dimension parameter space of the double power-law function at $z=5$. The blue lines and squares indicate the best-fit values of the global fit A at this redshift. The black dashed lines indicate $1\sigma$ dispersions. Similar behaviour of the posterior distribution is seen at other redshifts.}
    \label{fig:posterior}
\end{figure*}

In Figure~\ref{fig:posterior}, we show the posterior distribution of the four double-power-law parameters at $z=5$ in our global fit A~(see Table~\ref{tab:fits} for details). The global fit A is originally done in a $11$ dimension parameter space of the QLF evolution model. Here, we project the posterior distribution onto the $4$ dimension parameter space of the double power-law function at $z=5$. 

\section{Code and data}
\label{app:code}
The code and data used in this work are publicly available at 
\href{https://bitbucket.org/ShenXuejian/quasarlf/src/master/}{https://bitbucket.org/ShenXuejian/quasarlf/src/master/}. It includes the compiled observational datasets of the QLF, the mean SED model, the pipeline for bolometric and extinction corrections, the global best-fit bolometric QLF models and all other code for the analysis done in this paper. 

%%%%%%%%%%%%%%%%%%%%%%%%%%%%%%%%%%%%%%%%%%%%%%%%%%

% Don't change these lines
\bsp	% typesetting comment
\label{lastpage}
\end{document}